\documentclass[preprint,12pt]{elsarticle}
\usepackage{amssymb,times,graphicx}

\usepackage{ulem}

\journal{Journal of Computational and Applied Mathematics}

\begin{document}

\begin{frontmatter}

\title{A self-adaptive moving mesh method for the Camassa-Holm equation}


\author{Bao-Feng Feng$^{1}$, 
Ken-ichi Maruno$^1$, and Yasuhiro Ohta$^2$}

\address{$^1$~Department of Mathematics,
The University of Texas-Pan American,
Edinburg, TX 78539-2999, USA
}
\address{$^2$~Department of Mathematics,
Kobe University, Rokko, Kobe 657-8501, Japan
}

\begin{abstract}
A self-adaptive moving mesh method is proposed for the numerical simulations
of the Camassa-Holm equation. It is an integrable scheme in the sense that
it possesses the exact $N$-soliton solution.
It is named a self-adaptive moving mesh method, because the non-uniform mesh is
driven and adapted automatically by the solution. Once the non-uniform
mesh is evolved, the solution is determined by solving a tridiagonal linear
system. Due to these two superior features of the method,
several test problems give very satisfactory results
even if by using a small number of grid points.
\par
\kern\bigskipamount\noindent
\today
\end{abstract}

\begin{keyword}
The Camassa-Holm equation \sep integrable semi-discretization
\sep peakon and cupson solutions
\sep self-adaptive moving mesh method

\MSC 65M06 \sep 35Q58 \sep 37K40
\end{keyword}

\end{frontmatter}

\section{Introduction}
Since its discovery \cite{CH}, the Camassa-Holm (CH) equation
\begin{equation}
w_T+2\kappa w_X-w_{TXX}+3ww_X=2w_Xw_{XX}+ww_{XXX}.\label{CH-eq}
\end{equation}
has attracted considerable interest because it describes
unidirectional propagation of shallow water waves on
a flat bottom.
%!
It also appeared in a mathematical search of recursion operators
connected with the integrable partial differential
equations \cite{FF}.
By virtue of asymptotic procedures, the CH equation was
reconfirmed as a valid approximation to
the governing equation for shallow water waves
\cite{Johnson02,Constantin08}.
The CH equation also arises as a model for
water waves moving over an underlying shear flow \cite{Johnson03},
in the study of a certain
non-Newtonian fluids \cite{Busuioc}, and as a model for
nonlinear waves in cylindrical hyperelastic rods \cite{Dai}.
The CH equation is completely integrable (see \cite{CH} for
the Lax pair formulation and \cite{Constantin,Constantin2} for the inverse
scattering transform), and it has various exact
solutions such as solitons, peakons, and cuspons.
When $\kappa=0$, the CH equation
admits peakon solutions which are represented by piecewise
functions \cite{CH,CHH,Beals}.
When $\kappa\not=0$, cusped soliton (cuspon) solutions, as well as
smooth soliton solutions, were found by several authors.
\cite{Schiff,Kraenkel2,Johnson,Li,Parker,Parker2,DaiLi,Matsuno}.

Several numerical schemes have been proposed for the
CH equation in the literature. These include
a pseudospectral method \cite{Kalisch},
finite difference schemes \cite{Holden1,Coclite1}, a finite volume
method \cite{Artebrant},
finite element methods \cite{Chiwang,Matsuo,Matsuo2},
multi-symplectic methods
\cite{Cohen}, and
a particle method in terms of characteristics
based on the multi-peakon solution \cite{Camassa1,Lee1,Lee2,Lee3,Holden2}.
We comment that the schemes in \cite{Holden1,Coclite1} and
in \cite{Cohen} can handle peakon-antipeakon interactions.
%!
However, it still remains a challenging problem for the
numerical integration of the CH equation due to
the singularities of cuspon and
peakon solutions.

In the present paper, we will study an integrable difference scheme for
the CH equation (\ref{CH-eq}) based on an integrable semi-discrete
CH equation proposed by the authors \cite{Ohta}.
The scheme consists of an algebraic equation for the solution and
the non-uniform mesh for a fixed time, and a time evolution
equation for the mesh. Since the mesh is automatically driven and
adapted by the solution, we name it a self-adaptive moving mesh method hereafter.

As a matter of fact, Harten and Hyman has proposed a self-adjusting grid method for
one-dimensional hyperbolic problems \cite{HH83}. Since
then, there has been significant progress in developing adaptive
mesh methods for PDEs \cite{Miller, Dorfi,Brackbill, Huang,
Stockie,TaoTang}. These methods have been successfully
applied to a variety of physical and engineering problems with
singular or nearly singular solutions developed in fairly localized
regions, such as shock waves, boundary layers, detonation waves,
etc. Recently, an adaptive unwinding method was proposed for the
CH equation \cite{Artebrant}.
The method is high resolution and stable. However, in order to achieve
a good accuracy,
a large number of grid points ($=4096$) has to be used.
In addition, the designed method is
only suitable for the single peakon propagation and peakon-peakon
interactions, not for the
%!
peakon-antipeakon interaction. As shown subsequently, the self-adaptive
moving mesh method gives
accurate results by using a small number of grid points ($\approx 100$)
for some challenging test problems.

The remainder of this paper is organized as follows.
In Section 2,
%!
we present the self-adaptive moving mesh method and
show it is consistent with the CH equation as the mesh size
approaches to zero. Two time advancing methods in implementing
the self-adaptive moving mesh method
are presented in Section 3. In Section 4,
several numerical experiments,
including the propagations of ``peakon'' and ``cuspon'' solutions,
cuspon-cuspon and soliton-cuspon collisions,
are shown.
The concluding remarks are addressed in Section 5.

\section{A self-adaptive moving mesh method for the Camassa-Holm equation}
\noindent
It is shown in \cite{Ohta} that the CH equation can be derived from the bilinear equations of a deformation of the modified KP hierarchy
\begin{equation}
\left\{\begin{array}{l}\displaystyle
-\left(\frac{1}{2}D_tD_x-1\right)f\cdot f=gh\,,
\\[5pt]
2cff=(D_x+2c)g\cdot h\,,
\\[5pt]
-2ff=(D_tD_x+2cD_t-2)g\cdot h\,,
\end{array}\right.\label{ch-bilinear}
\end{equation}
through the hodograph transformation
\begin{equation}
\left\{\begin{array}{l}
X=2cx+\log \frac{g}{h}\,,
\\
T=t\,,
\end{array}\right.
\end{equation}
and the dependent variable transformation
$$
w=\left(\log\frac{g}{h}\right)_t\,.
$$
Here $c=1/\kappa$, $D_x$ and $D_t$ are Hirota's D-operator defined as
$$
D^n_x f\cdot g = \left.\left(\frac{\partial}{\partial x} - \frac{\partial}{\partial y}\right)^n f(x)g(y)\right|_{y=x}.
$$

It is proved in \cite{Ohta} that the bilinear equations (\ref{ch-bilinear}) admit a determinant solution $f=\tau_0$, $g=\tau_{-1}$, $h=\tau_1$, where $\tau_n$ is a Casorati-type determinant of any size. By discretizing the $x$-direction with an uniform mesh size $a$, the following bilinear equations
\begin{equation}
\left\{\begin{array}{l}\displaystyle
-2\left(\frac{1}{a}D_t-1\right)f_{k+1}\cdot f_k=g_{k+1}h_k+g_kh_{k+1}\,,
\\[5pt]
2acf_{k+1}f_k=(1+ac)g_{k+1}h_k-(1-ac)g_kh_{k+1}\,,
\\[5pt]
-2af_{k+1}f_k=((1+ac)D_t-a)g_{k+1}\cdot h_k-((1-ac)D_t+a)g_k\cdot h_{k+1}\,,
\end{array}\right. \label{semi-ch-bilinear}
\end{equation}
admits Casorati-type determinant solution with discrete index which is presented afterwards.
Starting from Eq.(\ref{semi-ch-bilinear}), a semi-discrete CH equation
\begin{equation}
\left\{\begin{array}{l}\displaystyle
-2\left(\frac{w_{k+1}-w_k}{\delta_k}-\frac{w_k-w_{k-1}}{\delta_{k-1}}\right)
+\delta_k\frac{w_{k+1}+w_k}{2}
+\frac{\delta_k}{c}
\frac{\displaystyle 1-\frac{4a^2c^2}{\delta_k^2}}{1-a^2c^2}
\\[5pt]\displaystyle
\hskip100pt
+\delta_{k-1}\frac{w_k+w_{k-1}}{2}
+\frac{\delta_{k-1}}{c}
\frac{\displaystyle 1-\frac{4a^2c^2}{\delta_{k-1}^2}}{1-a^2c^2}=0\,,
\\[5pt]\displaystyle
\frac{d\,\delta_k}{d\,t}=\left(1-\frac{\delta_k^2}{4}\right)(w_{k+1}-w_k)\,
\end{array}\right.\label{d-ch-final}
\end{equation}
was proposed (see the details in \cite{Ohta}). Here the solution $w(X_k,t)$ is approximated by $w_k(t)$ at the grid points $X_k$ ($k=1, \cdots, N$). The mesh $\delta_k=X_{k+1}-X_k$ is a discrete analogue of the hodograph transformation from the $x$-domain with uniform mesh size $a$ to $X$-domain. As is seen, it is non-uniform and time-dependent.

The semi-discrete CH equation (\ref{d-ch-final}) can be rewritten as
\begin{equation}
\left\{\begin{array}{l}\displaystyle \Delta^2
w_k=\frac{1}{\delta_k}M\left(\delta_kMw_k+ \frac{1}{c\delta_k}
\frac{\displaystyle \delta_k^2/c^2-4a^2}{1/c^2-a^2}\right)\,,
\\[5pt]\displaystyle
\frac{d\,\delta_k}{d\,t}=\left(1-\frac{\delta_k^2}{4}\right)\delta_k\Delta w_k\,
\end{array}\right. \label{semi-d-ch}
\end{equation}
by introducing a forward difference
operator and an average operator $\Delta$ and $M$
$$
\Delta F_k=\frac{F_{k+1}-F_{k}}{\delta_{k}}\,, \qquad
MF_k=\frac{F_{k}+F_{k+1}}{2}\,.
$$

%!
In the present paper, Eq. (\ref{d-ch-final}) or Eq. (\ref{semi-d-ch}) is used as a numerical scheme for the CH
equation (\ref{CH-eq}). It is shown to be integrable in \cite{Ohta} in the sense that it possesses $N$-soliton solution which, in the continuous limit, approaches $N$-soliton solution of the CH equation.
The $N$-soliton solution is of the form
\begin{equation}\label{Dept_tra}
    w_k=\left(\log\frac{g_k}{h_k}\right)_t\,,
\end{equation}
with
$$
f_k=\tau_0(k)\,, \qquad g_k=\tau_1(k)\,, \qquad h_k=\tau_{-1}(k)\,,
$$
$$
\tau_n(k)=\left|\matrix{ \psi_1^{(n)} &\psi_1^{(n+1)} &\cdots
&\psi_1^{(n+N-1)} \cr \psi_2^{(n)} &\psi_2^{(n+1)} &\cdots
&\psi_2^{(n+N-1)} \cr \vdots            &\vdots              &
&\vdots                \cr \psi_N^{(n)} &\psi_N^{(n+1)} &\cdots
&\psi_N^{(n+N-1)}}\right|\,
$$
where
$$
\psi_i^{(n)}=a_{i,1}(p_i-c)^n(1-ap_i)^{-k}e^{\xi_i}
+a_{i,2}(-p_i-c)^n(1+ap_i)^{-k}e^{\eta_i}\,,
$$
$$
\xi_i=\frac{1}{p_i-c}t+\xi_{i0}\,, \qquad
\eta_i=-\frac{1}{p_i+c}t+\eta_{i0}\,.
$$

Next, let us show that in the continuous limit, $a\to0$
($\delta_k\to0$), the proposed scheme is consistent with the CH
equation. To this end, the equation (\ref{semi-d-ch}) is rewritten as

$$
\left\{\begin{array}{l}\displaystyle
\frac{-2}{\delta_k+\delta_{k-1}} \left(\Delta w_{k}-\Delta
w_{k-1}\right) +\frac{\delta_kM w_{k}}{\delta_k+\delta_{k-1}}
+\frac{\delta_{k-1}M w_{k-1}}{\delta_k+\delta_{k-1}}
+\frac{1}{c(1-a^2c^2)}\\[15pt]\displaystyle
\qquad \qquad =\frac{4a^2c}{1-a^2c^2}\frac{1}{\delta_k\delta_{k-1}}\,,
\\[15pt]\displaystyle
\partial_t\delta_k=\left(1-\frac{\delta_k^2}{4}\right)(w_{k+1}-w_k)\,.
\end{array}\right.
$$

By taking logarithmic derivative of the first equation, we get
$$
\left\{\begin{array}{l}\displaystyle
\frac{\displaystyle\partial_t\left\{\frac{2}{\delta_k+\delta_{k-1}}
\left(\Delta w_{k}-\Delta w_{k-1}\right) -\frac{\delta_k M
w_{k}}{\delta_k+\delta_{k-1}} -\frac{\delta_{k-1}M
w_{k-1}}{\delta_k+\delta_{k-1}}\right\}}
{\displaystyle\frac{2}{\delta_k+\delta_{k-1}} \left(\Delta
w_{k}-\Delta w_{k-1}\right) -\frac{\delta_k M w_{k}
}{\delta_k+\delta_{k-1}}-\frac{\delta_{k-1}M
w_{k-1}}{\delta_k+\delta_{k-1}} -\frac{1}{c(1-a^2c^2)}}
\\[15pt]\displaystyle
\qquad \qquad
=-\frac{\partial_t\delta_k}{\delta_k}
-\frac{\partial_t\delta_{k-1}}{\delta_{k-1}}\,,
\\[30pt]\displaystyle
\partial_t\delta_k=\left(1-\frac{\delta_k^2}{4}\right)(w_{k+1}-w_k)\,.
\end{array}\right.
$$
Thus, we have
$$
\begin{array}{l}
\frac{\displaystyle\partial_t\left\{\frac{2}{\delta_k+\delta_{k-1}}
\left(\Delta w_{k}-\Delta w_{k-1}\right) -\frac{\delta_kM
w_{k}}{\delta_k+\delta_{k-1}} -\frac{\delta_{k-1}M
w_{k-1}}{\delta_k+\delta_{k-1}}\right\}}
{\displaystyle\frac{2}{\delta_k+\delta_{k-1}} \left(\Delta
w_{k}-\Delta w_{k-1}\right) -\frac{\delta_k M
w_{k}}{\delta_k+\delta_{k-1}} -\frac{\delta_{k-1}M
w_{k-1}}{\delta_k+\delta_{k-1}} -\frac{1}{c(1-a^2c^2)}}
\\[30pt]\displaystyle\qquad
=-\left(1-\frac{\delta_k^2}{4}\right)\Delta w_{k}
-\left(1-\frac{\delta_{k-1}^2}{4}\right)\Delta w_{k-1}\,.
\end{array}
$$

The dependent variable $w$ is a function of $k$ and $t$, and we regard
them as a function of $X$ and $T$, where $X$ is the space coordinate of
the $k$-th lattice point and $T$ is the time, defined by
$$
X=X_0+\sum_{j=0}^{k-1}\delta_j\,,\qquad T=t\,.
$$
Then in the continuous limit, $a\to0$ ($\delta_k\to0$), we have
$$
\Delta w_{k} \to w_X\,,
\quad \Delta w_{k-1} \to w_X\,, \quad Mw_k \to w\,, \quad \Delta w_{k-1} \to w_X \,,
$$
and
$$
\frac{2}{\delta_k+\delta_{k-1}}
\left(\Delta w_{k}-\Delta w_{k-1}\right) \to w_{XX}.
$$
Further, from
$$
\frac{\partial X}{\partial t}
=\frac{\partial X_0}{\partial t}
+\sum_{j=0}^{k-1}\frac{\partial\delta_j}{\partial t}
=\frac{\partial X_0}{\partial t}
+\sum_{j=0}^{k-1}\left(1-\frac{\delta_j^2}{4}\right)(w_{j+1}-w_j)
\to w\,,
$$
we have
$$
\partial_t=\partial_T+\frac{\partial X}{\partial t}\partial_X
\to\partial_T+w\partial_X\,,
$$
where the origin of space coordinate $X_0$ is taken so that
$\displaystyle\frac{\partial X_0}{\partial t}$ cancels $w_0$.
Then the above semi-discrete CH equation converges to the CH equation
$$
    \frac{(\partial_T+w\partial_X)(w_{XX}-w)} {\displaystyle
w_{XX}-w-\frac{1}{c}}=-2w_X\,,
$$
i.e.
\begin{equation}
(\partial_T+w\partial_X)(w_{XX}-w)
=-2w_X\left({\displaystyle
w_{XX}-w-\frac{1}{c}}\right)\,.
\end{equation}
Setting $c=1/\kappa$, we obtain the the CH equation (\ref{CH-eq}).

Note that, in our previous paper \cite{Ohta},
we put $c=1/\kappa^2$ which gives an alternative form of the CH equation
\begin{equation}
w_T+2\kappa^2 w_X-w_{TXX}+3ww_X=2w_Xw_{XX}+ww_{XXX}.
\end{equation}
It is shown that they are equivalent under the scaling transformation $w\to \kappa w$, $T\to T/\kappa$. In the present paper,
for the convenience in comparing of our results with other papers \cite{Johnson,Li,Parker,Parker2,DaiLi,Matsuno,MatsunoPeakon}, we
set $c=1/\kappa$.

\section{Implementation of the self-adaptive moving mesh method}
%!
In this Section, we will discuss how to implement the self-adaptive moving mesh method in
actual computations. Generally, given an arbitrary initial
condition $w(X,0)=w_0(X)$, the initial non-uniform mesh $\delta_k$
can be obtained by solving the nonlinear algebraic equations by
Newton's iteration method. However, for the propagation or
interaction of solitons or cuspons, which are challenging
problems numerically, the initial condition $w_k$ can be calculated
by (\ref{Dept_tra}) from $g_k$ and $h_k$ by putting $t=0$,
which are obtainable from the
corresponding determinant solutions. The initial non-uniform mesh
$\delta^0_k$ can also be calculated by \cite{Ohta}
\begin{equation}\label{initial_delta}
\delta^0_k=2\frac{(1+ac)g_{k+1} h_{k}-(1-ac)g_k h_{k+1}
}{(1+ac)g_{k+1} h_{k}+(1-ac)g_k h_{k+1}}.
\end{equation}

On the other hand, once the non-uniform mesh $\delta_k$ is known, the
solution $w_k$ can be easily obtained by solving a tridiagonal linear system
based on the first equation of the scheme.
\begin{equation}\label{tri-diagonal}
    a_l w^{n+1}_{l-1}+ b_l w^{n+1}_{l} + c_l w^{n+1}_{l+1} = d_l,
\end{equation}
where
$$
a_l= 0.5\delta^{n+1}_{k-1}-\frac{2}{\delta^{n+1}_{k-1}}; \ \
b_l=0.5(\delta^{n+1}_{k-1}+\delta^{n+1}_{k})+
\frac{2}{\delta^{n+1}_{k-1}}+\frac{2}{\delta^{n+1}_{k}}; \ \  c_l=
0.5\delta^{n+1}_{k}-\frac{2}{\delta^{n+1}_{k}};
$$
and
$$
d_l= \frac{4a^2c}{1-a^2c^2}
\left(\frac{1}{\delta^{n+1}_k}+\frac{1}{\delta^{n+1}_{k-1}}\right) -
\frac{\delta^{n+1}_{k-1}+\delta^{n+1}_{k}}{c(1-a^2c^2)}\,.
$$

In regard to the evolution of $\delta_k$, we propose two time advancing
methods. The first is the
modified forward Euler method, where we assume $w_k$ remains unchanged
in one
time step.
Integrating once, we have
\begin{equation}\label{delta}
\delta^{n+1}_k=2\frac{c^n_k e^{(w^n_{k+1}-w^n_k)\Delta t}-1}{c^n_k
e^{(w^n_{k+1}-w^n_k)\Delta t}+1},
\end{equation}
where $c^n_k=(2+\delta^n_k)/(2-\delta^n_k)$.
The second is the classical 4th-order Runge-Kutta method, where $w_k$
can be viewed as a function of
$\delta_k$ by solving the above tridiagonal linear system. Therefore, in
one time step,
we have to solve tridiagonal linear system four times.

In summary, the numerical computation in one time-step only
involves a ODE solver for non-uniform mesh and
a tridiagonal linear system solver. Hence, the computation cost
is much less than other
existing numerical methods.
%As shown subsequently, numerical solutions with high accuracy are
%%obtained by using small number of grid points.
A Matlab code is made to perform all the computations. Iterative
methods, for instance, the bi-conjugate gradient method {bicg in
Matlab} are used to solve the tridiagonal system.

%!
For the sake of numerical experiments in the subsequent section,
we list exact one- and two- soliton/cuspon and peakon solutions.

{\bf (1). One soliton/cuspon solution}: The $\tau$-functions for the one
soliton/cuspon solution are
\begin{equation}
g \propto 1 \pm \left(\frac{c-p_1}{c+p_1} \right)  e^{\xi_1}\,, \qquad
  h \propto 1 \pm \left(\frac{c+p_1}{c-p_1}\right) e^{\xi_1}\,,
\end{equation}
with $\xi_1=p_1(2x-v_1t-x_{10})$, $v_1=2/(c^2-p_1^2)$.
This leads to a solution
\begin{equation}\label{1-soliton-cuspon}
    w(x,t)= \frac{2p_1^2cv_1}{(c^2+p_1^2)\pm (c^2-p_1^2) \cosh \xi_1}\,,
\end{equation}
\begin{equation}\label{Hodo1}
   X=2cx + \log\left(\frac{g}{h}\right)\,,\quad T=t\,,
\end{equation}
where the positive case in Eq.(\ref{1-soliton-cuspon}) stands for
the one smooth soliton solution when $p_1 < c$, while the negative case in
Eq.(\ref{1-soliton-cuspon}) stands for the
one-cuspon solution when
$p_1>c$. Otherwise, the solution is singular. Thus
Eq.(\ref{1-soliton-cuspon}) for nonsingular cases can be expressed
by
\begin{equation}
    w(x,t)= \frac{2p_1^2cv_1}{(c^2+p_1^2)+ |c^2-p_1^2| \cosh \xi_1}\,.
\end{equation}

Similarly, for the semi-discrete case, we have
\begin{equation}
g_k \propto 1 + \left| \frac{c-p_1}{c+p_1} \right| \left(
\frac{1+ap_1}{1-ap_1} \right)^k e^{\xi_1}\,, \qquad
 h_k \propto 1 + \left| \frac{c+p_1}{c-p_1} \right|
\left( \frac{1+ap_1}{1-ap_1}
\right)^k e^{\xi_1}\,,
\end{equation}
with $\xi_1=p_1(-v_1t-x_{10})$, resulting in a solution of the form
\begin{equation}\label{1-discrete}
    w_k(t)= \frac{2p_1^2cv_1}
{(c^2+p_1^2)+\frac{|c^2-p_1^2|}{2}
\left[
\left( \frac{1+ap_1}{1-ap_1}
\right)^{-k}e^{-\xi_1}+\left( \frac{1+ap_1}{1-ap_1}
\right)^{k}e^{\xi_1}
 \right]
},
\end{equation}
in conjunction with a transform between an uniform mesh $a$
and a non-uniform mesh
$$
\delta_k=2\frac{(1+ac)g_{k+1}h_k-(1-ac)g_kh_{k+1}}
{(1+ac)g_{k+1}h_k+(1-ac)g_kh_{k+1}}.
$$
%Equation (\ref{1-discrete}) corresponds to the 1-soliton solution when $p<c$,
%the 1-cuspon solution when $p>c$.

{\bf (2). Two soliton/cuspon solutions}:  The $\tau$-functions for the
two
soliton/cuspon solution are
\begin{eqnarray*}
  g \propto 1 + \left| \frac{c-p_1}{c+p_1} \right|e^{\xi_1} +
  \left| \frac{c-p_2}{c+p_2} \right|e^{\xi_2} +
  \left| \frac{(c-p_1)(c-p_2)}{(c+p_1)(c+p_2)} \right|
\left( \frac{p_1-p_2}{p_1+p_2}\right)^2
  e^{\xi_1+\xi_2}\,,
\end{eqnarray*}
\begin{eqnarray*}
  h \propto 1 + \left| \frac{c+p_1}{c-p_1} \right|e^{\xi_1} +
  \left| \frac{c+p_2}{c-p_2} \right|e^{\xi_2} +
  \left| \frac{(c+p_1)(c+p_2)}{(c-p_1)(c-p_2)} \right|
\left( \frac{p_1-p_2}{p_1+p_2}\right)^2
  e^{\xi_1+\xi_2}\,,
\end{eqnarray*}
with $\xi_1=p_1(2x-v_1t-x_{10})$, $\xi_2=p_2(2x-v_2t-x_{20})$,
$v_1=2/(c^2-p_1^2)$, $v_2=2/(c^2-p_2^2)$. The parametric
solution can be calculated through
\begin{equation}\label{2-soliton-cuspon}
    w(x,t) = \left (\log\frac{g}{h} \right)_t, \qquad X=2cx +
    \log\left(\frac gh\right), \quad T=t\,,
\end{equation}
whose form is complicated and is omitted here. Note that the above
expression includes the two-soliton solution ($p_1<c$, $p_2<c$),
the two-cuspon solution ($p_1>c$, $p_2>c$), or the soliton-cuspon
solution ($p_1<c$, $p_2>c$).

Similarly, for the semi-discrete case, we have
\begin{eqnarray*}
&&   g_k \propto 1 + \left| \frac{c-p_1}{c+p_1} \right| \left(
\frac{1+ap_1}{1-ap_1} \right)^ke^{\xi_1} +
  \left| \frac{c-p_2}{c+p_2} \right|
\left( \frac{1+ap_2}{1-ap_2} \right)^k
e^{\xi_2}\\
&&\quad  + \left| \frac{(c-p_1)(c-p_2)}{(c+p_1)(c+p_2)}
\right| \left( \frac{p_1-p_2}{p_1+p_2}\right)^2 \left(
\frac{1+ap_1}{1-ap_1} \right)^k \left( \frac{1+ap_2}{1-ap_2}
\right)^k
  e^{\xi_1+\xi_2}\,,
\end{eqnarray*}
\begin{eqnarray*}
&&
  h_k \propto 1 + \left| \frac{c+p_1}{c-p_1} \right|
\left( \frac{1+ap_1}{1-ap_1} \right)^k e^{\xi_1} +
  \left| \frac{c+p_2}{c-p_2} \right|
\left( \frac{1+ap_2}{1-ap_2} \right)^k
e^{\xi_2} \\
&& \quad +
  \left| \frac{(c+p_1)(c+p_2)}{(c-p_1)(c-p_2)} \right|
\left( \frac{p_1-p_2}{p_1+p_2}\right)^2 \left( \frac{1+ap_1}{1-ap_1}
\right)^k \left( \frac{1+ap_2}{1-ap_2} \right)^k
  e^{\xi_1+\xi_2}\,,
\end{eqnarray*}
with $\xi_1=p_1(-v_1t-x_{10})$, $\xi_2=p_2(-v_2t-x_{20})$.
The solution can be calculated through
\begin{equation}
    w(x,t) = \left (\log\frac{g_k}{h_k} \right)_t,
\end{equation}
with a transform
\begin{equation}
\delta_k=2\frac{(1+ac)g_{k+1}h_k-(1-ac)g_kh_{k+1}}
{(1+ac)g_{k+1}h_k+(1-ac)g_kh_{k+1}}\,.
\end{equation}
Again, the explicit form of the solution is complicated and is omitted
here.

{\bf (3). Peakon solutions}:
In the continuous CH equation, it is possible to
construct peakon solutions from soliton solutions by taking
the peakon limit \cite{CHH,LiOlver,Schiff,Johnson,
Parker,PKMatsuno,MatsunoPeakon}.

\begin{figure}[htbp]
\centerline{
\includegraphics[scale=0.5]{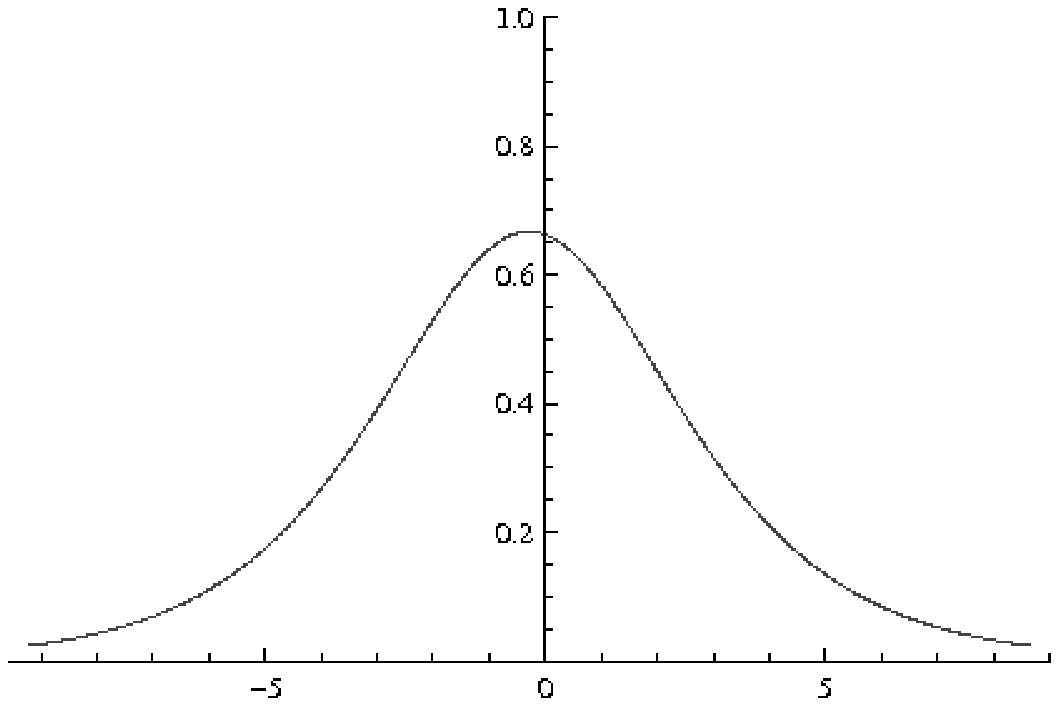}\quad
\includegraphics[scale=0.5]{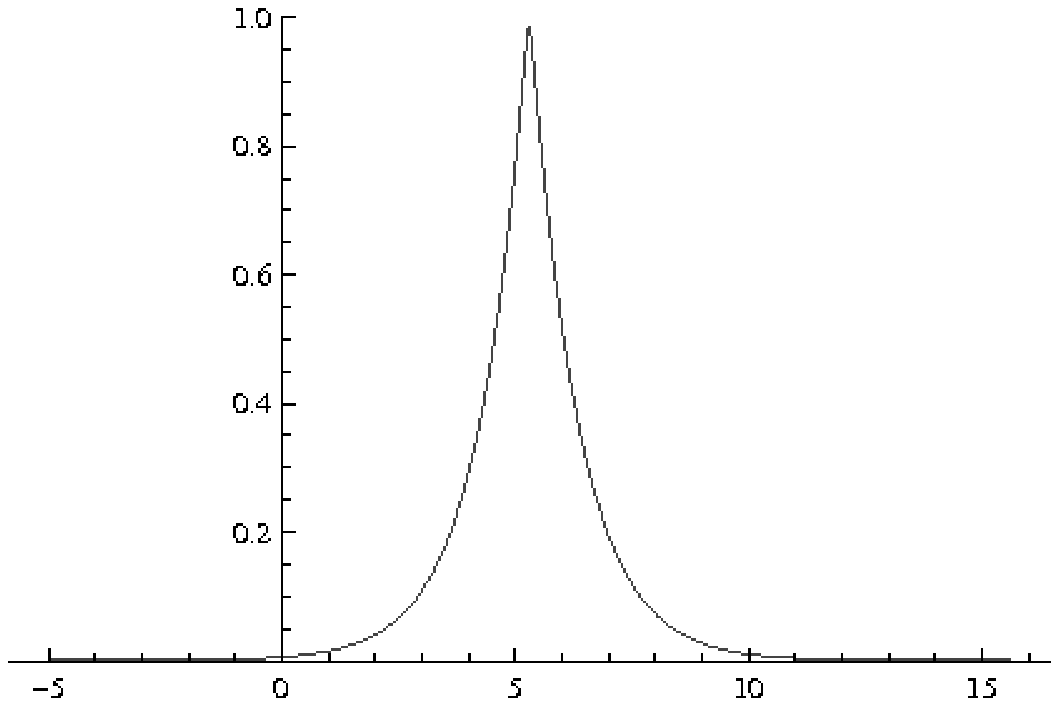}}
\caption{1-soliton solution for the CH equation:
the left: $p_1=0.5, c=1$; the right (close to the peakon limit):
$p_1=99, c=100$.} \label{peakon-limit}
\end{figure}

For the continuous case, we can express
the 1-soliton solution as
\[
w=\frac{2p_1^2\kappa v_1}
{1+p_1^2\kappa^2+(1-p_1^2\kappa^2){\rm cosh}\xi_1}\,,
\]
where $\kappa=\frac{1}{c}$, $v_1=2\kappa^2/(1-p_1^2\kappa^2)$,
$\xi_1=p_1\kappa (2x/\kappa-(v_1/\kappa)t-x_{10}/\kappa)$.
Taking the peakon limit
$\kappa\to 0$, $p_1\kappa\to 1$, $v_1={\rm const.}$,
%%%%To perform the peakon limit, we can set
%%%%\[
%%%%p_1\kappa \sim 1-\frac{\kappa^2}{v_1}.
%%%%\]
%%%%After some computation,
%%%%we will get
%%%%\[
%%%%u=\kappa w\sim \frac{v_1}
%%%%{1+\frac{\kappa^2}{v_1}{\rm cosh}\xi_1}\,,???
%%%%\]
%%%%where $\xi_1=2x/\kappa-v_1\tilde{t}-\tilde{x}_{10}$,
%%%%$\tilde{t}=t/\kappa$, $\tilde{x}_{10}=x_{10}/\kappa$,
the solution $(X(x,t),w(x,t))$, where
$X(x,t)=2x/\kappa+\log \frac{g}{h}$, gives the 1-peakon solution
\cite{MatsunoPeakon}.
In Fig.\ref{peakon-limit},
one can see that the 1-soliton solution approaches to the 1-peakon
solution as $\kappa$ approaches to 0.

\begin{figure}[htbp]
\centerline{
\includegraphics[scale=0.5]{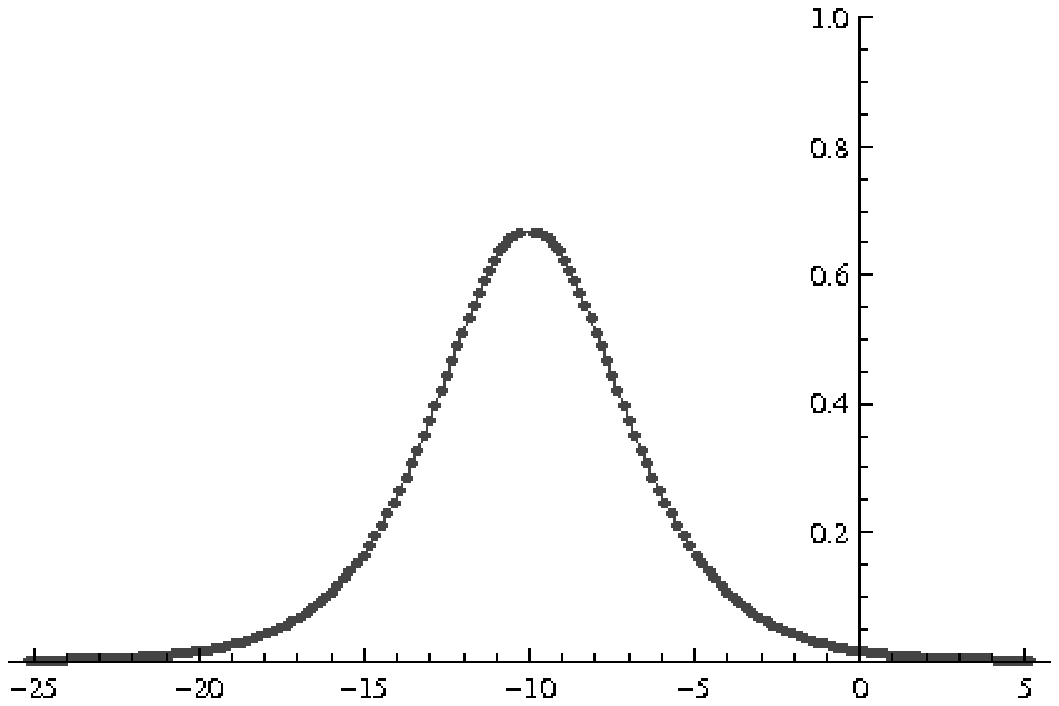}\quad
\includegraphics[scale=0.5]{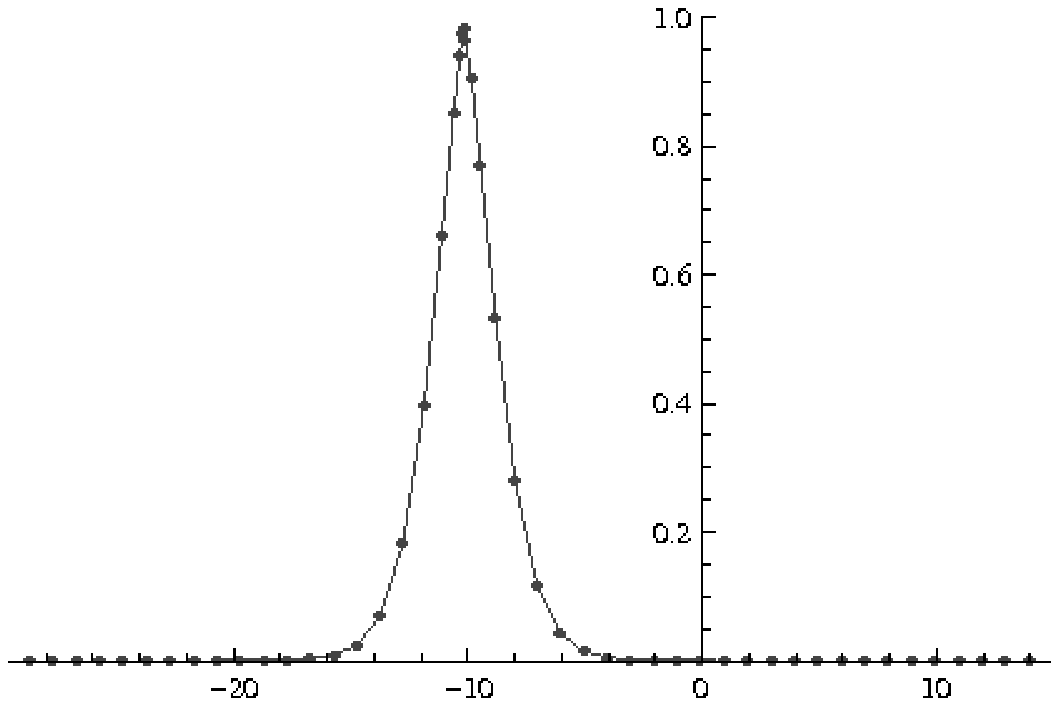}}
\caption{1-soliton solution for the semi-discrete CH equation:
the left: $p_1=0.5, c=1, a=0.1$; the right (close to the peakon limit):
$p_1=99, c=100, a=0.005$.} \label{peakon-limit-discrete}
\end{figure}

We can also consider the peakon limit for the semi-discrete CH
equation. For the semi-discrete case,
we can express
the 1-soliton solution as
\[
w_k=\frac{2p_1^2\kappa v_1}
{1+p_1^2\kappa^2+\frac{1-p_1^2\kappa^2}{2}
\left[\left( \frac{1+ap_1}{1-ap_1}
\right)^{-k}e^{-\xi_1}+\left( \frac{1+ap_1}{1-ap_1}
\right)^{k}e^{\xi_1}
 \right]}
\,,
\]
where $\kappa=\frac{1}{c}$, $v_1=2\kappa^2/(1-p_1^2\kappa^2)$,
$\xi_1=p_1\kappa (-(v_1/\kappa)t-x_{10}/\kappa)$.
The peakon limit for the semi-discrete CH equation
is
again $\kappa\to 0$, $p_1\kappa\to 1$, $v_1={\rm const.}$
Taking the peakon limit, the solution $(X_k(t),w_k(t))$, where
$X_k(t)=X_0+\sum_{j=0}^{k-1}\delta_j(t)$,
approaches to a solution which approaches to the peakon solution
of the CH equation as taking the continuous limit.
In Fig.\ref{peakon-limit-discrete},
one can see that the 1-soliton solution approaches to the 1-peakon
like solution as $\kappa$ approaches to 0.
Taking the continuous limit, this solution approaches to the 1-peakon
solution of the CH equation.

%%From the above observation, we expect that there is an explicit form
%%of the peakon solution for the semi-discrete CH equation.
%%The detail of the peakon solution for the
%%semi-discrete CH equation will be reported in the forthcoming paper.

\section{Numerical experiments}
In this section, we apply our scheme to several test problems. They
include: 1) propagation and interaction of nearly-peakon solutions;
2) propagation and interaction of cuspon solutions; 3)
interactions of soliton-cuspon solutions; 4) non-exact initial
value problems.

\subsection{Propagation and interaction of nearly-peakon solutions}
\begin{figure}[htbp] \centerline{
\includegraphics[scale=.4]{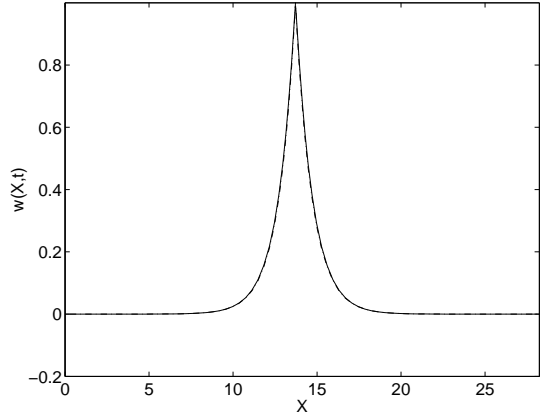}}\quad
\caption{Comparison between one peakon solution and one-soliton
solution with $c=200.0$.} \label{f:peakon}
\end{figure}

{\bf Example 1: One peakon propagation.}
It has been shown in \cite{PKMatsuno,MatsunoPeakon} that the
analytic $N$-soliton solution of the CH equation converges
to the nonanalytic $N$-peakon solution when $\kappa \to 0$ ($c \to
\infty $). To show this, we choose one soliton solution with
parameters $c=1000$, $p=998.9995$. Thus the speed of the soliton
($v_1/2$) is $1.0$.
%%$1.0$.
Its profile is plotted  and is compared with one peakon
solution $u(x,t)=e^{-|x-t|}$ in Fig. \ref{f:peakon}. These two solutions are
indistinguishable from the graph. The error in $L_{\infty}$, where
$L_{\infty} = \max|{w}_{l}-u_{l}|$, is calculated to be $O(10^{-3})$, and
the discrepancy for the first conserved quantity $I_1 = \int u\,dx$
is less than $0.7\%$. Therefore, this soliton solution can be viewed
as an approximate peakon solution with amplitude $1.0$.

\begin{figure}[htbp]
\centerline{
\includegraphics[scale=0.4]{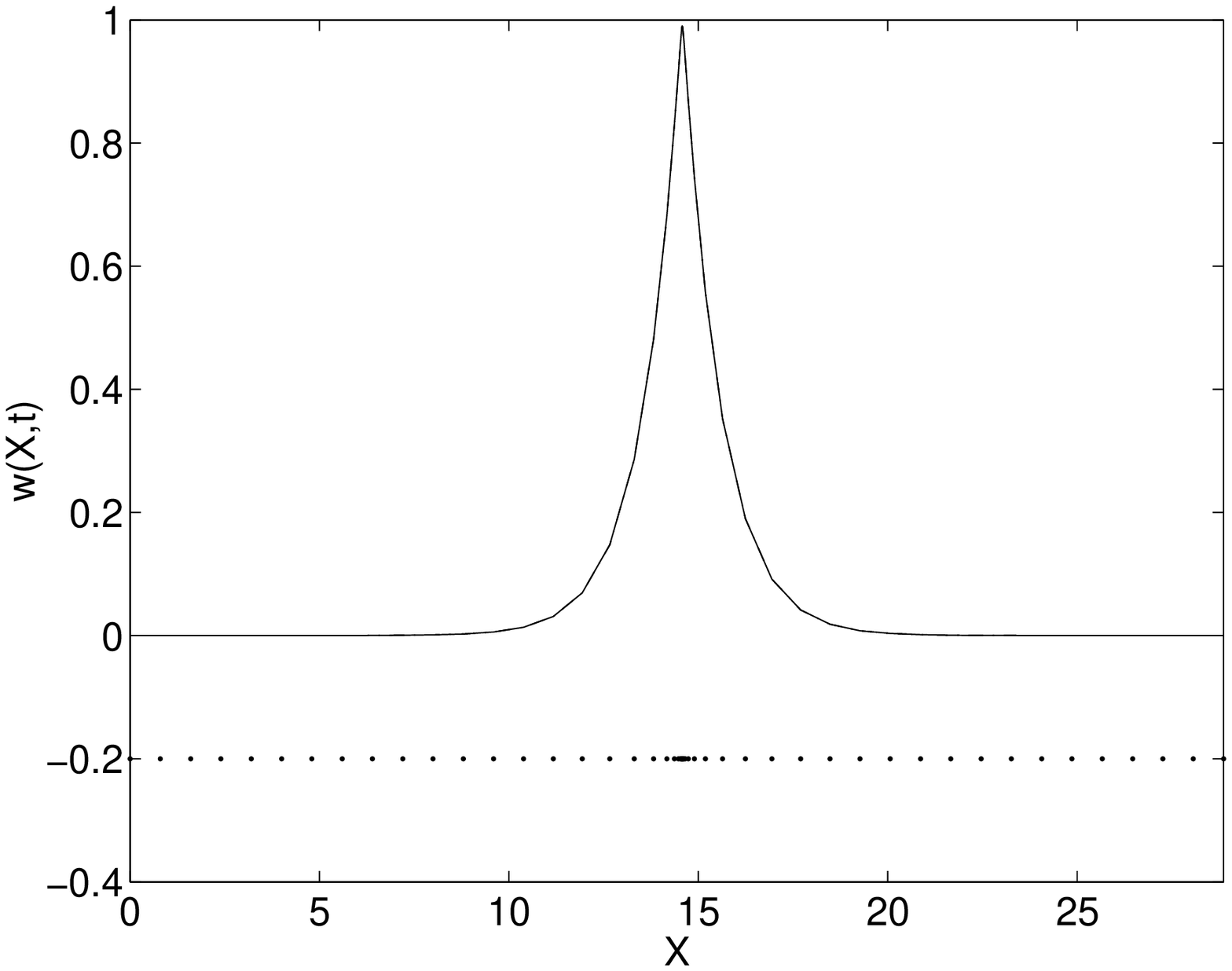}\quad
\includegraphics[scale=0.4]{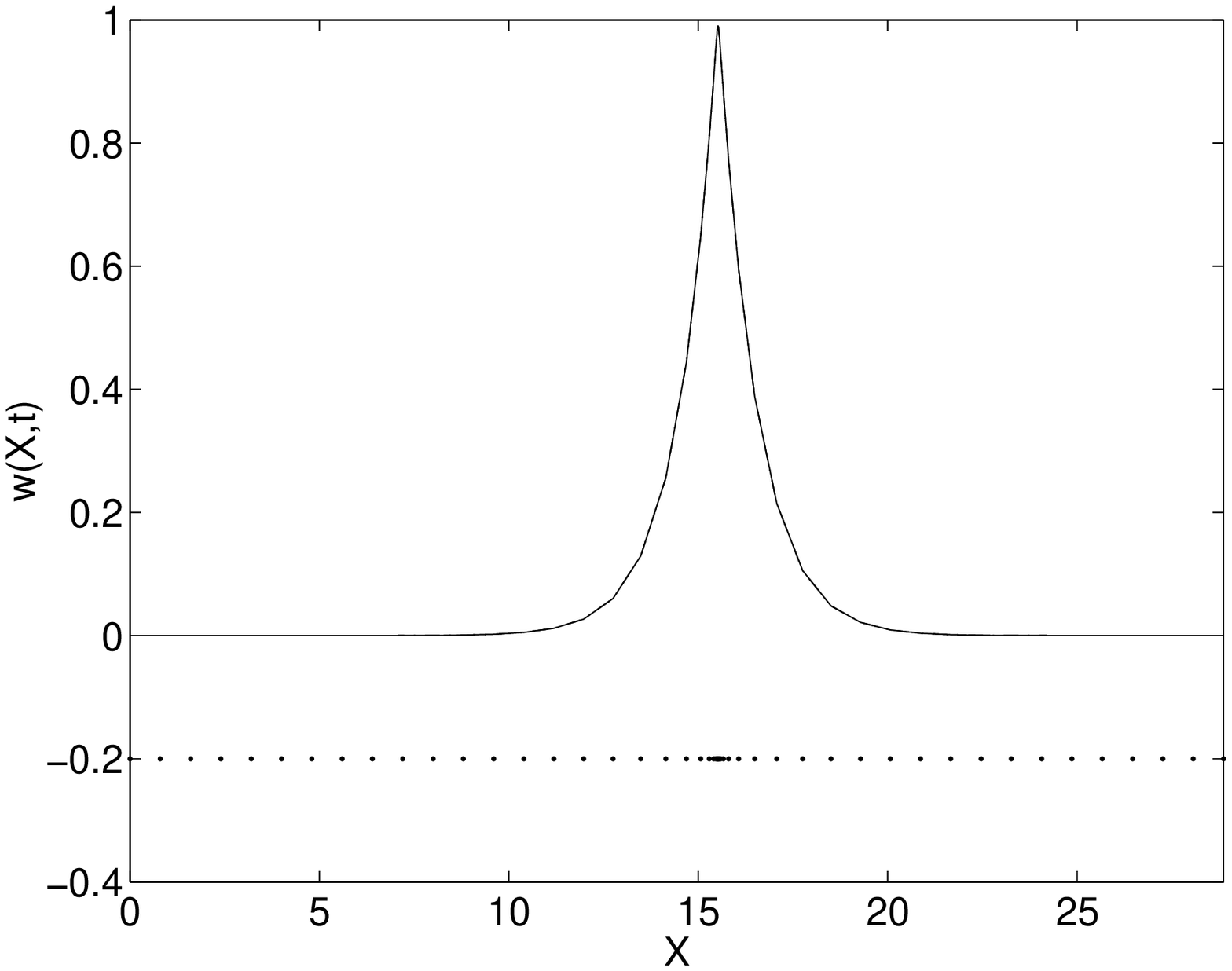}}
\kern-0.355\textwidth \hbox to
\textwidth{\hss(a)\kern2em\hss(b)\kern2em} \kern+0.355\textwidth
\centerline{
\includegraphics[scale=0.4]{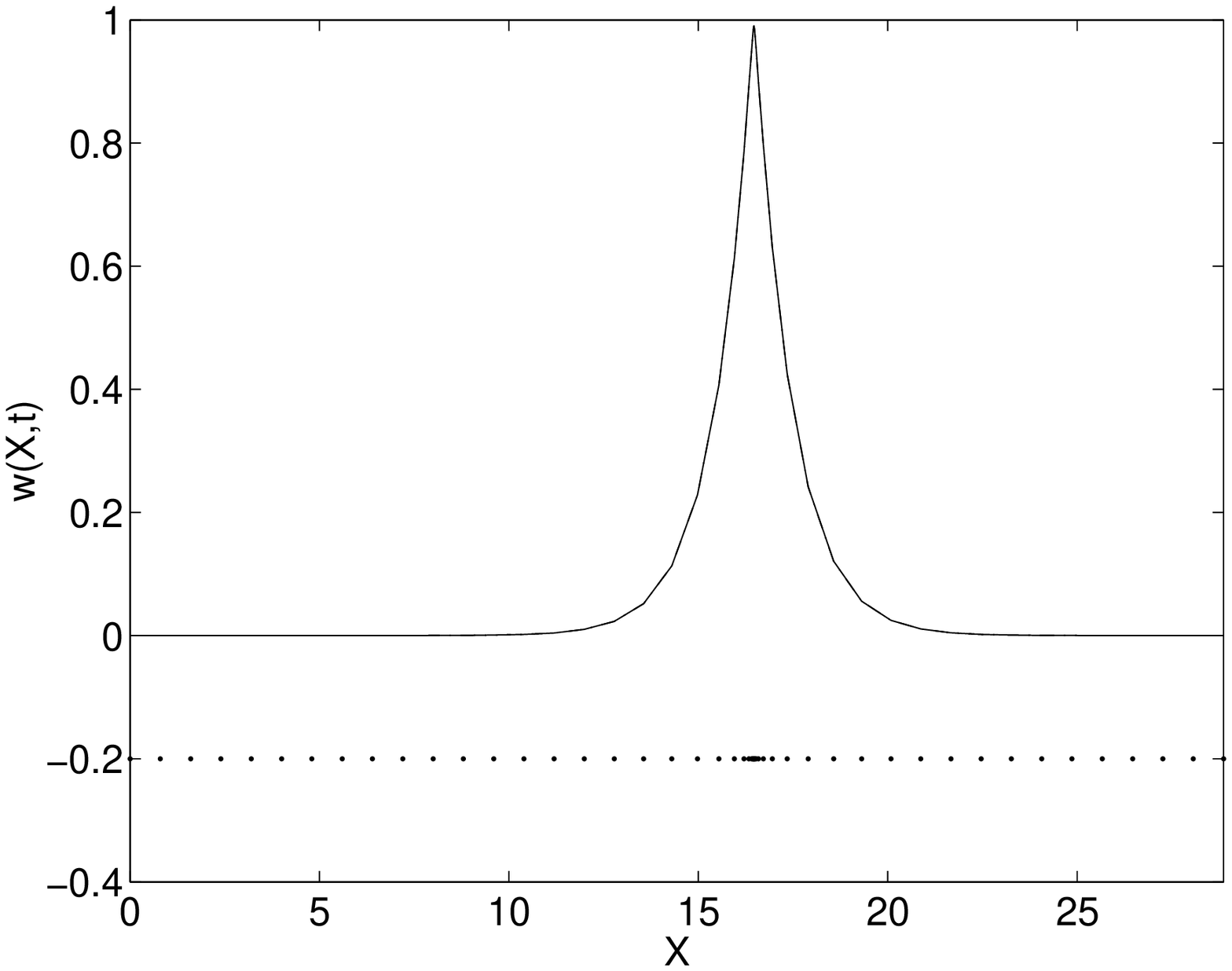}\quad
\includegraphics[scale=0.4]{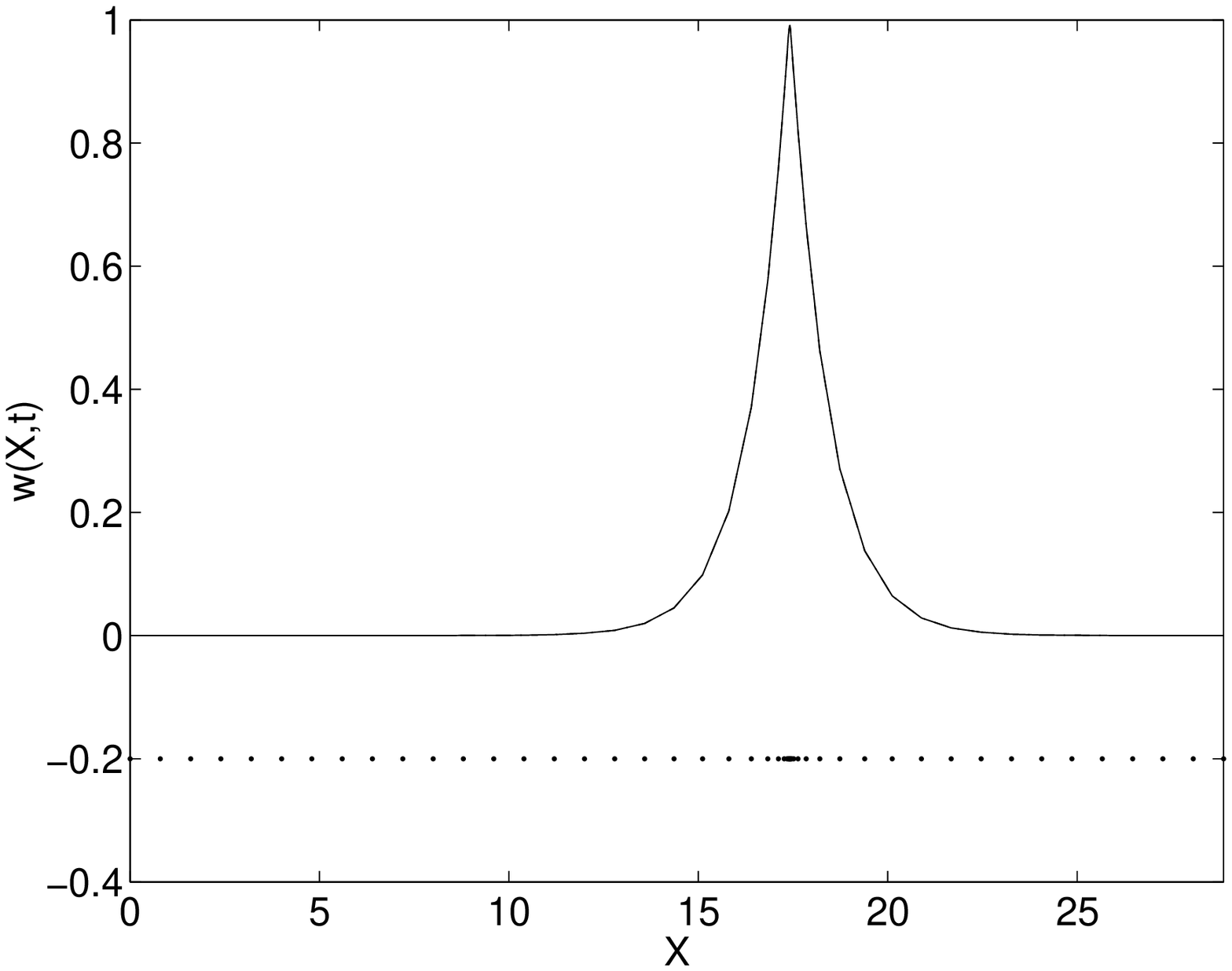}}
\kern-0.355\textwidth \hbox to
\textwidth{\hss(c)\kern2em\hss(d)\kern2em} \kern+0.355\textwidth
%\kern-0.355\textwidth \hbox to \textwidth{\hss(c)\kern0.0em}
\caption{Numerical solution of one single peakon solution: (a)
$t=1.0$; (b) $t=2.0$; (c) $t=3.0$; (c) $t=4.0$.} \label{f:1peakon}
\end{figure}

The propagation of the above designed approximate peakon solution
is solved by the self-adaptive mesh scheme with two different time advancing methods:
the modified forward Euler method (MFE) and the classical Runge-Kutta method
(RK4). The length of the interval in the $x$-domain is chosen to be $0.02$ and the number of grid is $N=101$.
For the above parameters of one-peakon solution, the length of
the computation domain turns out to be about $28.5$. %%% this value is correct??
Figures \ref{f:1peakon} (a)-(d)
display the numerical solutions at $t=1.0, 2.0, 3.0, 4.0$,
together with the
self-adjusted mesh. It can be seen that the non-uniform mesh is
dense around the crest. The most dense part of the non-uniform mesh
moves along with the peakon point with the same speed.
With the same grid points $N=101$, the relative errors in
$L_{\infty}$-norm and the first conserved quantity $I_1 = \int
w\,dx$ are computed and compared in Table \ref{t:peakon}.
Here, $L_{\infty} =
{\mbox{max}\left|\frac{\tilde w_l-w_l}{w_l}\right|}$, where $\tilde{w_l}$ and $w_l$
represent the numerical and analytical solutions at the
grid points $X_l$, respectively. $E_1 = |\bar I_1 -I_1|/|I_1|$ indicates the relative error in $I_1$, where
$\bar I_1$ stands for the counterpart of $I_1$
by the numerical solution. Trapezoidal rule on the non-uniform mesh
is employed for the evaluation of the integrals.

\begin{table}
\caption{Comparison of  $L_\infty$ and $I_1$ errors for one-soliton propagation.}
\begin{center}
\begin{tabular}{c c l l l} \hline
  & $\Delta t$ & T & $L_\infty$  & ${E_1}$
 \\ \hline
MFE & 0.001 & $2.0$ & $6.9(-4)$ & $8.1(-4)$
 \\
  & 0.001 & $4.0 $ & $1.4(-3)$ & $1.6(-3)$
 \\ \hline
 RK4 &  0.01 & $2.0$ & $1.7(-12)$ & $4.5(-13)$ \\
 &  0.01 & $4.0$ & $5.4(-12)$ & $8.4(-13)$
 \\ \hline
\label{t:peakon}
\end{tabular}
\end{center}
\end{table}

\begin{table}
\caption{$L_\infty$ and $I_1$ errors for two  approximate peakon
interaction by the self-adaptive moving mesh method}
\begin{center}
\begin{tabular}{c c ll l} \hline
  & $\Delta t$ & T & $L_\infty$  & ${E_1}$
 \\ \hline
MFE & 0.001 & $5.0$ & $2.2(-2)$ & $5.5(-3)$
 \\
  & 0.001 & $10.0 $ & $7.1(-2)$ & $1.2(-2)$
 \\ \hline
 RK4 &  0.01 & $5.0$ & $2.0(-9)$ & $1.5(-7)$ \\
 &  0.01 & $10.0$ & $3.2(-9)$ & $1.4(-5)$
 \\ \hline
\label{t:peakon2}
\end{tabular}
\end{center}
\end{table}

\begin{figure}[htbp]
\centerline{
\includegraphics[scale=0.35]{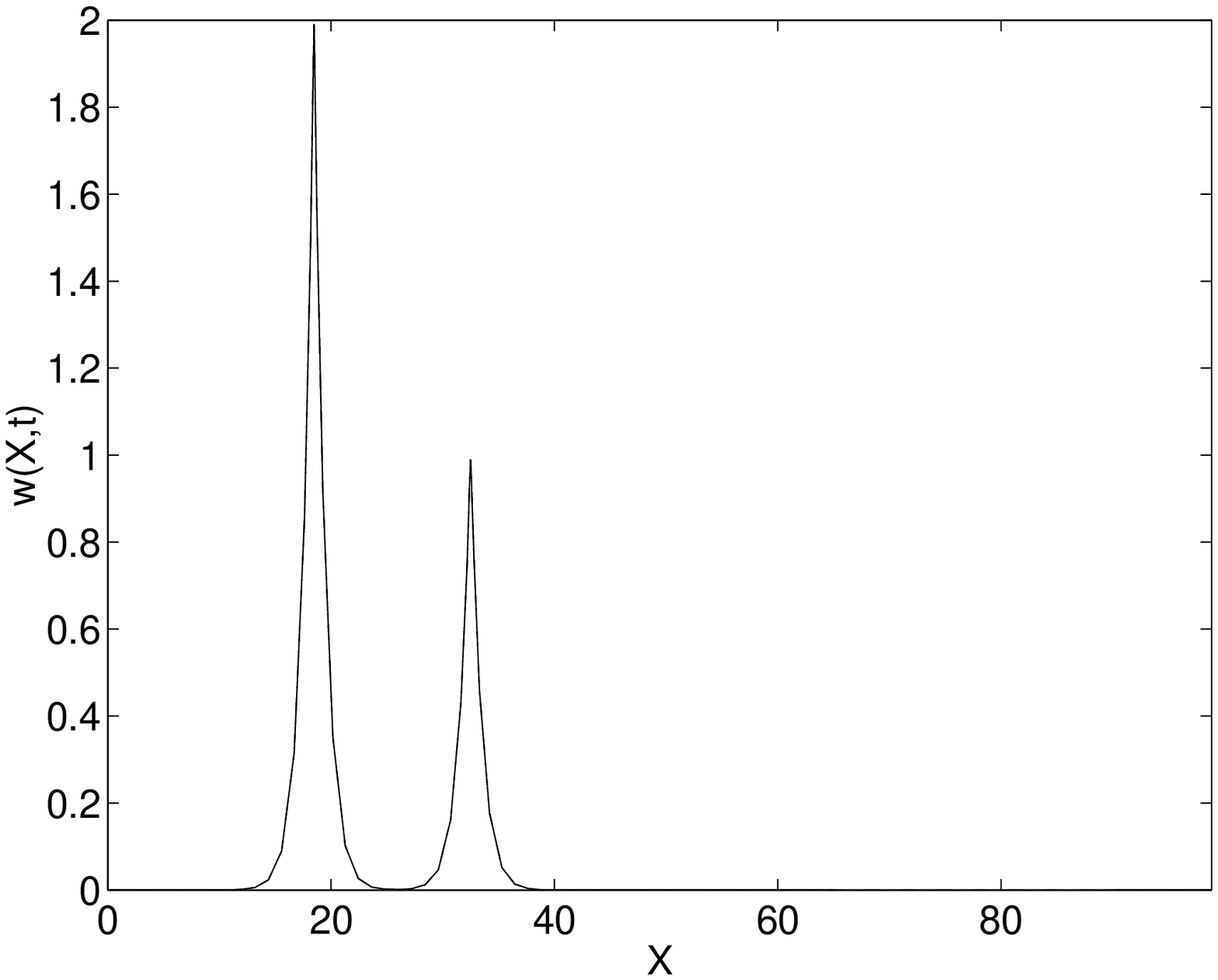}\quad
\includegraphics[scale=0.35]{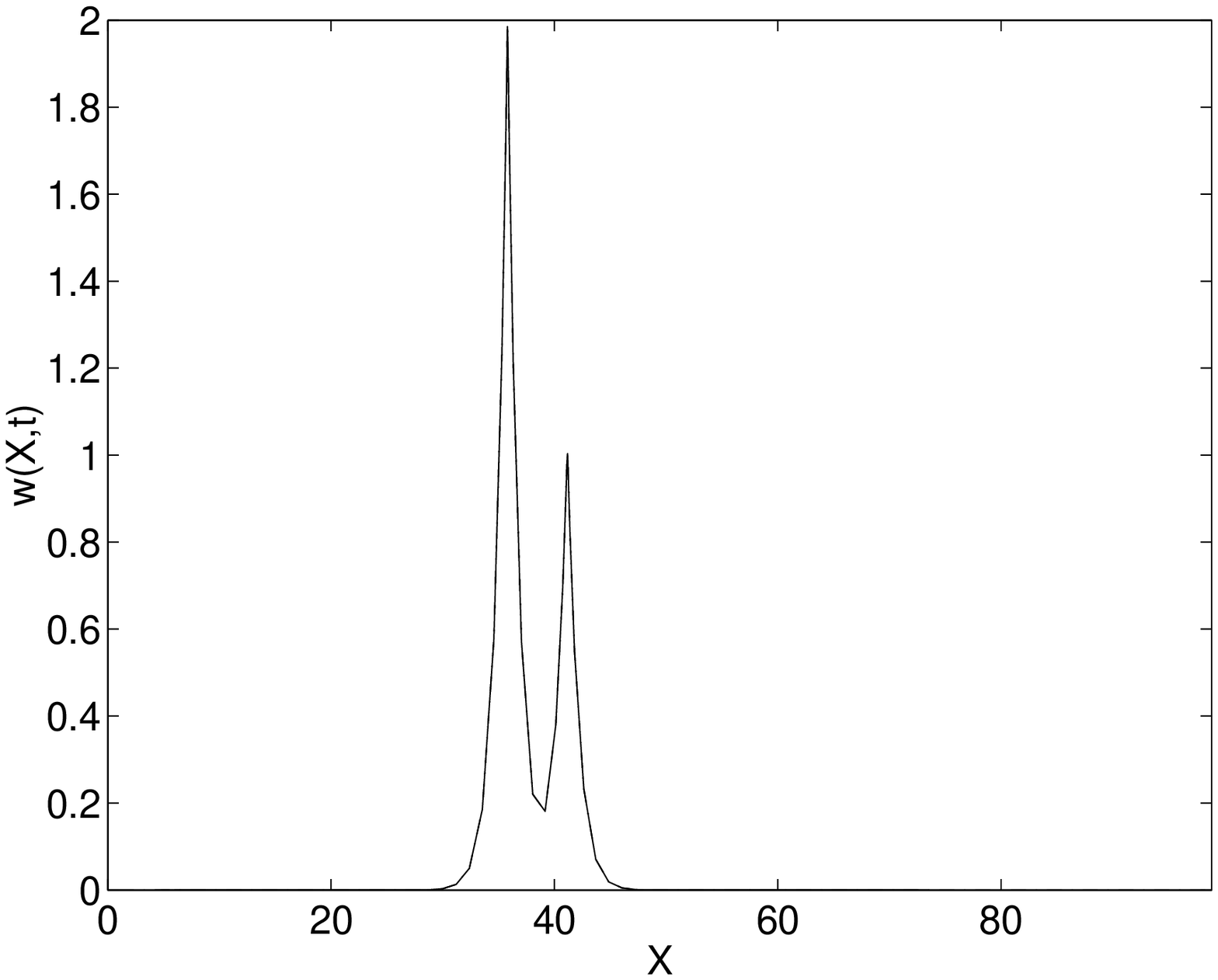}}
\kern-0.315\textwidth \hbox to
\textwidth{\hss(a)\kern0em\hss(b)\kern4em} \kern+0.315\textwidth
\centerline{
\includegraphics[scale=0.35]{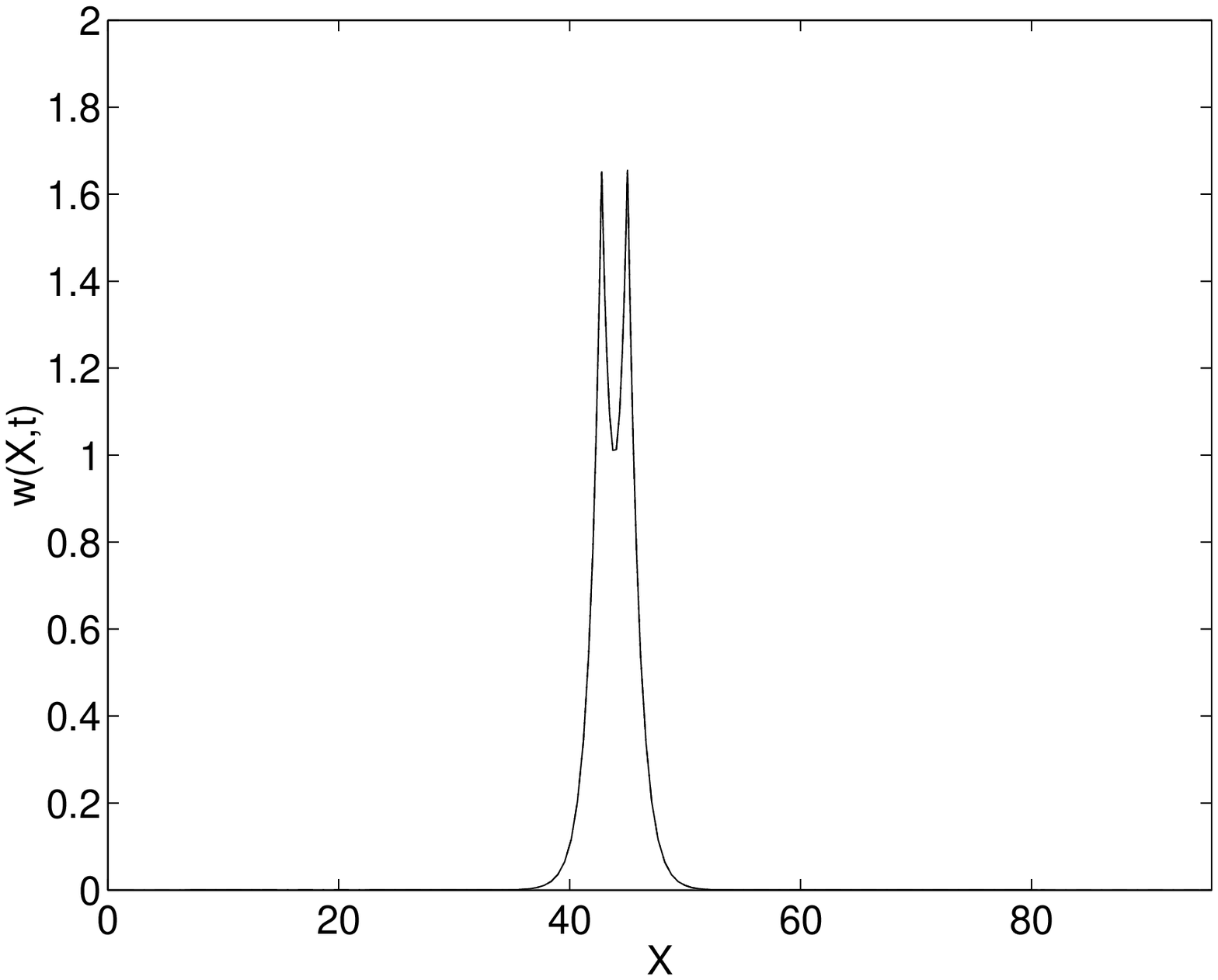}\quad
\includegraphics[scale=0.35]{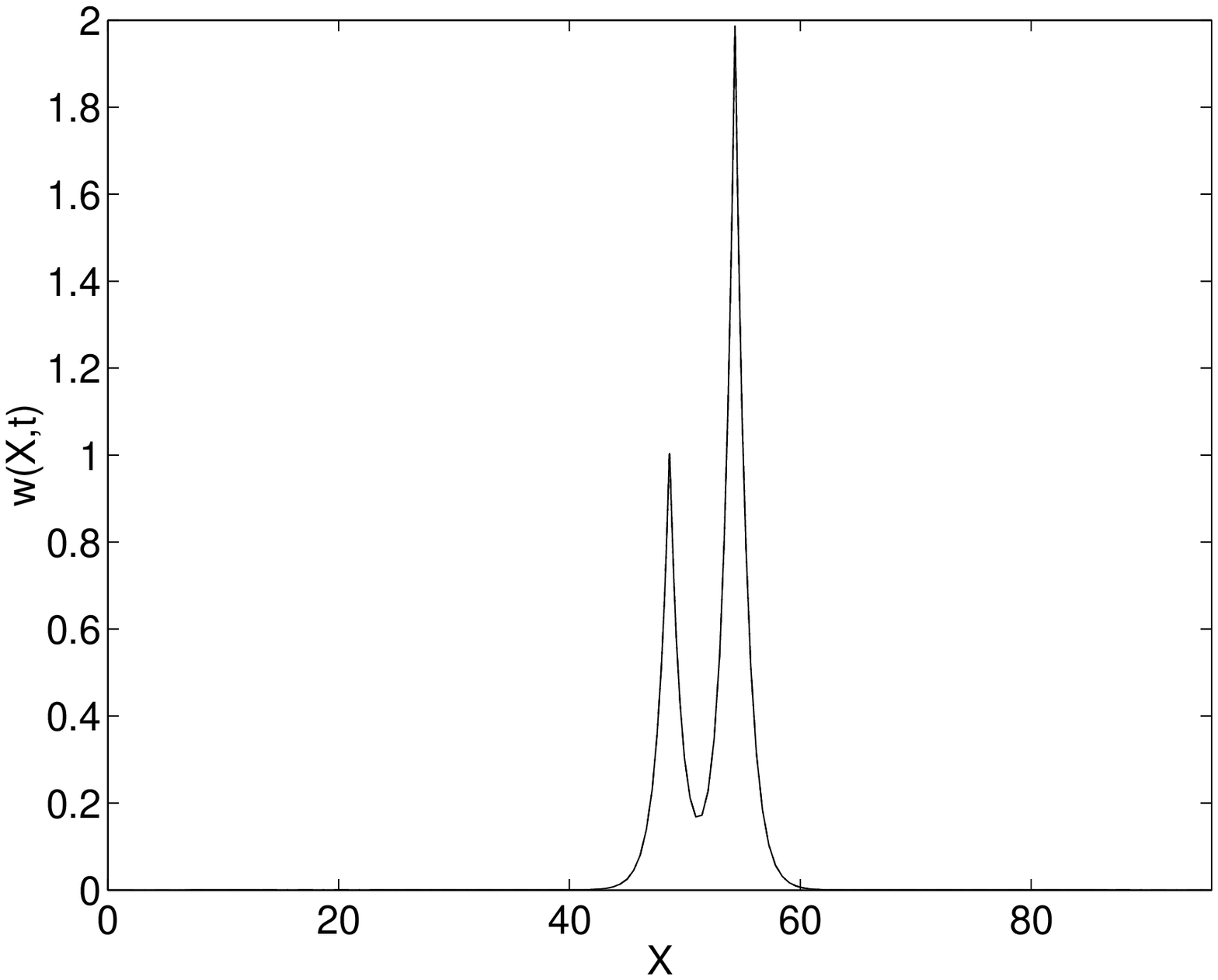}}
\kern-0.315\textwidth \hbox to
\textwidth{\hss(c)\kern0em\hss(d)\kern4em} \kern+0.315\textwidth
\centerline{
\includegraphics[scale=0.35]{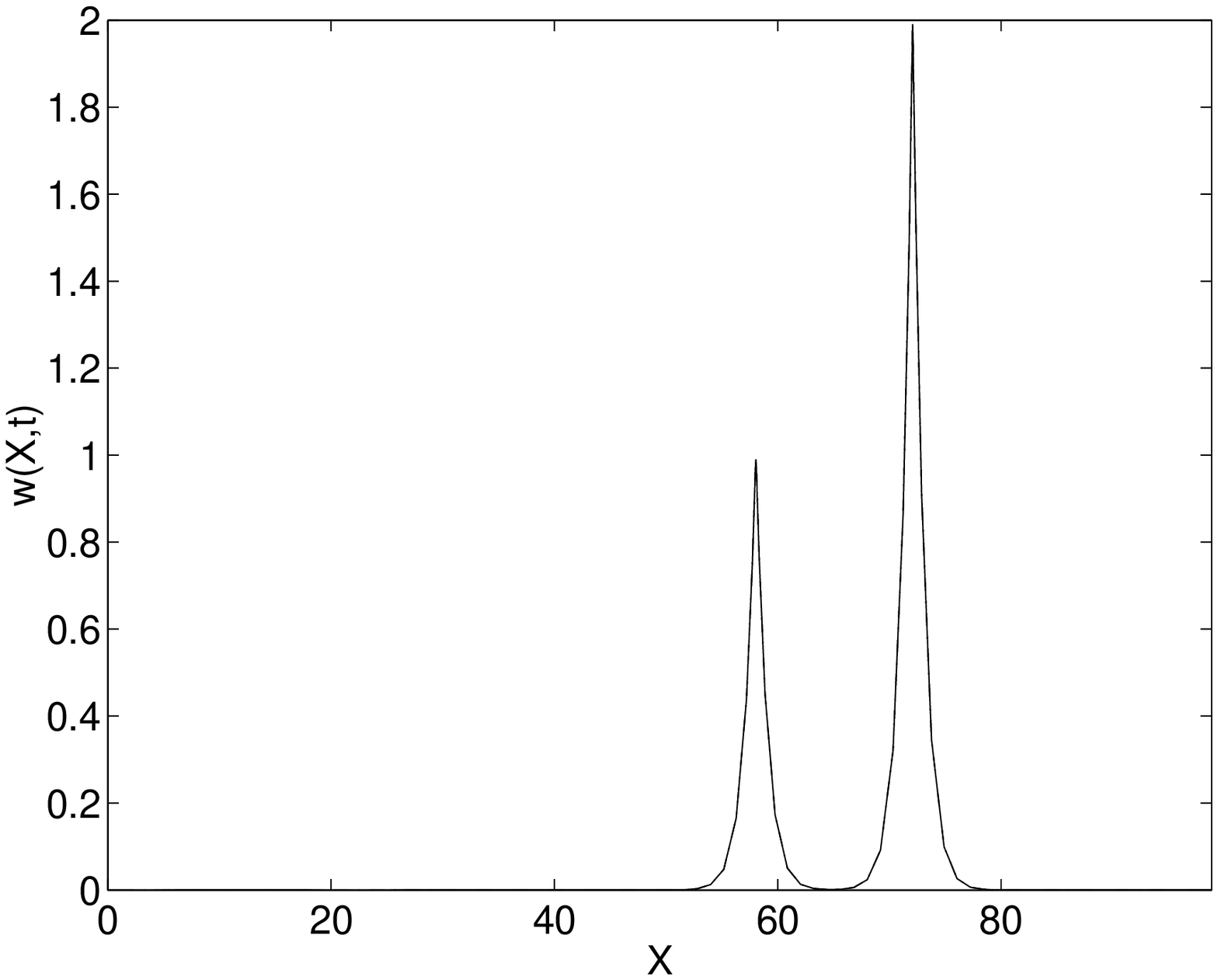}}
\kern-0.315\textwidth \hbox to \textwidth{\hss(e)\kern13em}
\kern+0.315\textwidth
%\quad
%\includegraphics[scale=0.4]{stcpp19p12p210p98t17.eps}}
%\kern-0.0\textwidth{\hss(g)\kern0em}
%\kern+0.355\textwidth
\caption{Numerical solution for collision of two nearly-peakon with
$p_1=198.9975$, $p_2=199.4995$ and $c=200.0$: (a) $t=0.0$; (b) $t=10.0$; (c)
 $t=15.0$; (d) $t=20.0$; (e) $t=30.0$. }
\label{f:soliton_soliton}
\end{figure}

{\bf Example 2: Two peakon interaction.}
For $c=1000$, we initially choose two approximate peakon solutions
moving with velocity
%%$v_1=2.0$,
$v_1/2=2.0$,
and
%%%$v_2=1.0$,
$v_2/2=1.0$,
respectively. Their interaction is
numerically solved by MFE and RK4, respectively, with a fixed grid
number
of $N=101$.
Figure \ref{f:soliton_soliton} displays the process of collision
at different times.
Table \ref{t:peakon2} presents the errors in $L_\infty$-norm and $E_1$.
It could be seen that, in spite of  a small number of grid points
and a large time step, RK4 simulates the collision of two
approximate peakons with good accuracy.

In regard to the propagation and interaction of approximate peakon solutions,
we summarize as follows:
\begin{enumerate}
  \item Due to the integrability of the scheme and the
  self-adaptive feature of the non-uniform mesh, the $L_\infty$-norm
is small and
  the first conserved quantity is preserved extremely well even for
  a small number of grid points.
  %!
  %Almost doubling the grid numbers
  %from $51$ to $101$ doesn't get the accuracy improved since a grid
  %number of $51$ is already good enough for the spatial resolution.
  \item The errors is mainly due to the time advancing methods. MFE is first
order in time, so it produces relatively large $L_\infty$ and $E_1$,
roughly changing in proportional with time. RK4 is fourth-order in
time, so up to $T=4.0$, $L_\infty$ and $E_1$ are of the orders
$10^{-12}$ and $10^{-13}$ for a grid number of $N=101$ and a time step
$\Delta t=0.01$.
\end{enumerate}

\subsection{Propagation and interaction of cuspon solutions}
The classical 4th-order Runge-Kutta method fails whenever the
cuspon solution is involved.
It seems that a kind of instability occurs in this case, whose
theoretical
reason is still unclear.
Therefore, only MFE is employed to conduct the numerical experiments
hereafter.

\begin{figure}[htbp]
\centerline{
\includegraphics[scale=0.4]{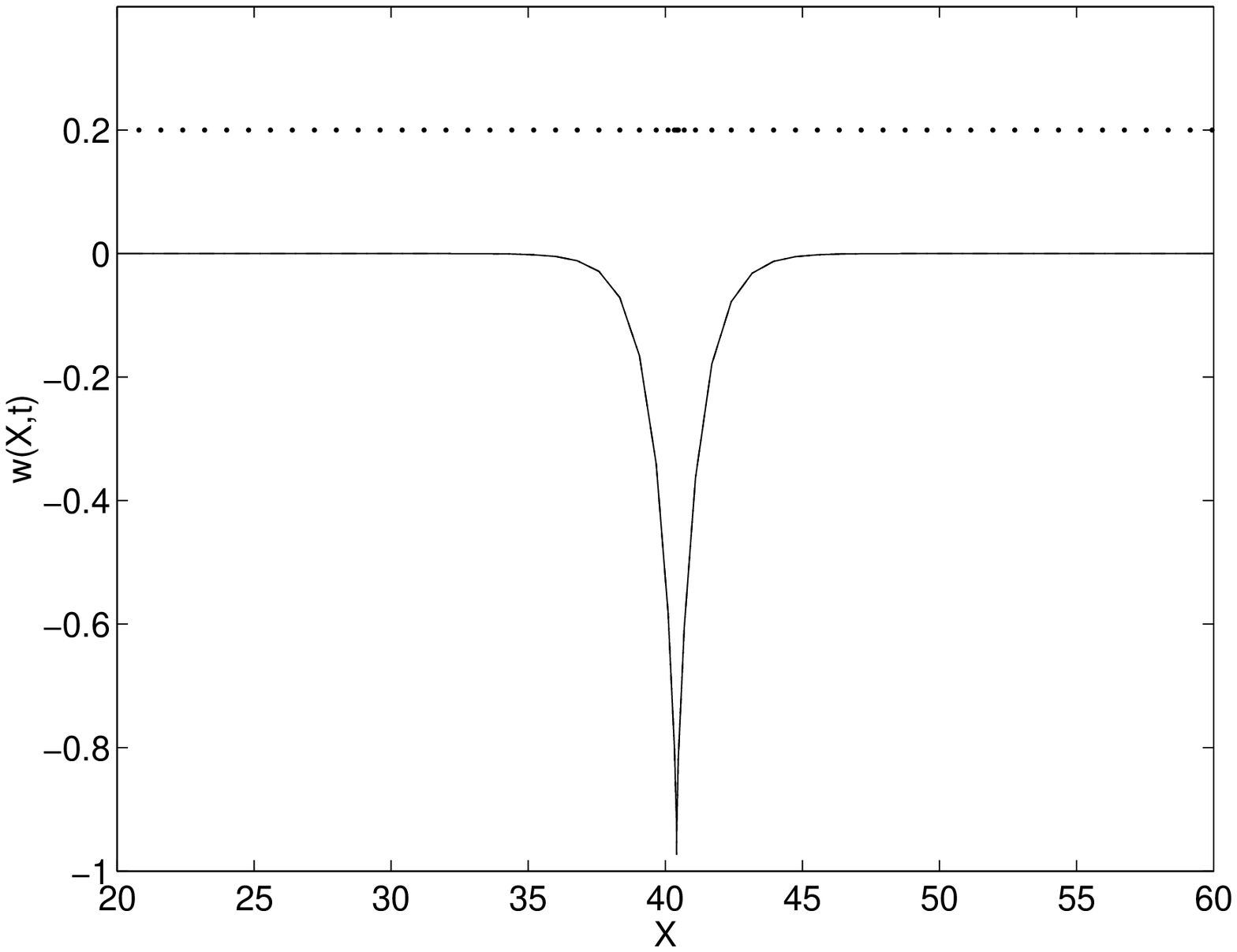}\quad
\includegraphics[scale=0.4]{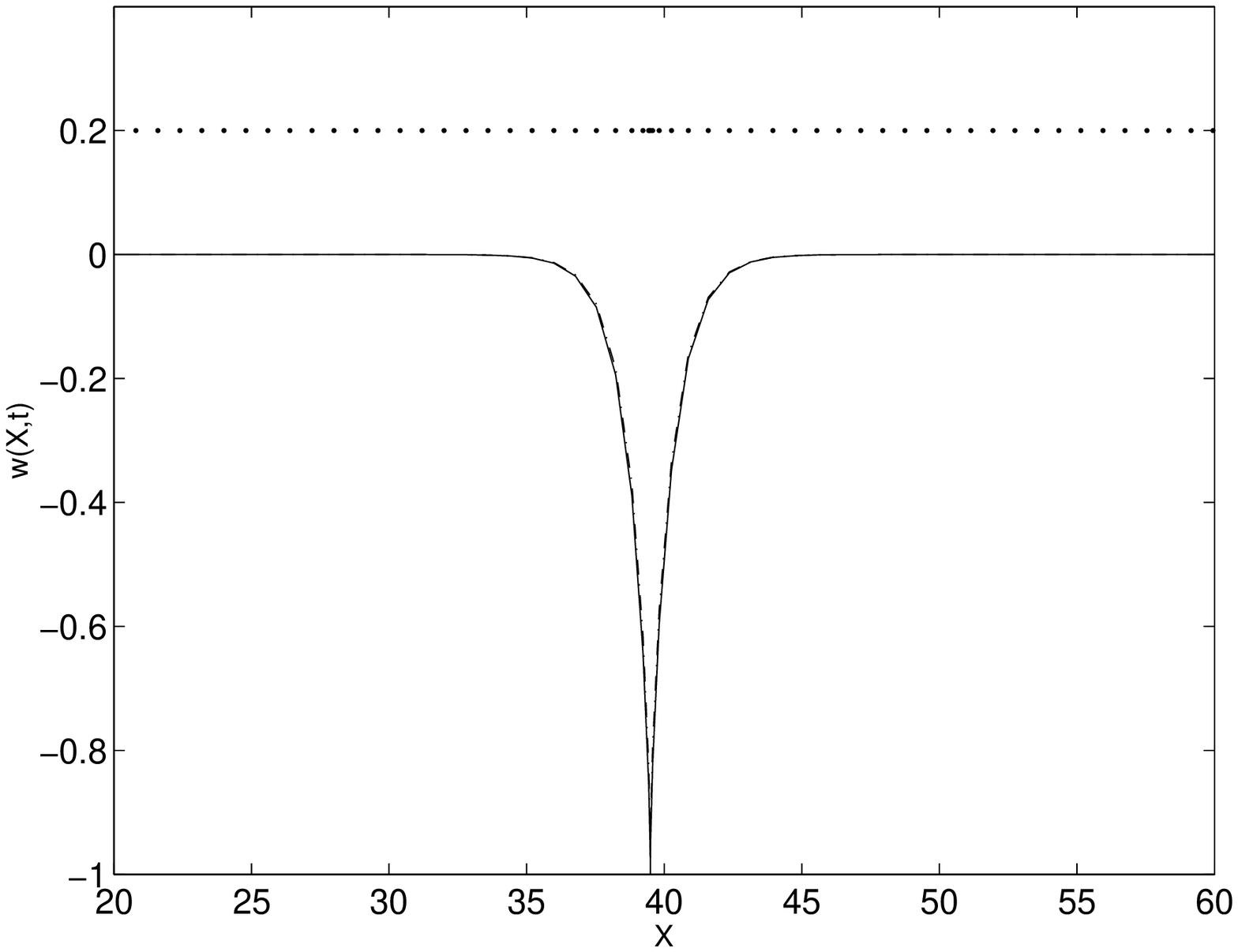}}
\kern-0.355\textwidth \hbox to
\textwidth{\hss(a)\kern0em\hss(b)\kern4em} \kern+0.355\textwidth
\centerline{
\includegraphics[scale=0.4]{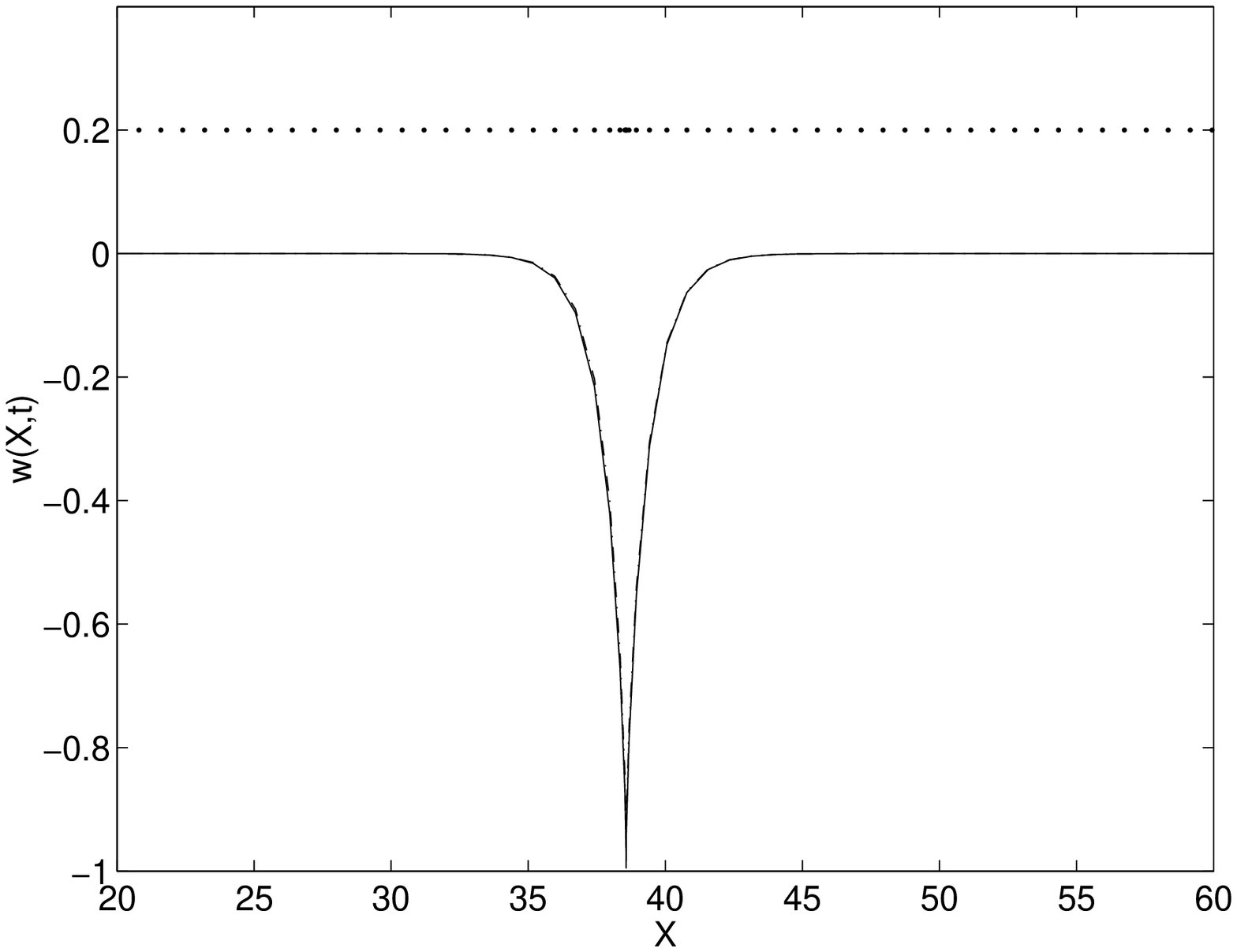}\quad
\includegraphics[scale=0.4]{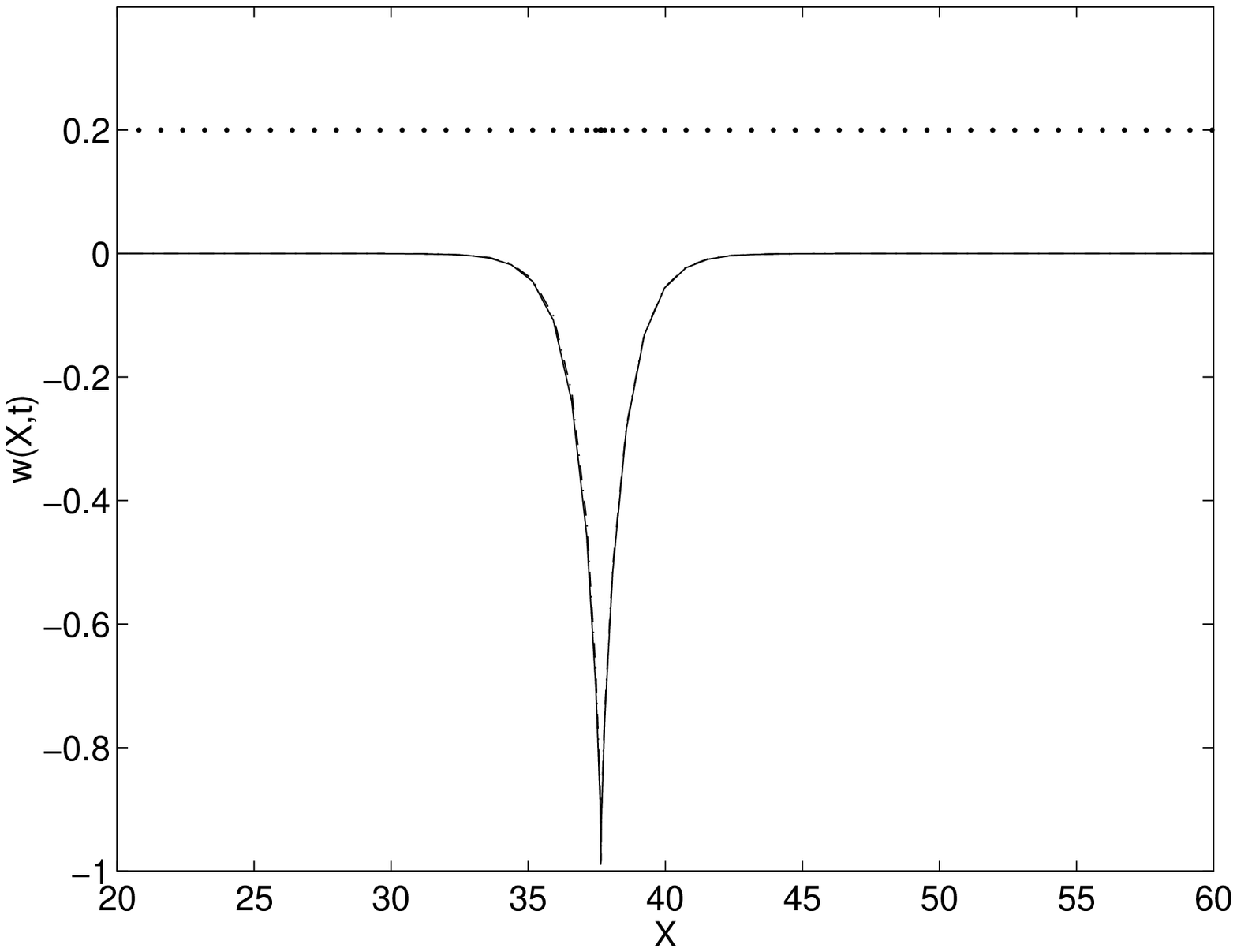}}
\kern-0.355\textwidth \hbox to
\textwidth{\hss(c)\kern0em\hss(d)\kern4em} \kern+0.355\textwidth
%\kern-0.355\textwidth \hbox to \textwidth{\hss(c)\kern0.0em}
\caption{Numerical solution of one single cuspon solution: (a)
$t=0.0$; (b) $t=2.0$; (c) $t=3.0$; (c) $t=4.0$.} \label{f:1cuspon}
\end{figure}

{\bf Example 3: One-cuspon propagation.} The parameters taken for
the one-cuspon solution are $p=10.98$, $c=10.0$. The number of grid is
taken as $101$ in an interval of width of $4$ in the $x$-domain.
Through the hodograph transformation, this corresponds to an interval of width
$74.34$ in the $X$-domain.
Figure \ref{f:1cuspon}(a) shows the initial profile and the
initial mesh.
Figures \ref{f:1cuspon}(b)-(d) display the numerical
solutions (solid
line) and exact solutions (dotted line) at $t=2, 3, 4$, together
with the self-adjusted mesh. It can be seen that the non-uniform
mesh is dense around the cuspon point, and moves to the left in
accordance with the
movement of the cuspon point. Table \ref{tbl:1cuspon}
exhibits the results of relative errors in
$L_\infty$-norm and $E_1$.

\begin{table}[hbtp]
%!
\caption{Relative errors in  $L_\infty$ norm and the first conservative
quantity for one-cuspon propagation}
\begin{center}
\begin{tabular} {l c l  l }
\hline
  $\Delta t$ & $t$
 &{\hfill $L_{\infty}$ \hfill} & $E_1$ \\
\hline \hline
  $0.005$ & 2.0  & $3.3(-2)$ & $4.7(-2)$     \\
  $0.005$ & 4.0  & $9.7(-2)$ & $1.2(-1)$     \\  \hline
  $0.001$ & 2.0  & $1.1(-2)$ & $1.2(-2)$     \\
  $0.001$ & 4.0  & $2.9(-2)$ & $3.7(-2)$     \\  \hline
\end{tabular}
\end{center}
\label{tbl:1cuspon}
\end{table}

\begin{figure}[htbp]
\centerline{
\includegraphics[scale=0.4]{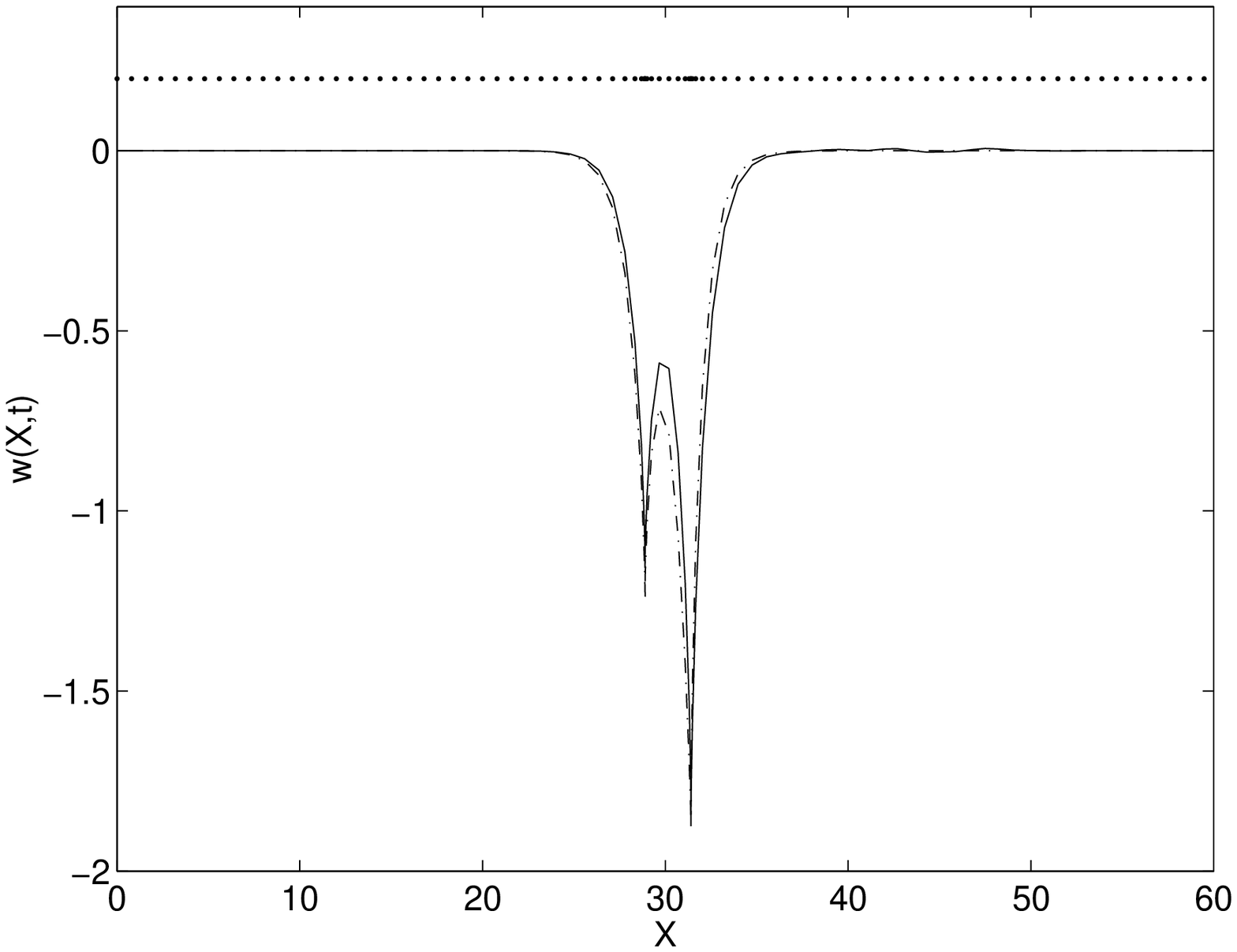}\quad
\includegraphics[scale=0.4]{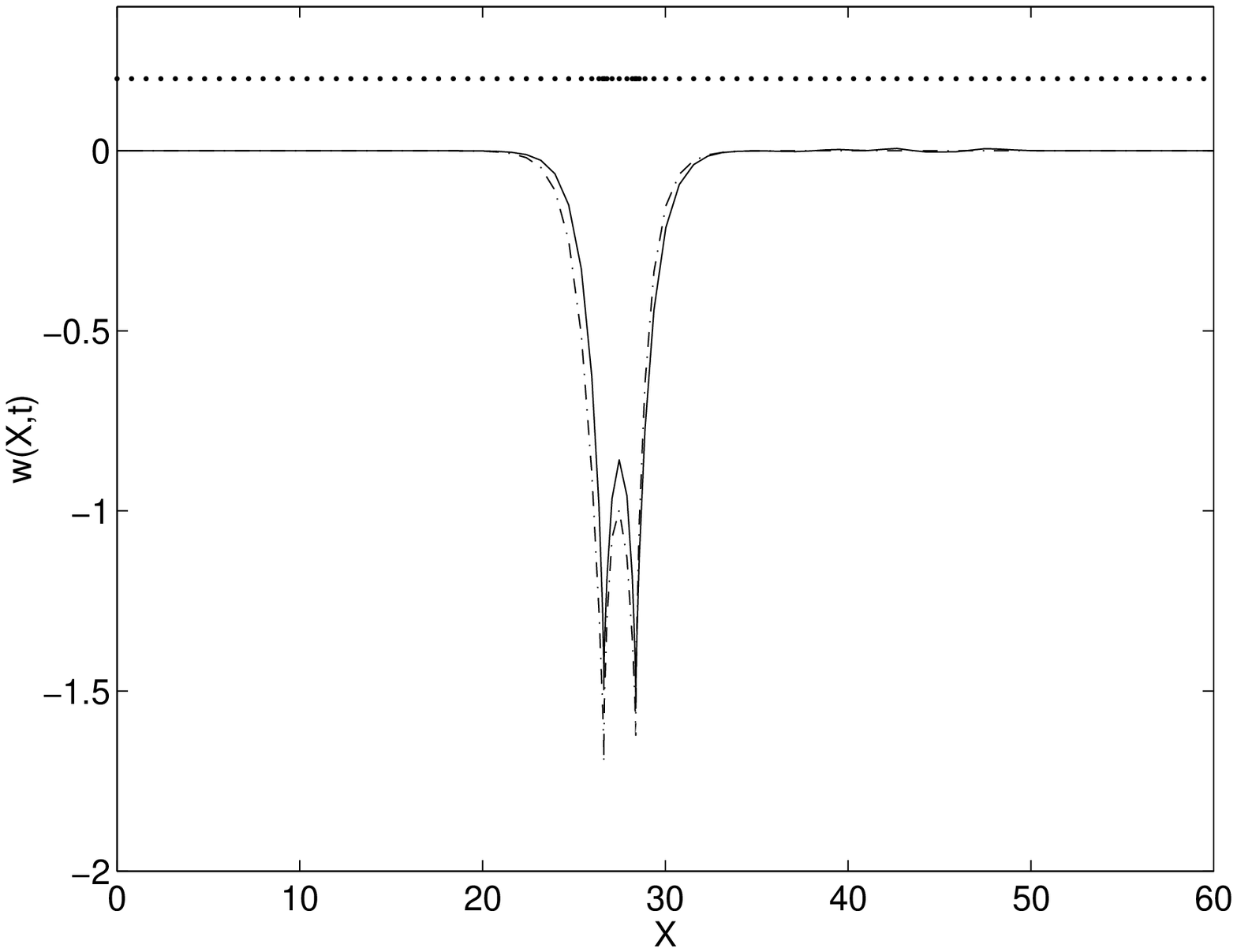}}
\kern-0.1\textwidth \hbox to
\textwidth{\hss(a)\kern0em\hss(b)\kern4em}
\kern+0.1\textwidth\centerline{
\includegraphics[scale=0.4]{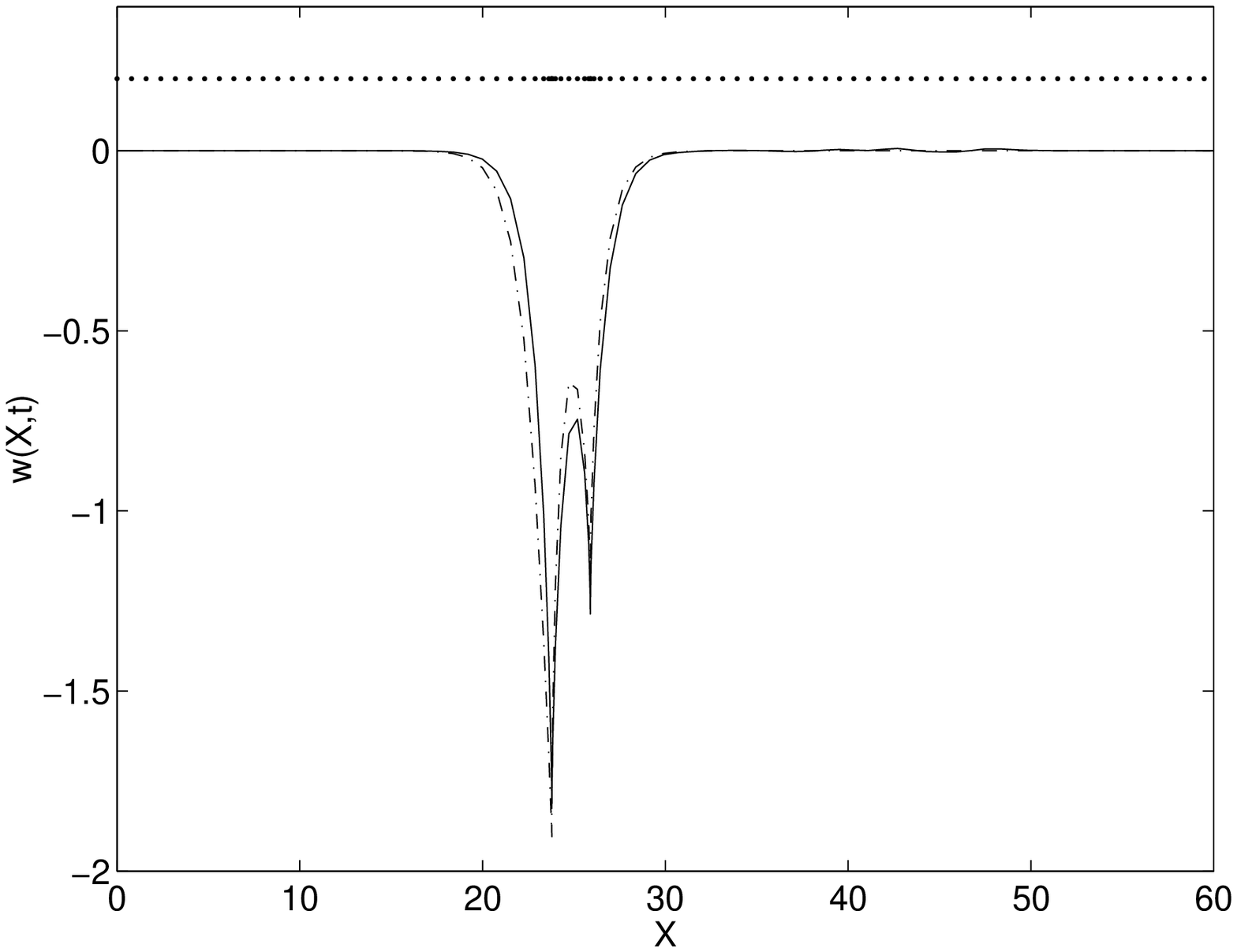}\quad
\includegraphics[scale=0.4]{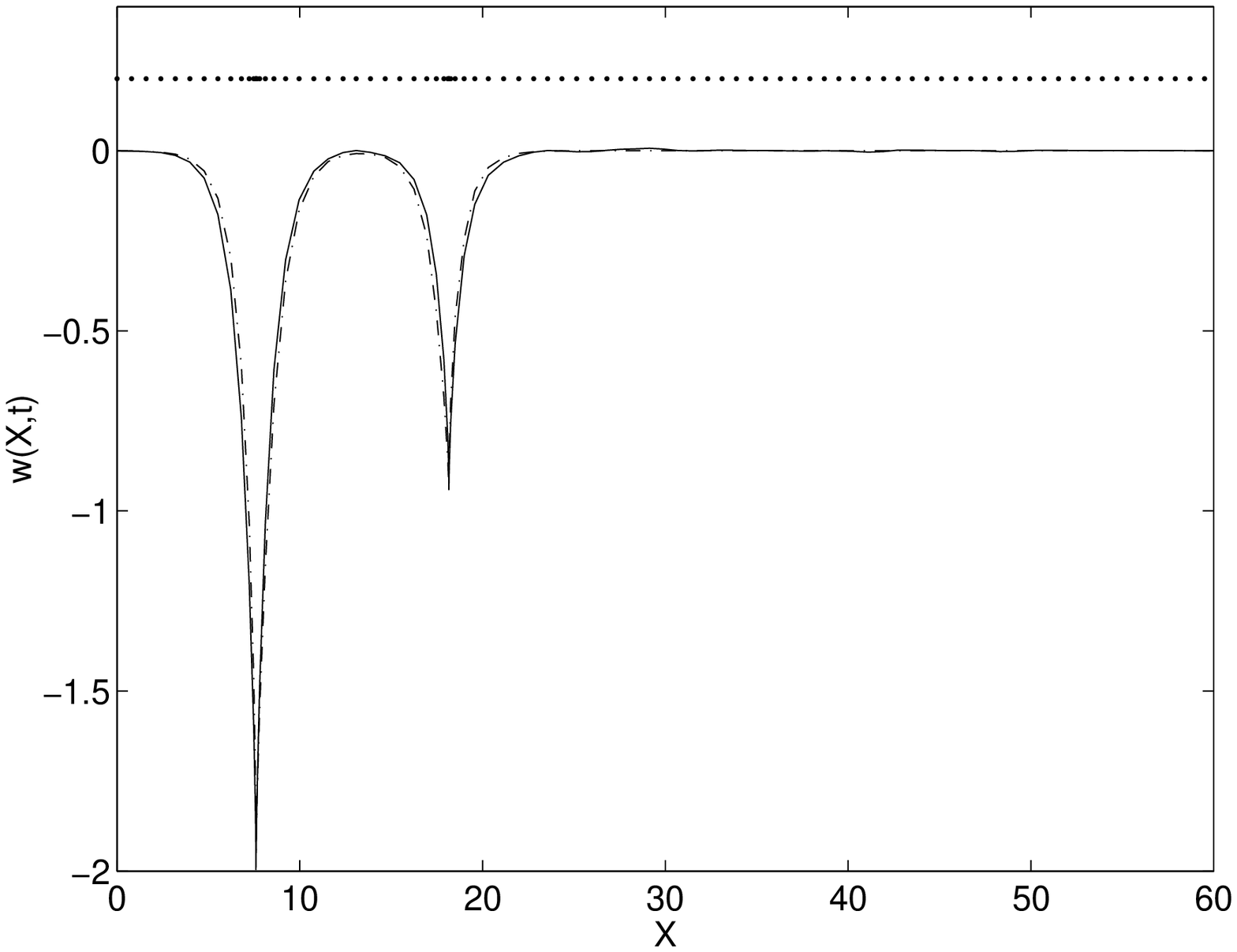}}
\kern-0.1\textwidth \hbox to
\textwidth{\hss(c)\kern0em\hss(d)\kern4em} \kern+0.1\textwidth
%\centerline{
%\includegraphics[scale=0.4]{Fig2e.eps}}
%\kern-0.1\textwidth \hbox to \textwidth{\hss(e)\kern13em}
%\kern+0.1\textwidth
%\kern0.0 \textwidth{\hss(e)\kern0em}
\caption{Numerical solution for the collision of two-cuspon solution
with $p_1=11.0$, $p_2=10.5$, $c=10.0$: (a) $t=13.0$; (b)
 $t=14.8$; (c) $t=16.6$; (d) $t=25.0$.} \label{f:2cuspon}
\end{figure}

{\bf Example 4: Two-cuspon interaction.} The parameters taken for
the two-cuspon solutions are $p_1=11.0$, $p_2=10.5$, $c=10.0$.
Figures \ref{f:2cuspon}(a)-(d)
display the process of collision at several different times,
along with the exact solution. Meanwhile, the self-adaptive mesh is
also shown in the graph. It can be seen that two cuspon solutions
undertake elastic collision, regaining their shapes after the
collision is complete. As mentioned in \cite{DaiLi}, the two cuspon
points are always present during the collision. The grid points are
automatically adapted with the movement of the cuspons, and are
always concentrated at the cuspon points. In compared with the exact
solutions, we can comment that the numerical solutions are in a good
agreement with exact solutions. As far as we know, what is shown
here is the first numerical demonstration for the cuspon-cuspon
interaction.

\subsection{Soliton-cuspon interactions}
\begin{figure}[htbp]
\centerline{
\includegraphics[scale=0.27]{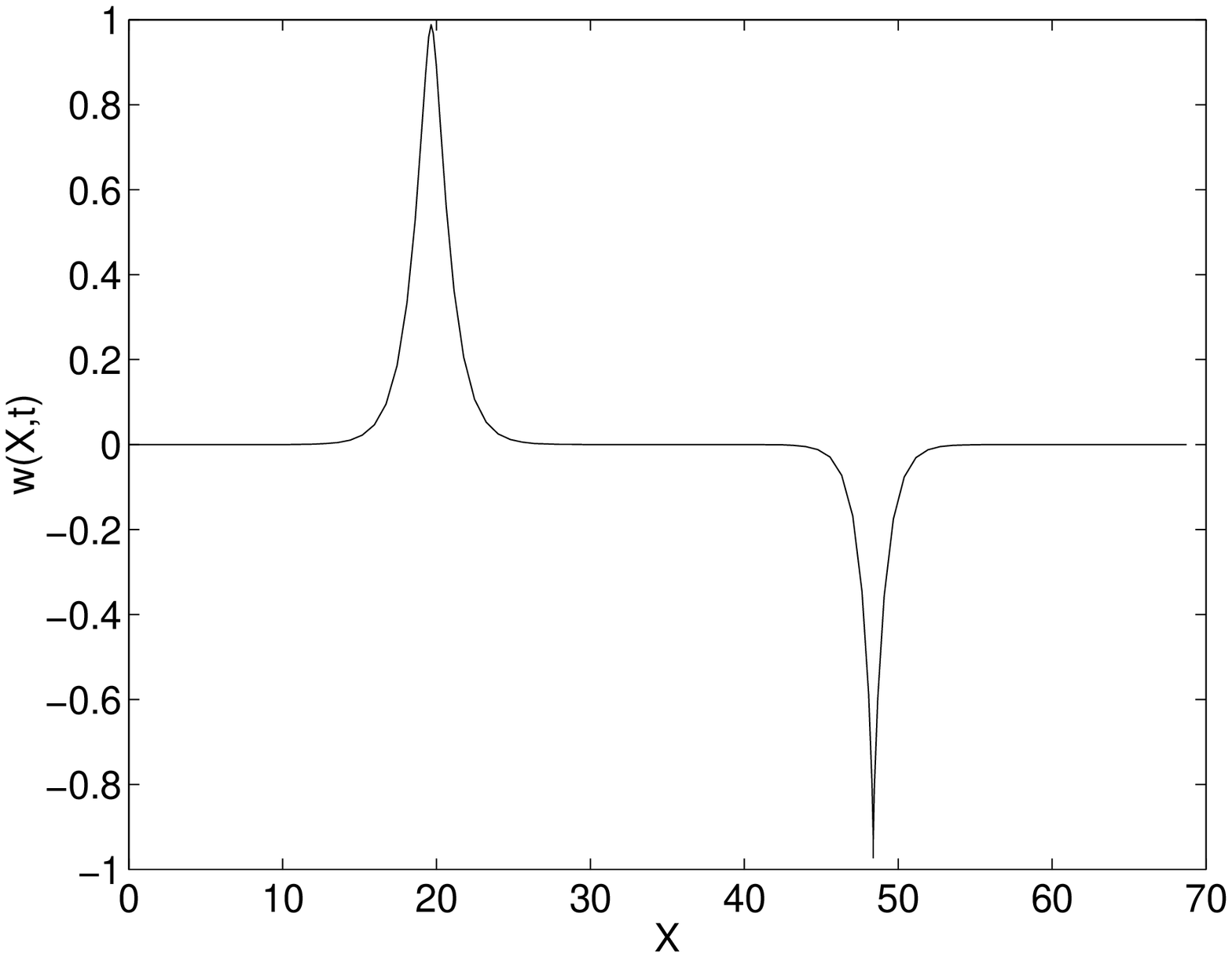}\quad
\includegraphics[scale=0.27]{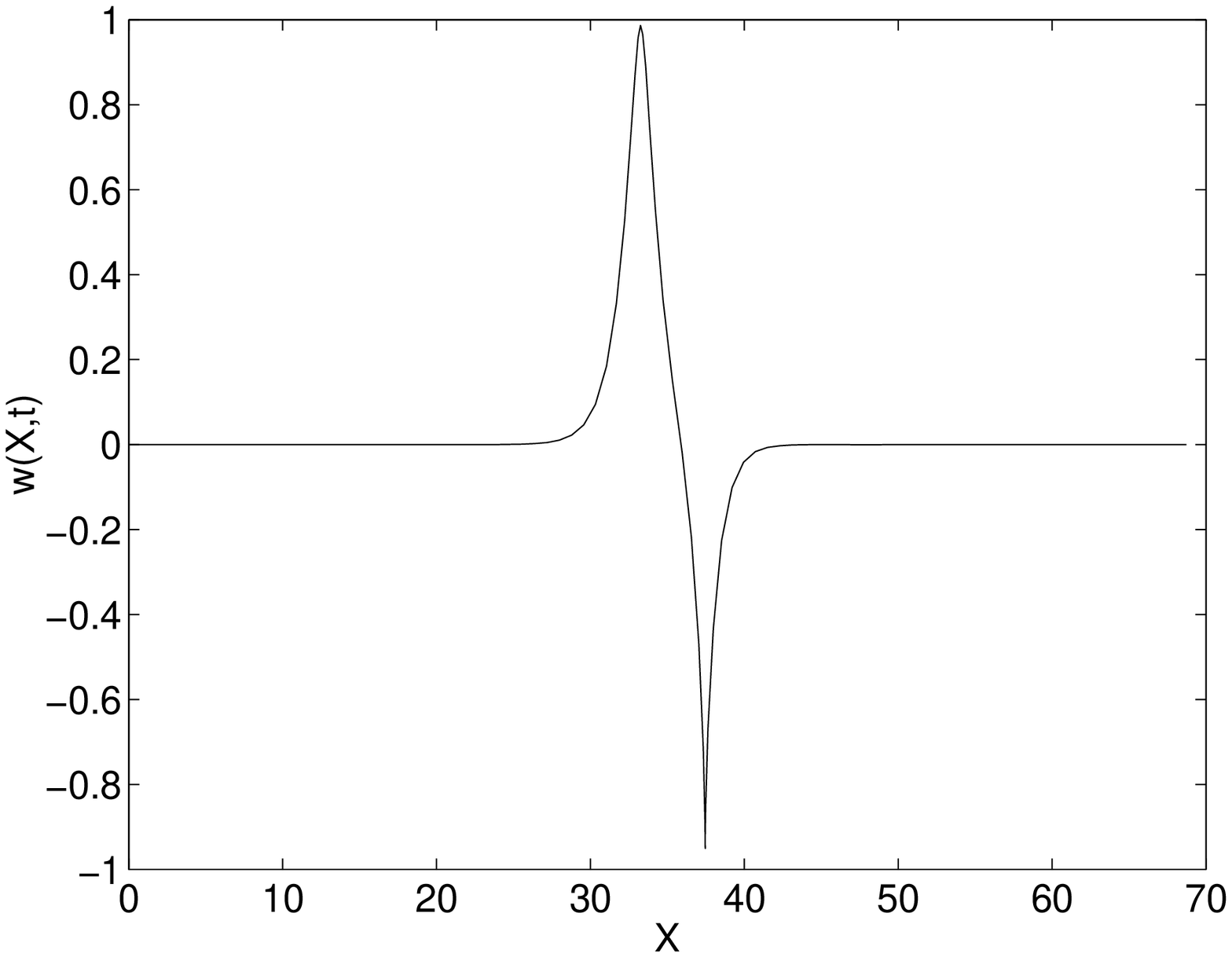}}
\kern-0.25\textwidth \hbox to
\textwidth{\hss(a)\kern0em\hss(b)\kern5em} \kern+0.25\textwidth
\centerline{
\includegraphics[scale=0.27]{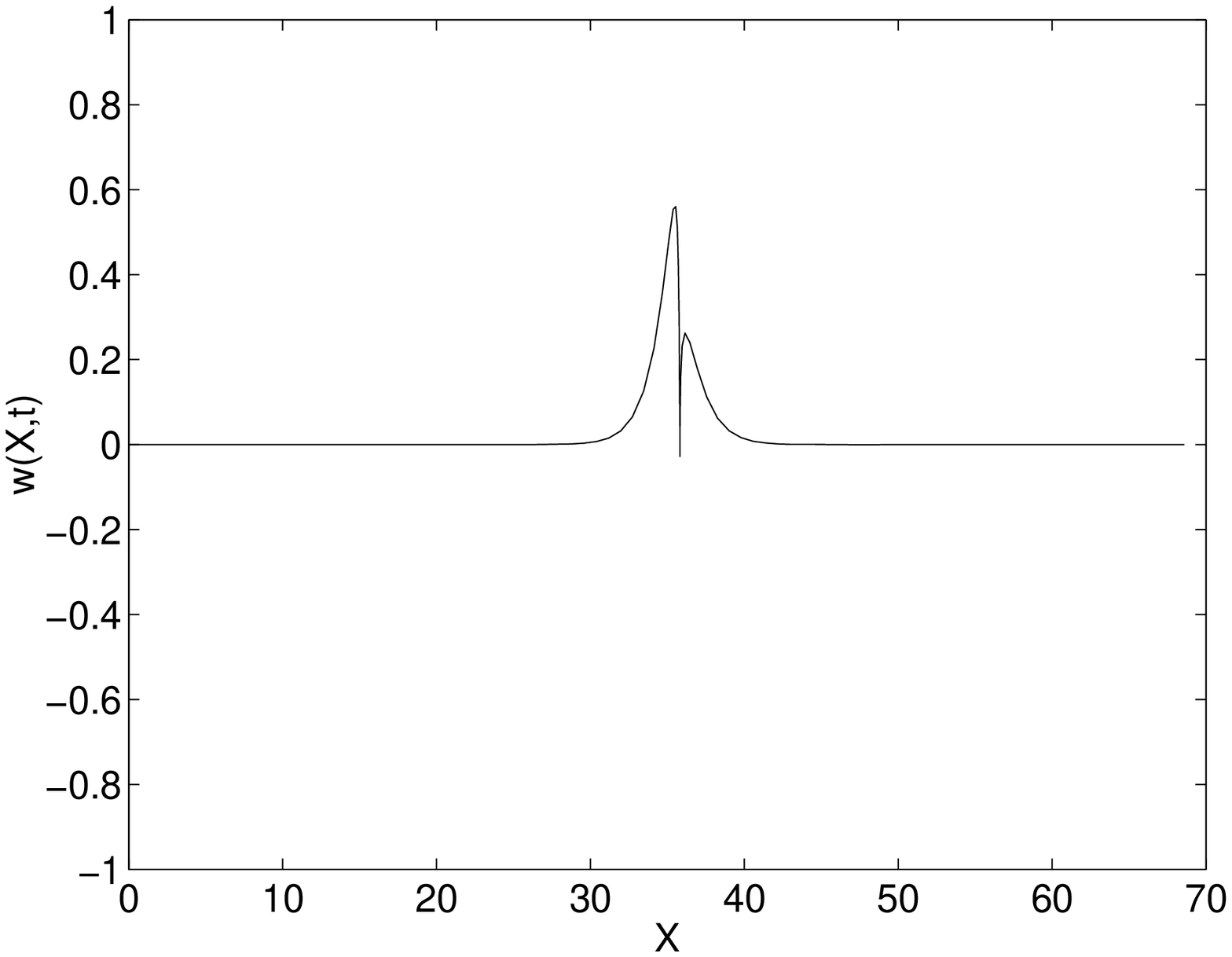}\quad
\includegraphics[scale=0.27]{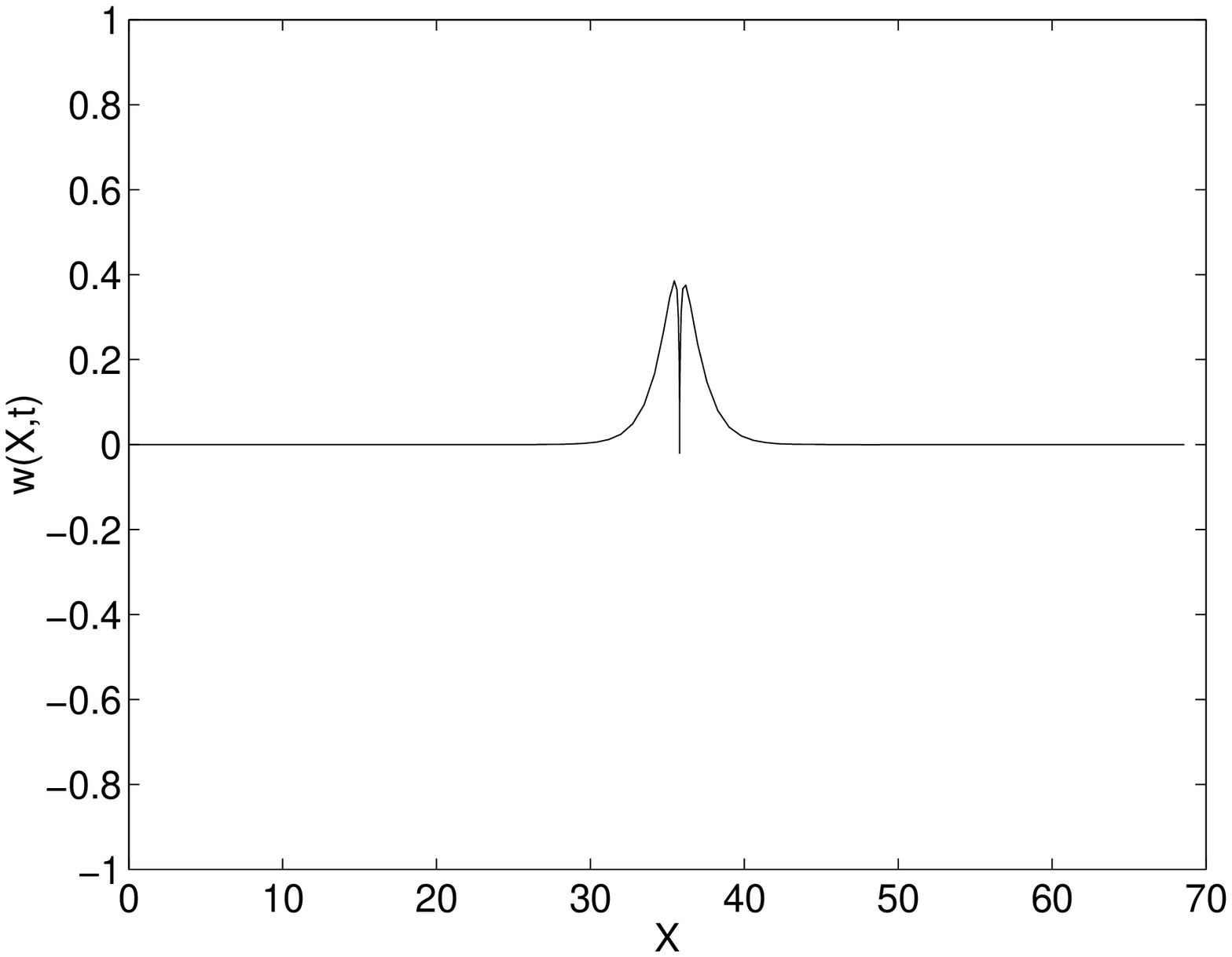}}
\kern-0.25\textwidth \hbox to
\textwidth{\hss(c)\kern0em\hss(d)\kern5em} \kern+0.25\textwidth
\centerline{
\includegraphics[scale=0.27]{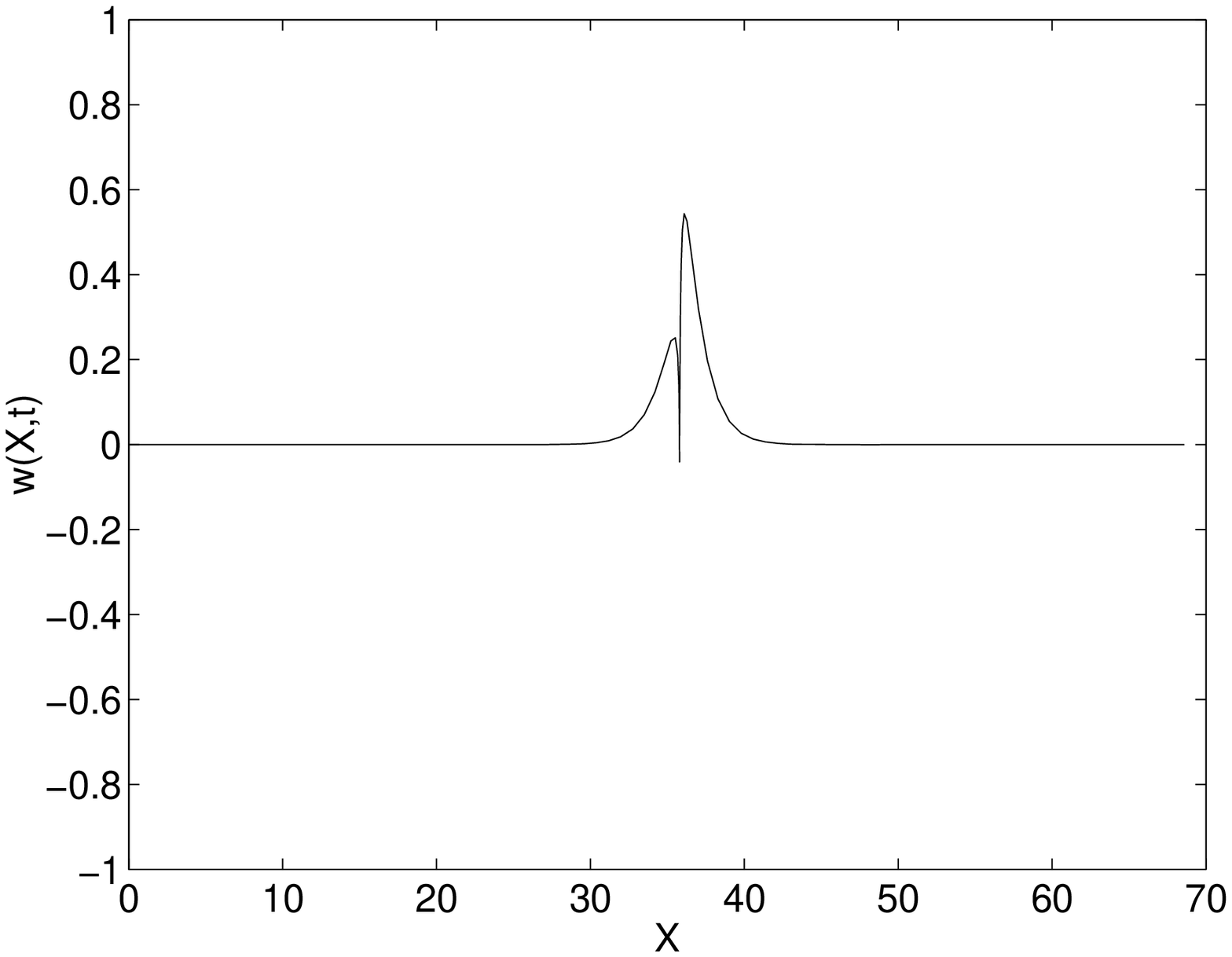}\quad
\includegraphics[scale=0.27]{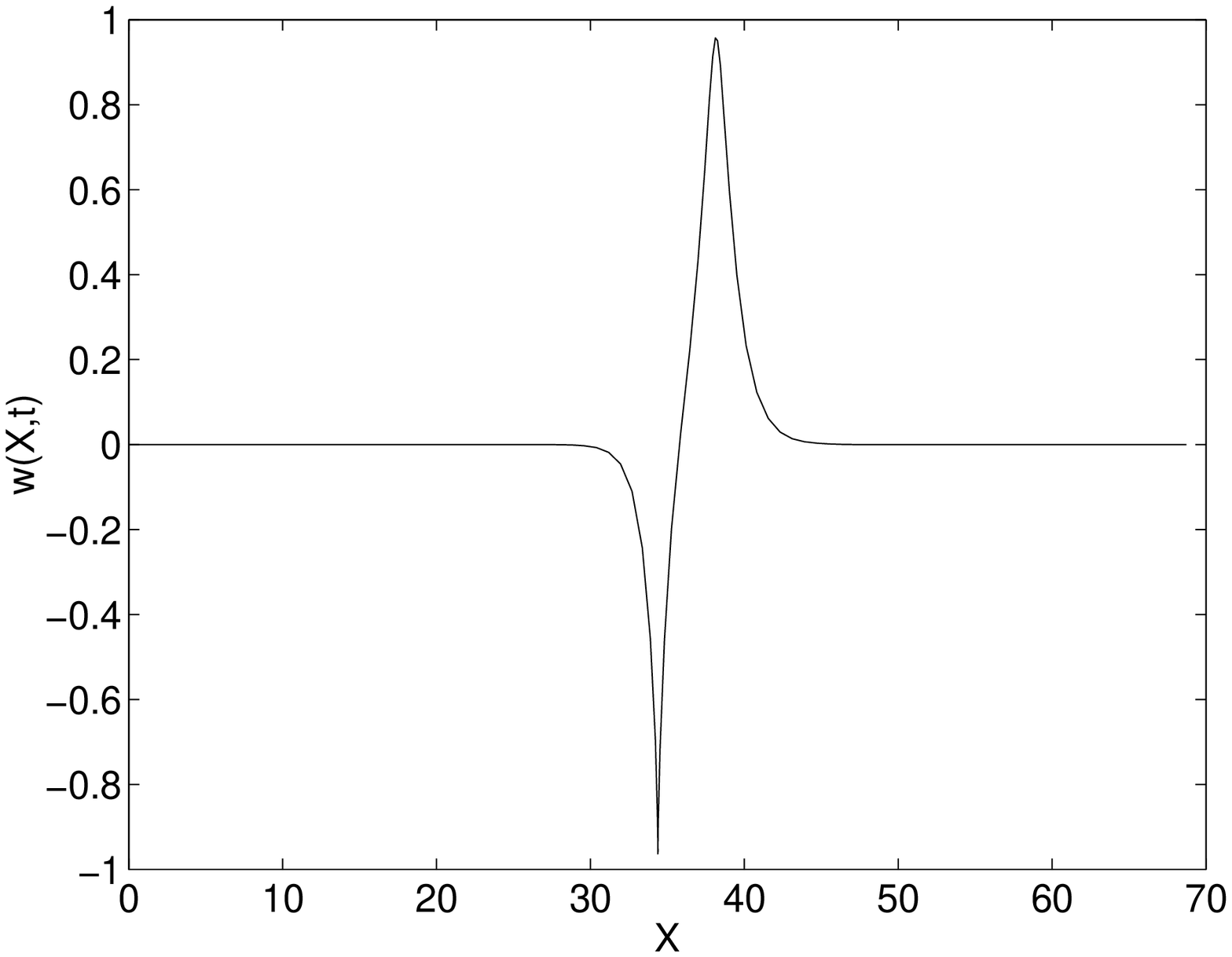}}
\kern-0.25\textwidth \hbox to
\textwidth{\hss(e)\kern0em\hss(f)\kern5em} \kern+0.25\textwidth
\centerline{
\includegraphics[scale=0.27]{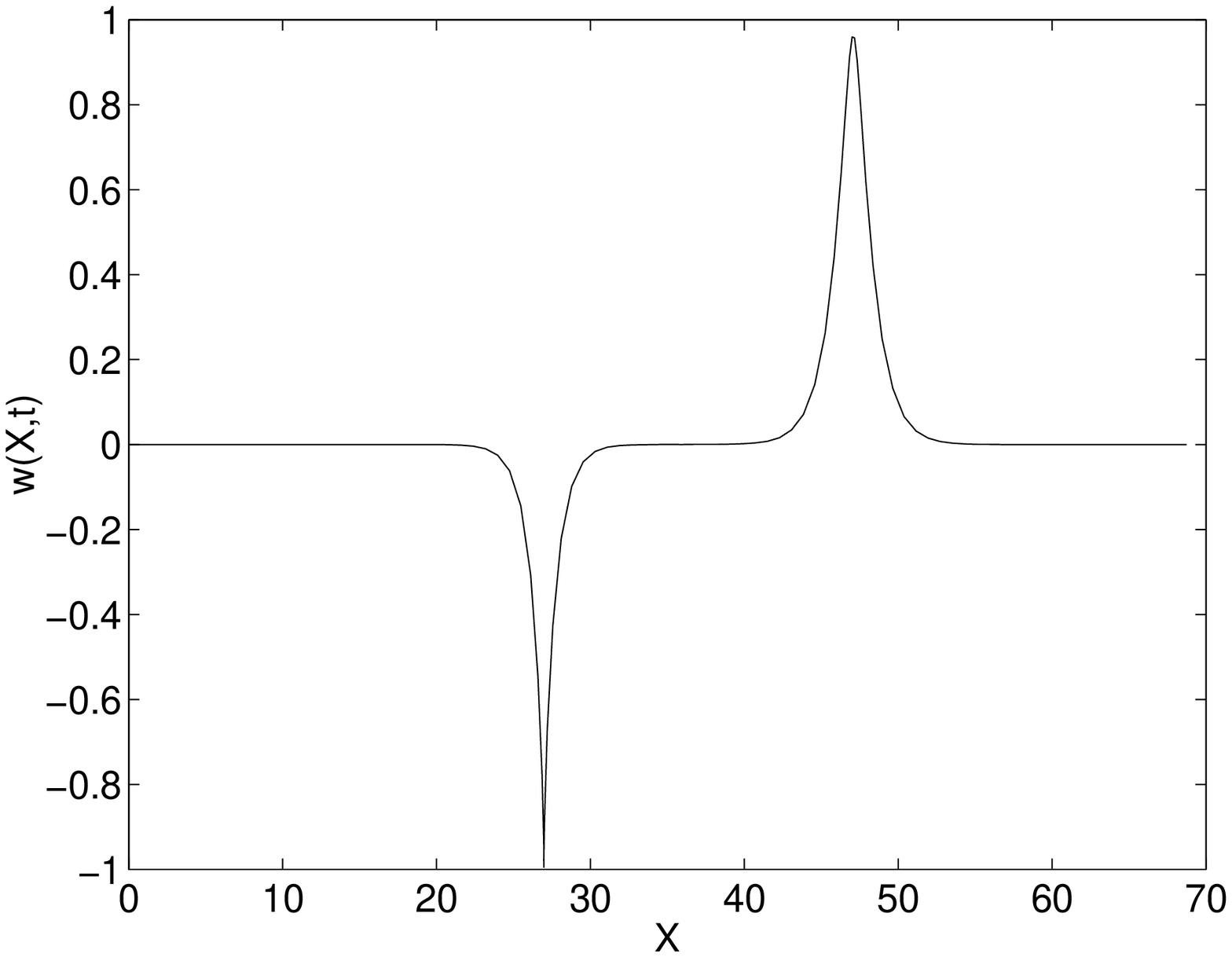}}
\kern-0.25\textwidth \hbox to \textwidth{\hss(g)\kern11em}
\kern+0.25\textwidth
\caption{Numerical solution for cuspon-soliton collision with
$p_1=9.12$, $p_2=10.98$ and $c=10.0$: (a) $t=0.0$; (b) $t=12.0$; (c)
 $t=14.4$; (d) $t=14.6$; (e) $t=14.8$; (f) $t=17.0$; (g) $t=25.0$.}
\label{f:cuspon_soliton}
\end{figure}

Here we show two examples for the soliton-cuspon interaction
with $c=10.0$. In Fig.\ref{f:cuspon_soliton},
we plot the interaction process between a soliton of $p_1=9.12$
and a cuspon of $p_2=10.98$ at several different times where the
soliton and the cuspon have almost the same amplitude. It can be
seen that when the collision starts ($t=12.0$), another singularity
point with infinite derivative ($w_x$) occurs. As collision goes on
($t=14.4, 14.6, 14.8$), the soliton seems 'eats up' the cuspon, and
the profile looks like a complete elevation. However, the cuspon
point exists at all times, especially, at $t=14.6$, the profile
becomes one symmetrical hump with a cuspon point in the middle of
the hump.

\begin{figure}[htbp]
\centerline{
\includegraphics[scale=0.27]{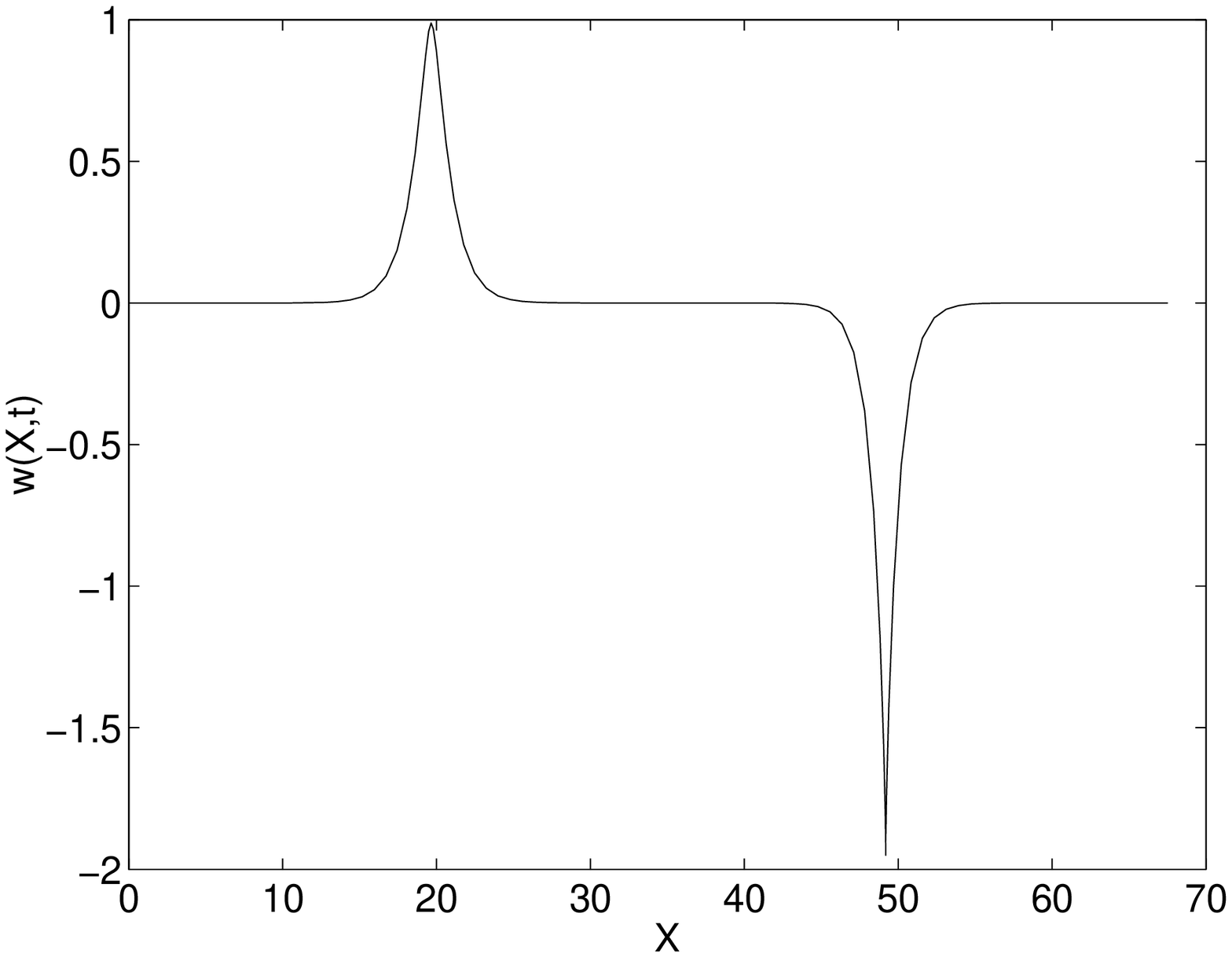}\quad
\includegraphics[scale=0.27]{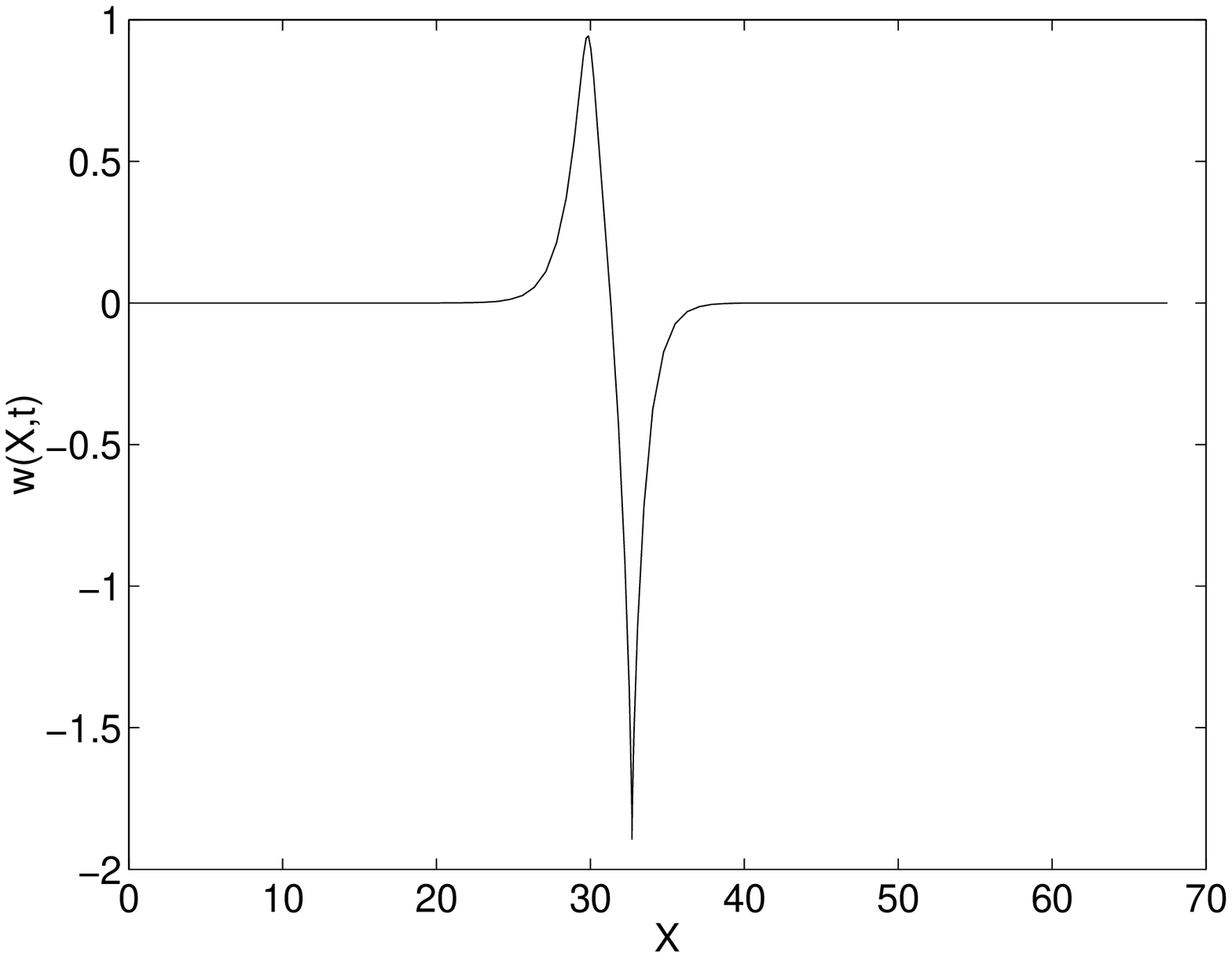}}
\kern-0.25\textwidth \hbox to
\textwidth{\hss(a)\kern0em\hss(b)\kern5em} \kern+0.25\textwidth
\centerline{
\includegraphics[scale=0.27]{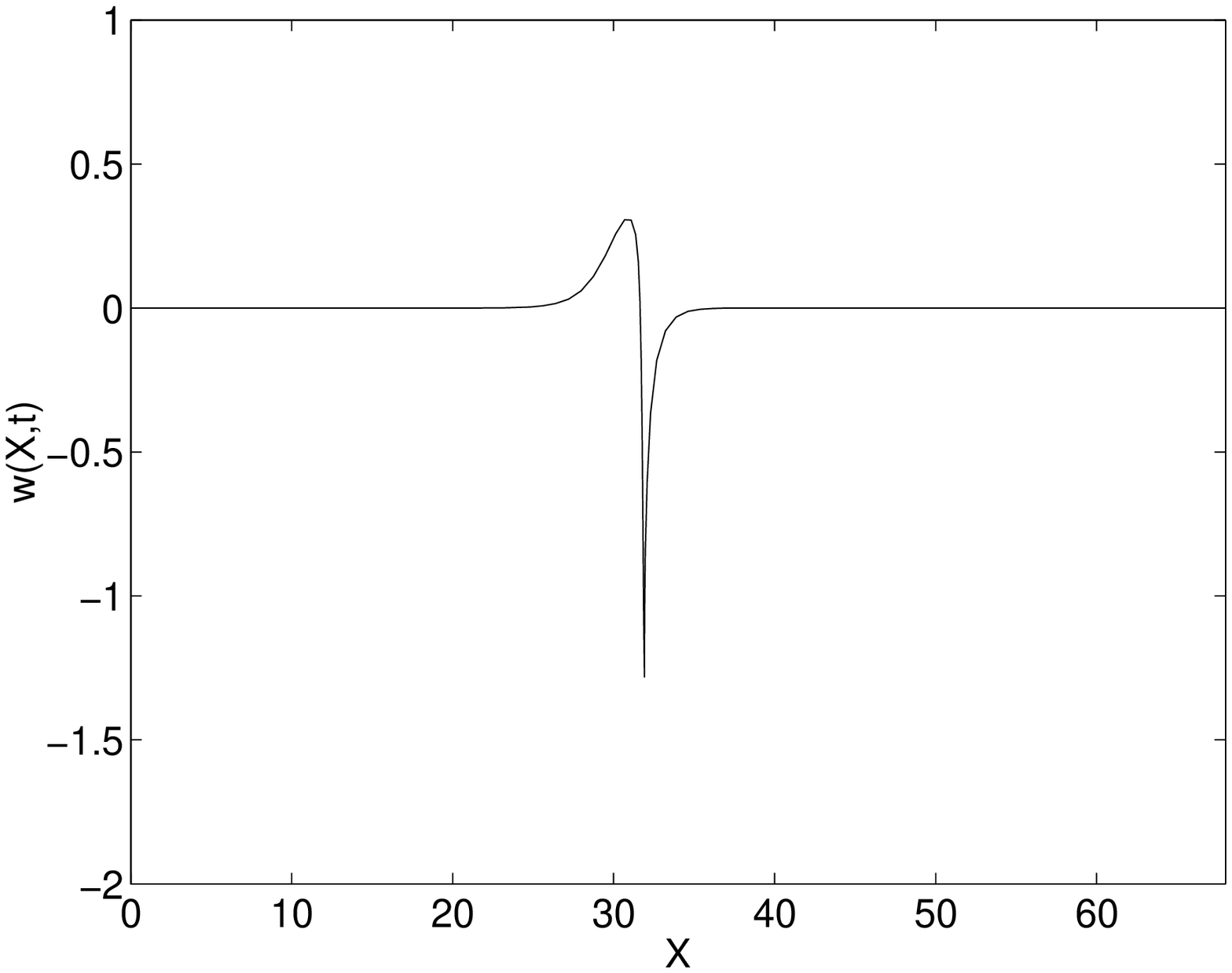}\quad
\includegraphics[scale=0.27]{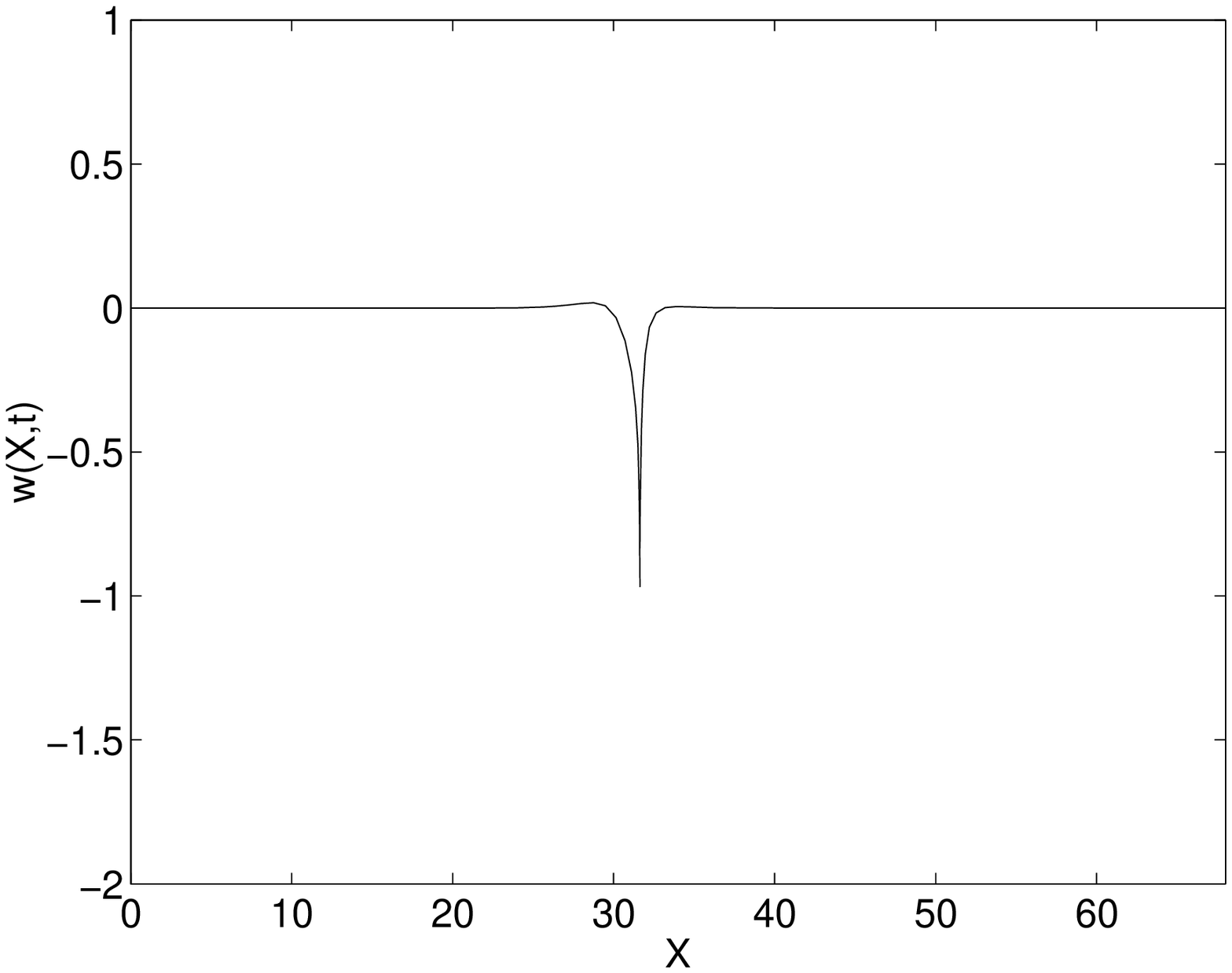}}
\kern-0.25\textwidth \hbox to
\textwidth{\hss(c)\kern0em\hss(d)\kern5em} \kern+0.25\textwidth
\centerline{
\includegraphics[scale=0.27]{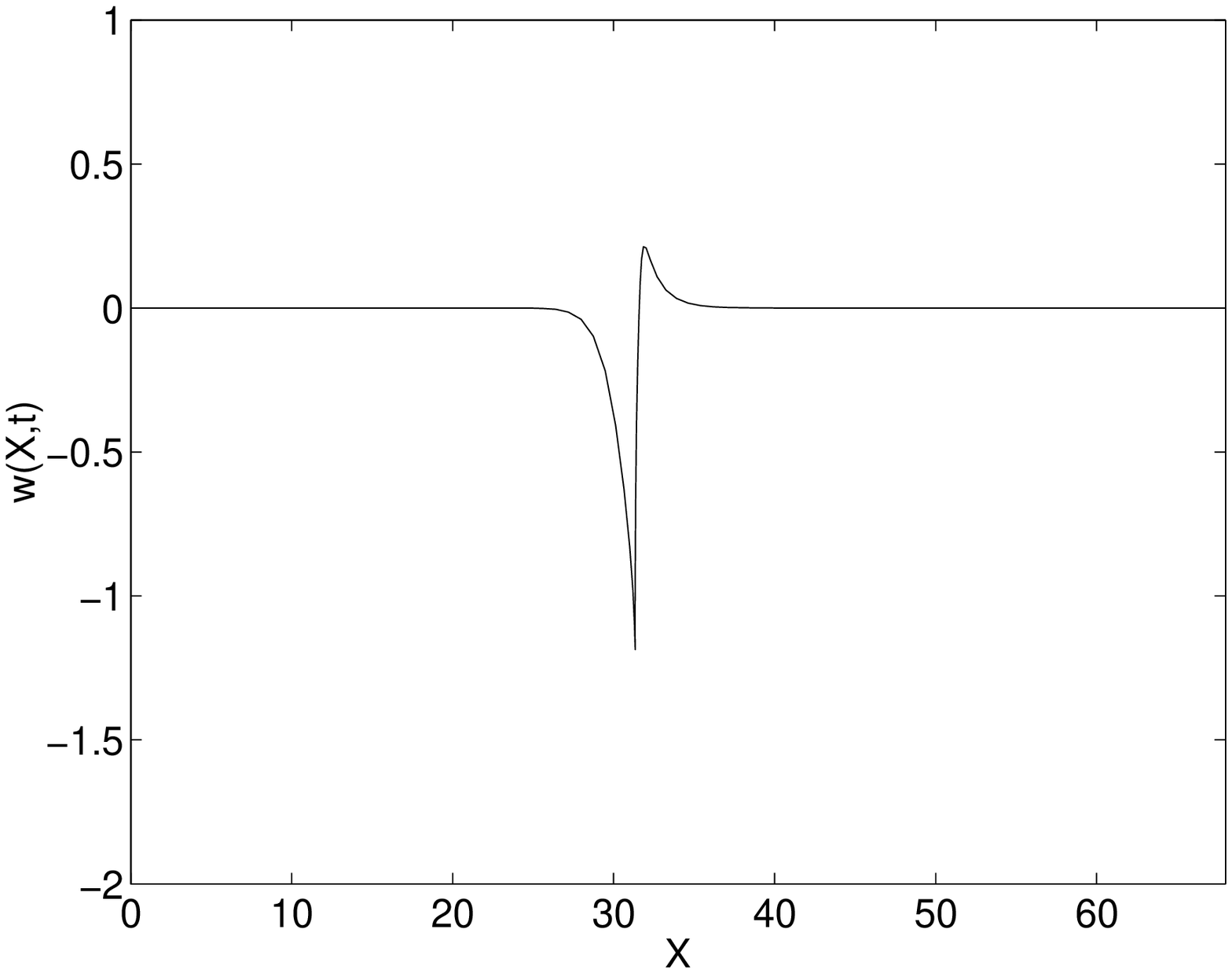}\quad
\includegraphics[scale=0.27]{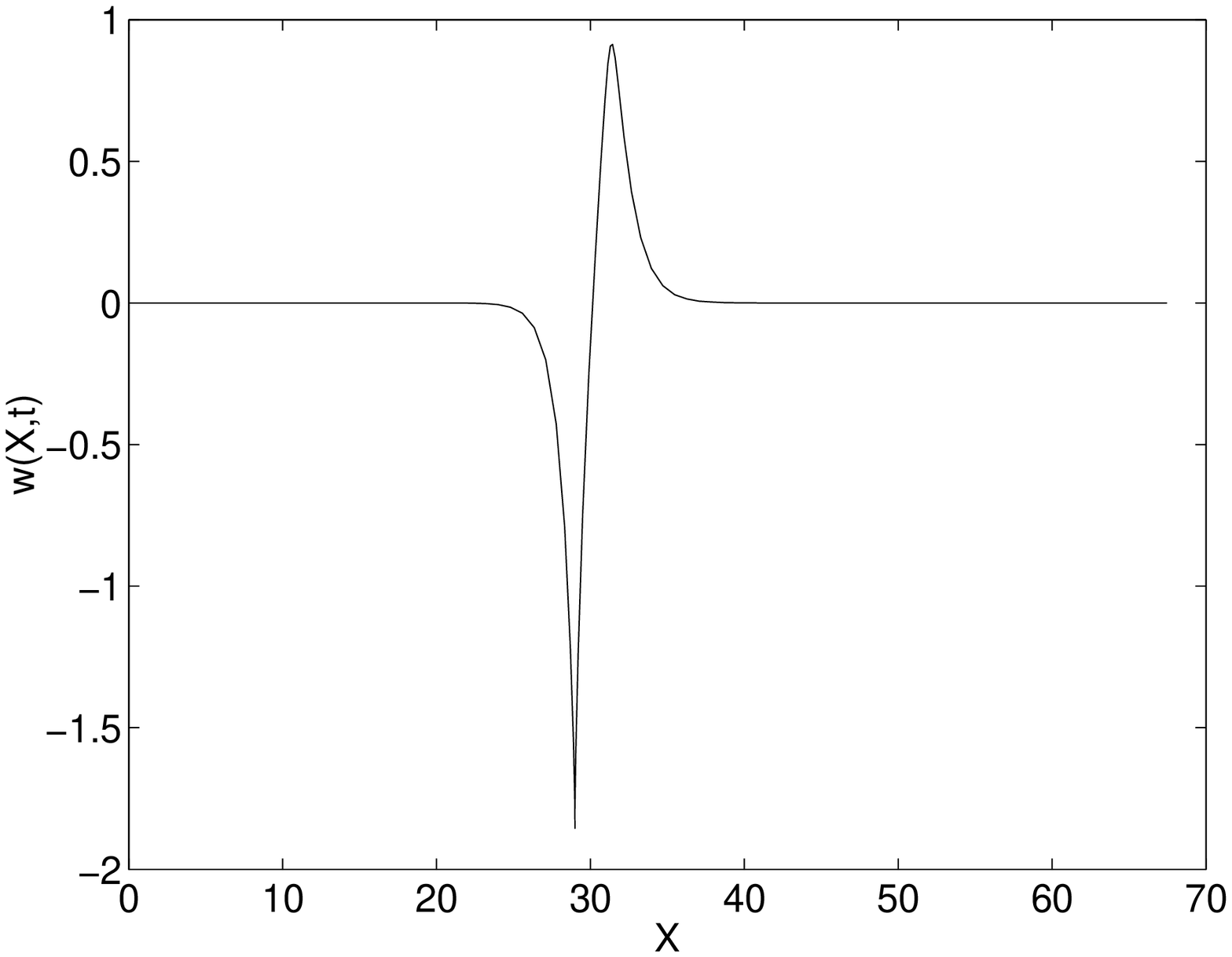}}
\kern-0.25\textwidth \hbox to
\textwidth{\hss(e)\kern0em\hss(f)\kern5em} \kern+0.25\textwidth
\centerline{
\includegraphics[scale=0.27]{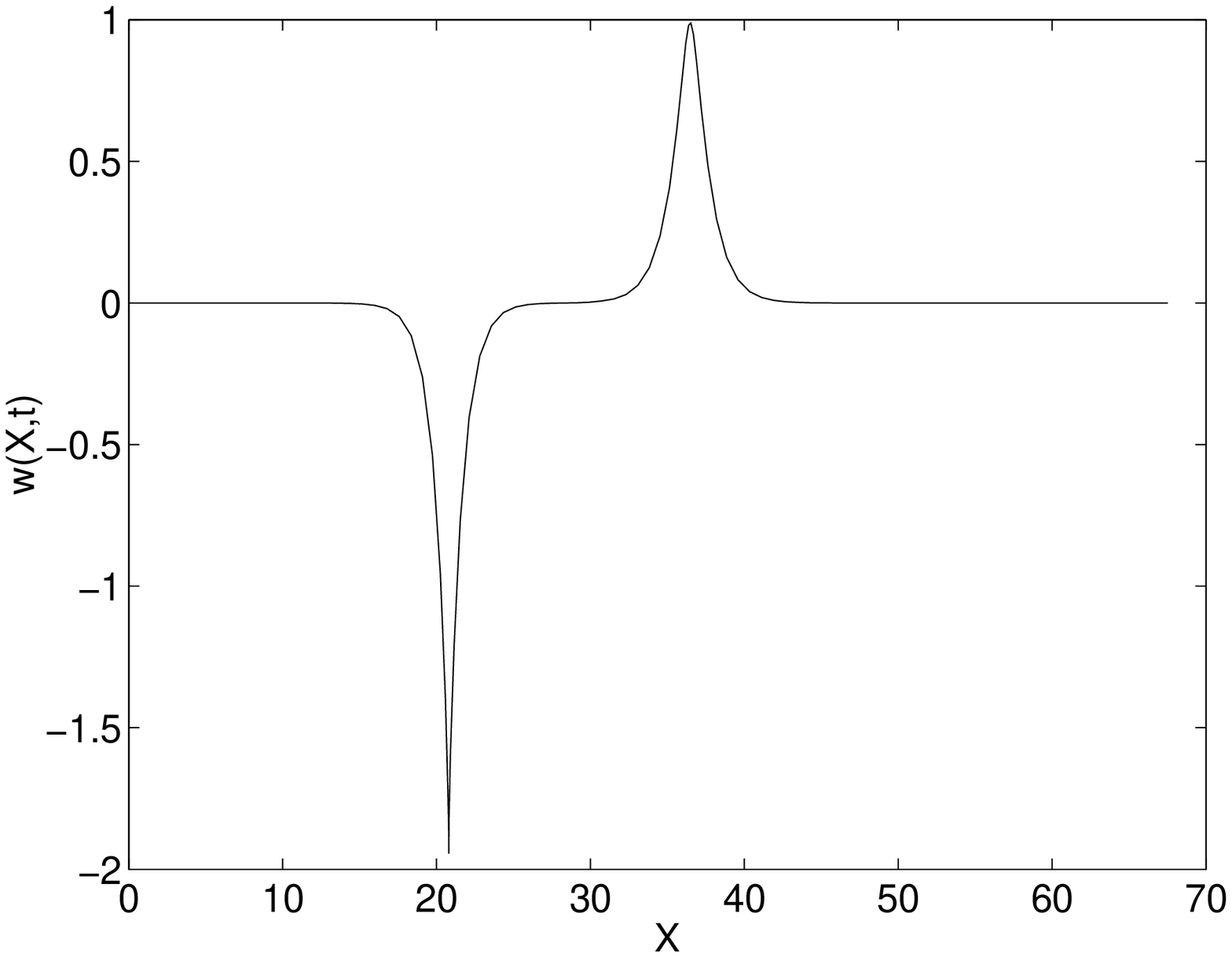}}
\kern-0.25\textwidth \hbox to \textwidth{\hss(g)\kern11em}
\kern+0.25\textwidth
%\quad
%\includegraphics[scale=0.4]{stcpp19p12p210p98t17.eps}}
%\kern-0.0\textwidth{\hss(g)\kern0em}
%\kern+0.355\textwidth
\caption{Numerical solution for cuspon-soliton collision with
$p_1=9.12$, $p_2=10.5$ and $c=10.0$: (a) $t=0.0$; (b) $t=9.0$; (c)
 $t=10.0$; (d) $t=10.3$; (e) $t=10.6$; (f) $t=11.5$; (g) $t=16.0$.}
\label{f:cuspon_soliton2}
\end{figure}

In Fig.\ref{f:cuspon_soliton2},
we present another example of a collision between a
soliton ($p_1=9.12$) and a cuspon ($p_2=10.5$) where the cuspon has
a larger amplitude ($2.0$) than the soliton ($1.0$). Again, when the
collision starts, another singularity point appears. As collision
goes on, the soliton is gradually absorbed by the cuspon. At
$t=10.3$, the whole profile looks like a single cuspon when the
soliton is completely absorbed. Later on, the soliton reappears from
the right until $t=16$, the soliton and cuspon recover their
original shapes except for a phase shift when the collision is
complete.

\begin{figure}[htbp]
\centerline{
\includegraphics[scale=0.35]{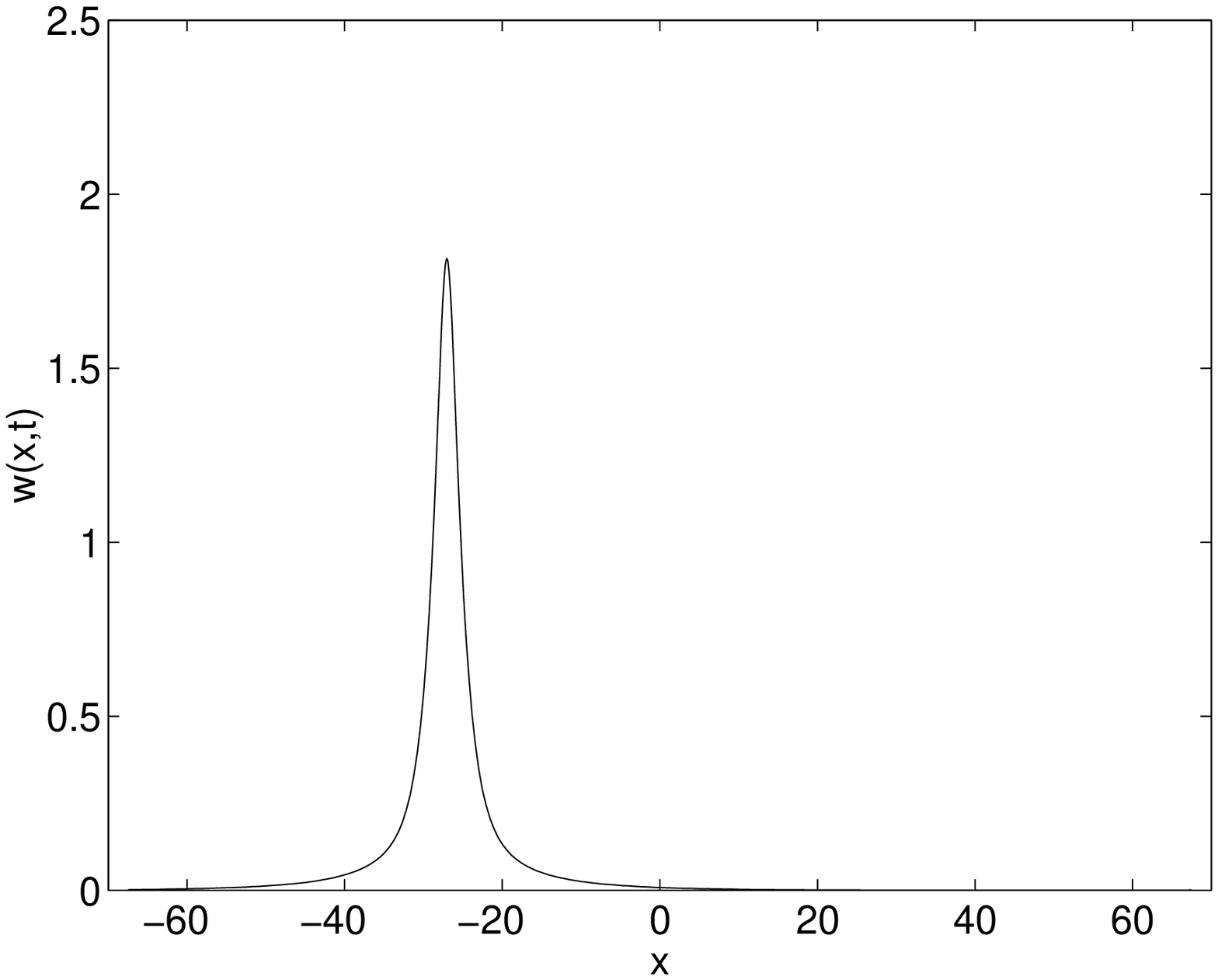}\quad
\includegraphics[scale=0.35]{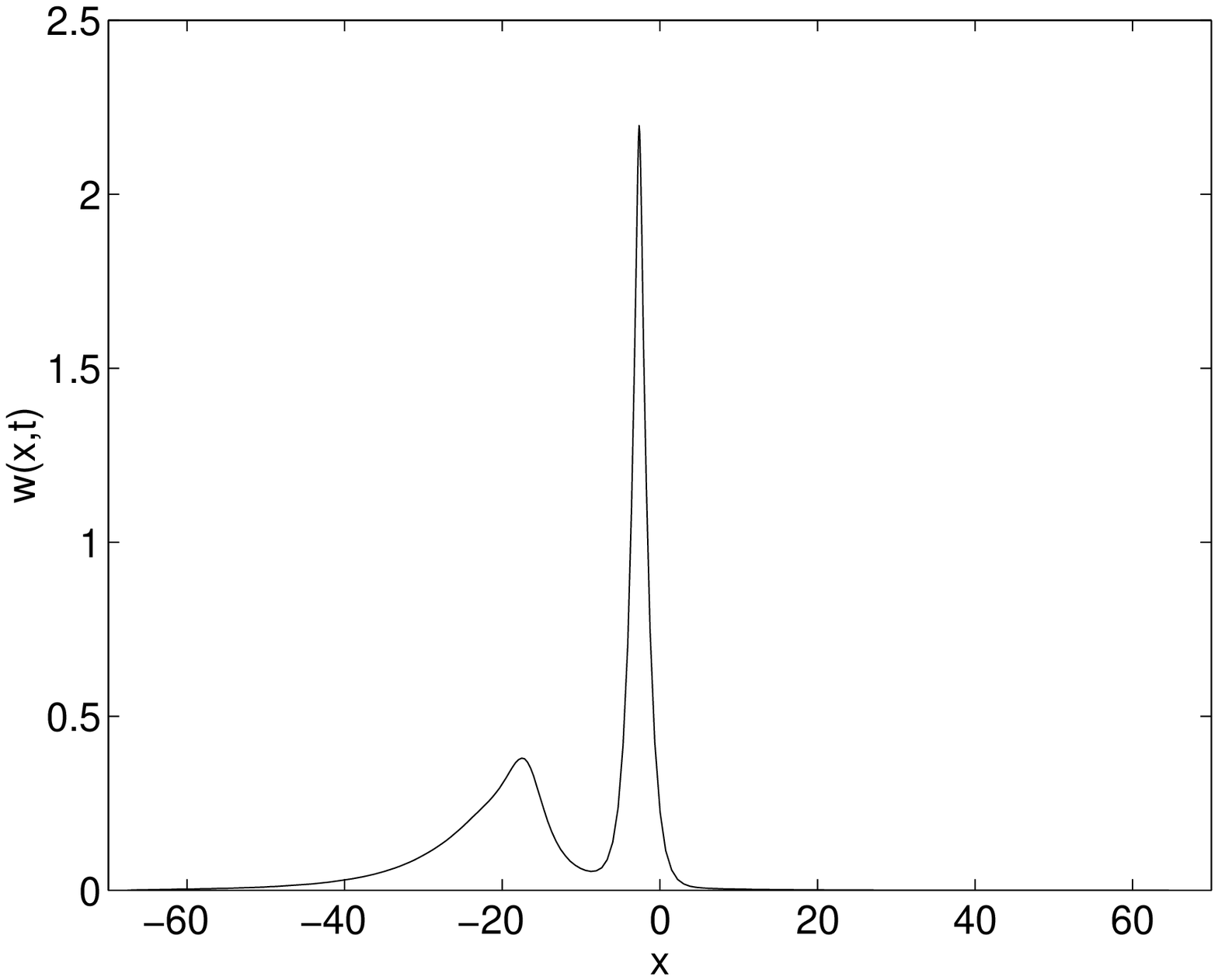}}
\kern-0.315\textwidth \hbox to
\textwidth{\hss(a)\kern0em\hss(b)\kern4em} \kern+0.315\textwidth
\centerline{
\includegraphics[scale=0.35]{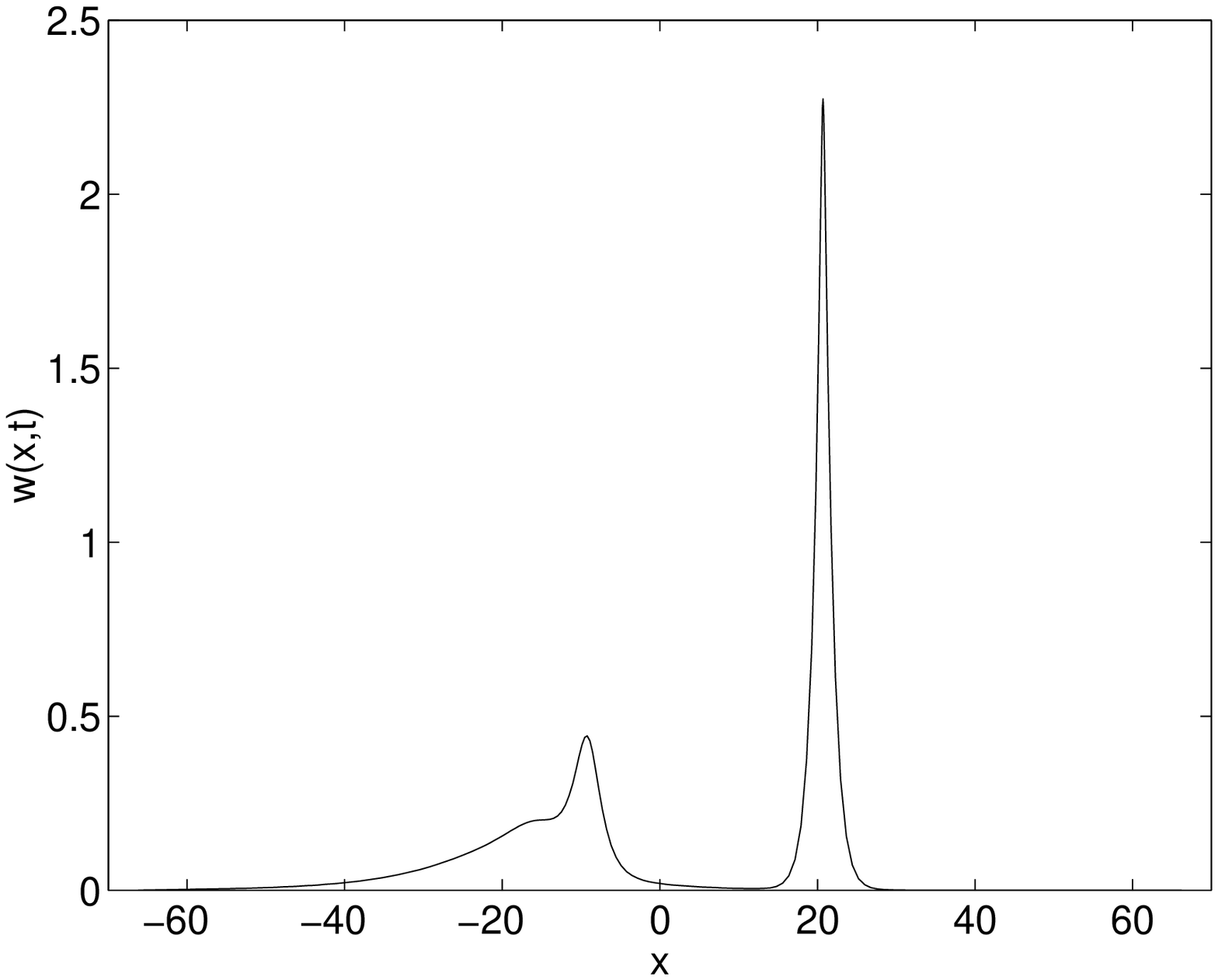}\quad
\includegraphics[scale=0.35]{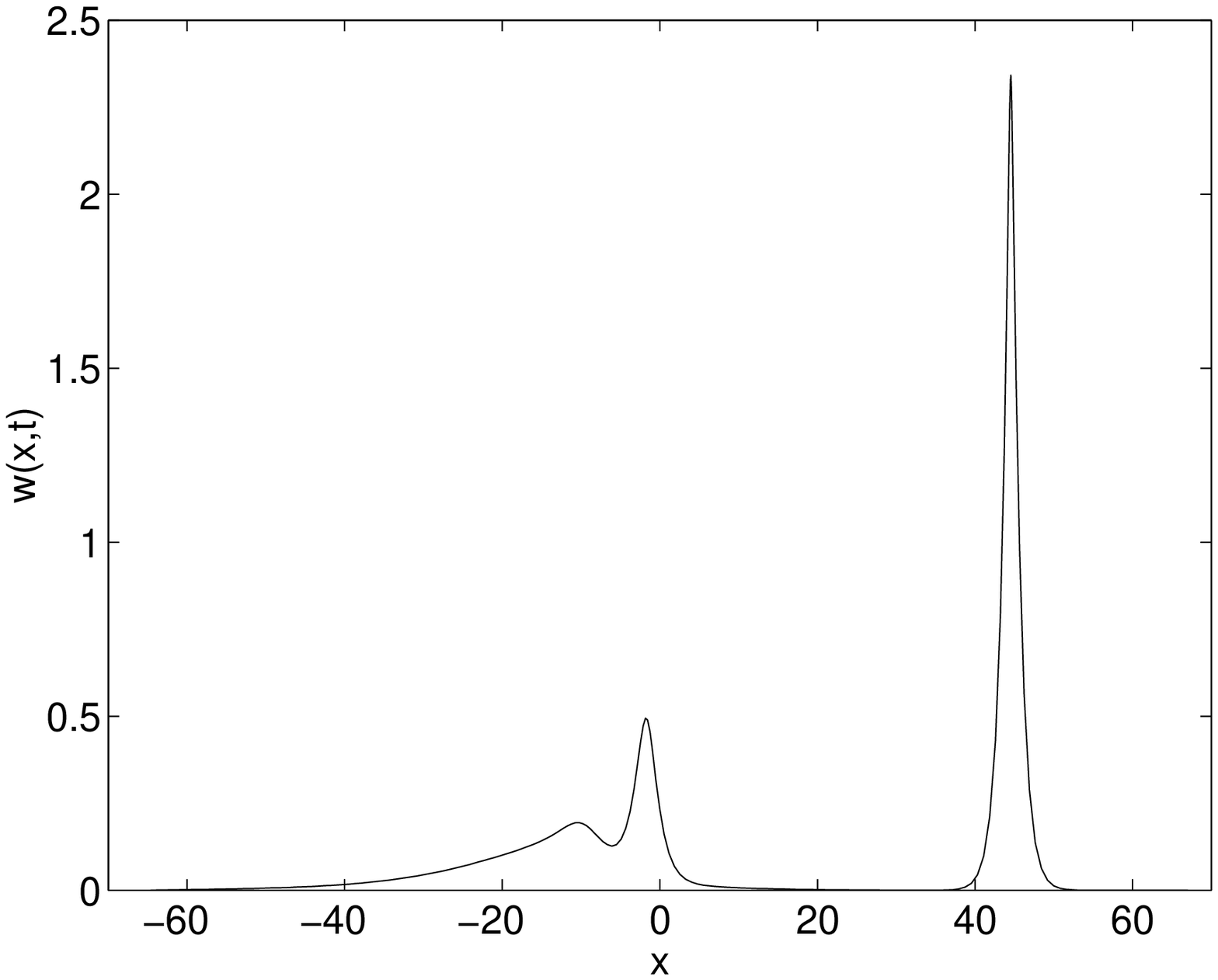}}
\kern-0.315\textwidth \hbox to
\textwidth{\hss(c)\kern0em\hss(d)\kern4em} \kern+0.315\textwidth
\caption{Numerical solution starting from an initial condition (19) with $c=10$: (a)
$t=0.0$; (b) $t=10.0$; (c)
 $t=20.0$; (d) $t=30.0$.}
\label{f:nonexact}
\end{figure}

\subsection{Non-exact initial value problems}
Here, we show that the integrable scheme can also be applied for the
initial value problem starting with non exact solutions. To the
end, we choose an initial condition whose mesh size is determined by
\begin{equation}\label{nonexact}
    \delta_k=2\,c\,h\,(1-0.8\, {\rm sech}(2kh-W_x/2)),
\end{equation}
then, the initial profile can be calculated through the second
equation of the semi-discretization, which is plotted in
Fig.\ref{f:nonexact} (a).
Figures \ref{f:nonexact} (b), (c) and (d) show the evolutions at $t=10,20,30$,
respectively.
Note that $c=10$ in this computation.
It can be seen that a soliton with large amplitude is
firstly developed, and moving fast to the right. By $t=30$, a second
soliton with small amplitude is to be developed.

\begin{figure}[htbp]
\centerline{
\includegraphics[scale=0.35]{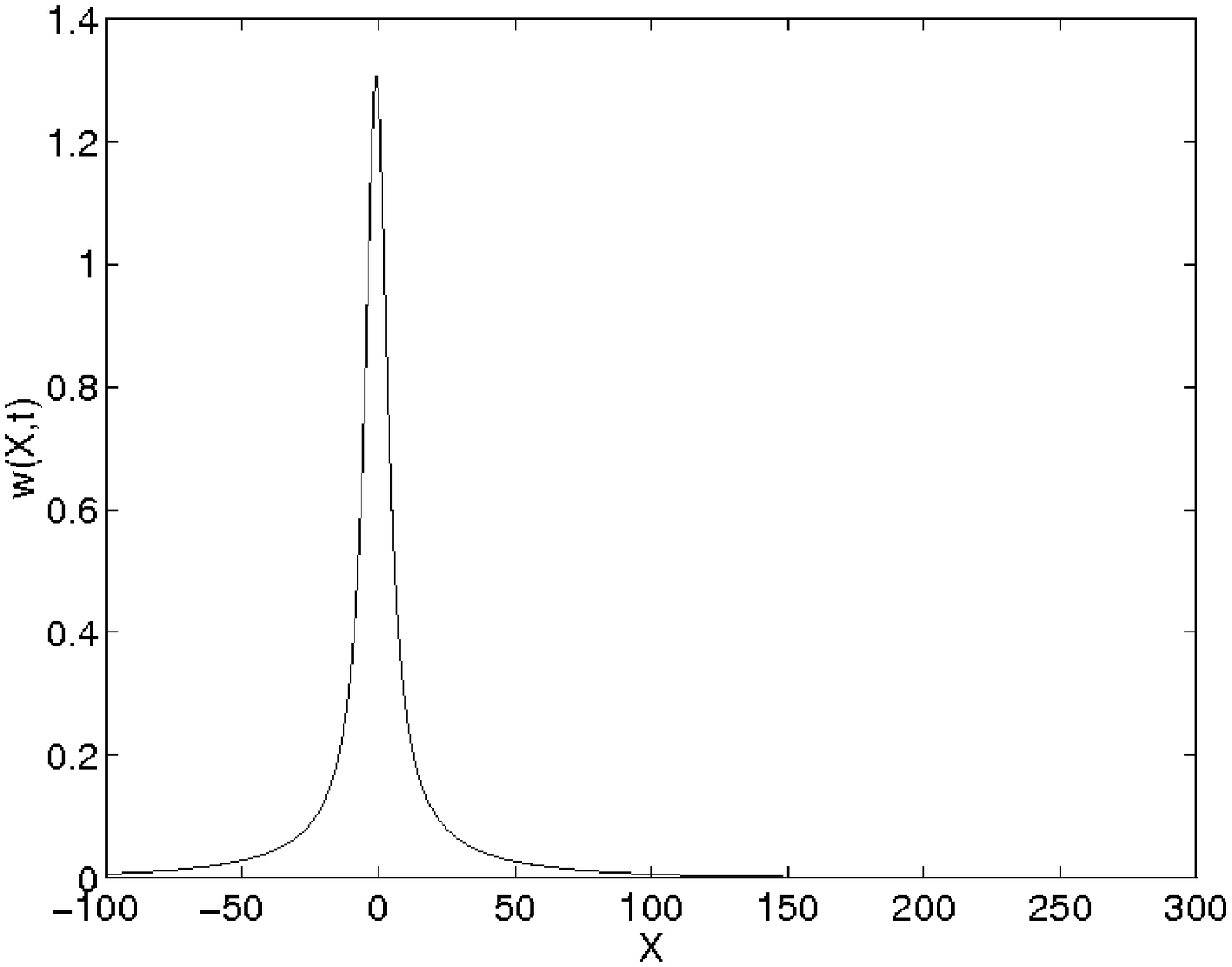}\quad
\includegraphics[scale=0.35]{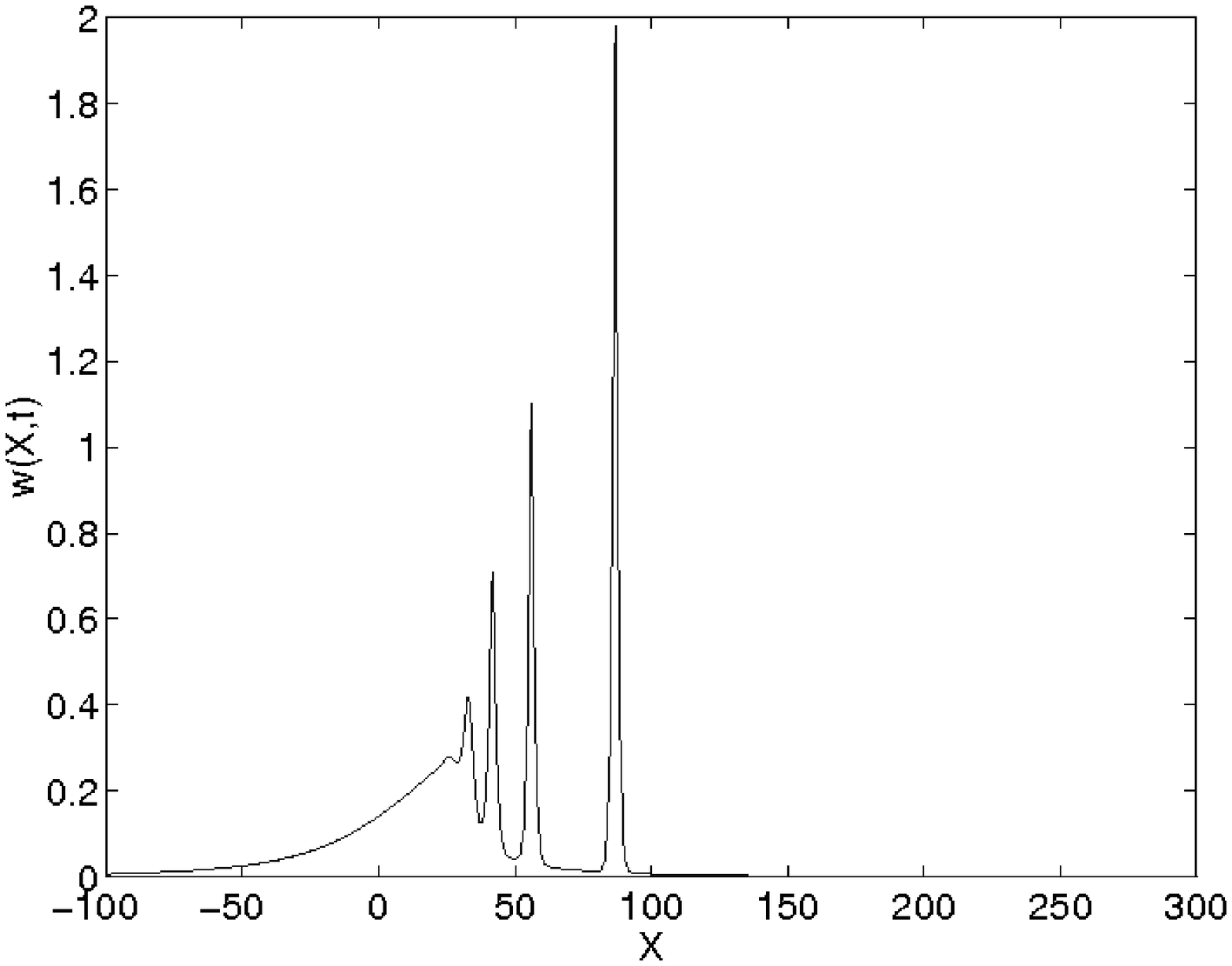}}
\kern-0.315\textwidth \hbox to
\textwidth{\hss(a)\kern0em\hss(b)\kern4em} \kern+0.315\textwidth
\centerline{
\includegraphics[scale=0.35]{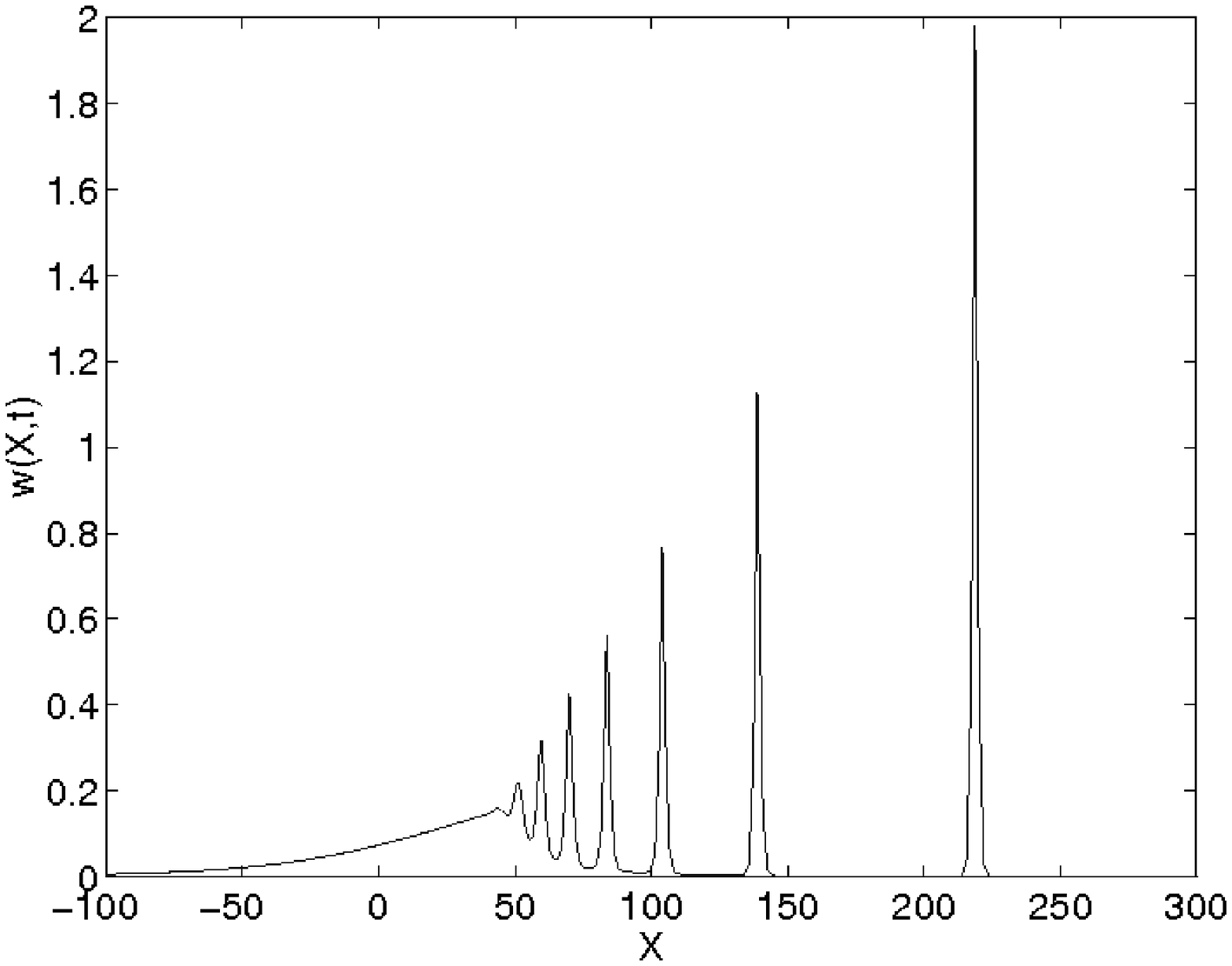}\quad
\includegraphics[scale=0.35]{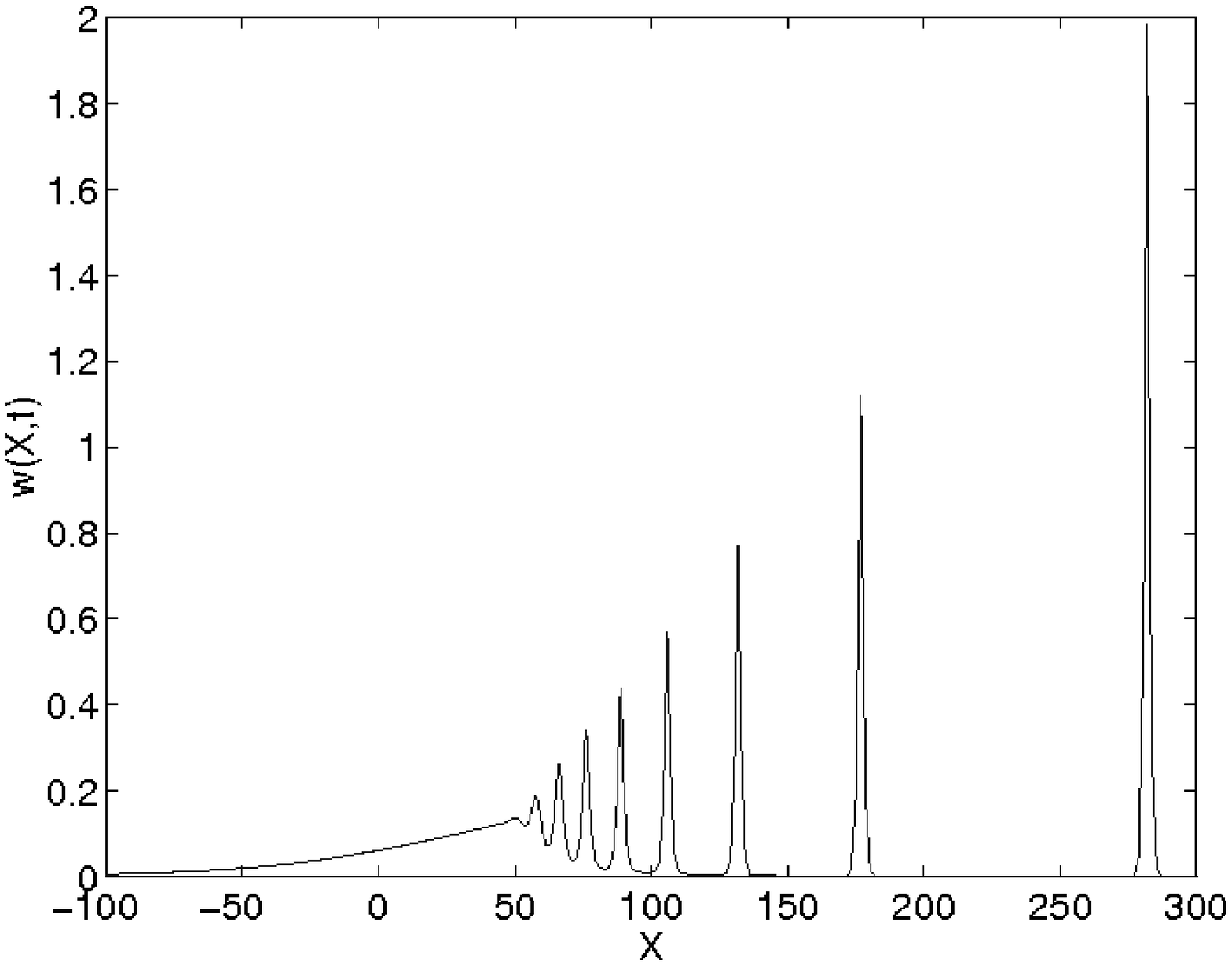}}
\kern-0.315\textwidth \hbox to
\textwidth{\hss(c)\kern0em\hss(d)\kern4em} \kern+0.315\textwidth
\caption{Numerical solution starting from an initial condition (19) with $c=90$: (a)
$t=0.0$; (b) $t=50.0$; (c)
 $t=150.0$; (d) $t=200.0$.}
\label{f:nonexactB}
\end{figure}

Next, we increase the value of $c$ to 90, which implies
a very small dispersion term, corresponding to the
dispersionless CH equation.
The initial profile and the evolutions at $t=50, 150, 200$
are shown in Fig.\ref{f:nonexactB}. It is seen that four
nearly-peakons are developed from the initial profile at $t=50$.
Later on, an array of nearly-peakons of seven and eight are developed
at $t=150, 200$, respectively. This result is similar to the
result for the KdV type equations with a small dispersion,
i.e. the peakon trains are generated.
(For the KdV type equations, soliton trains are generated.
For example, see \cite{Kamchatnov,El} for
numerical simulations and \cite{Karpman} for a theoretical analysis for
the KdV equation.)
A theoretical analysis
for the dispersionless CH equation to explain the above intriguing
numerical
result is called for.
%[Please add the numerical results here.]

\section{Concluding Remarks}
In the present paper, we have proposed a self-adaptive moving mesh method for the
CH equation, which based on an integrable
semi-discretization of the CH equation. It has the properties: (1)
it is integrable in the sense that the scheme itself admits
the $N$-soliton solution approaching to the $N$-soliton solution of the CH
equation in the limit of mesh size going to zero; (2) the mesh is
non-uniform and is automatically adjusted so that it is concentrated
in the region where the solution changed sharply, for example, the
cuspon point; (3) once the non-uniform mesh is evolved, the solution
is determined from the evolved mesh by solving a tridiagonal linear system.
Therefore, either from the accuracy or
from the computation cost, the proposed method is expected to be
superior than other existing numerical methods of the CH equation.
This is indeed true. The numerical results in this paper
indicate that a very good
accuracy is obtained.

Two time advancing methods, the modified forward Euler method and the
classical 4th-order Runge-Kutta method, are used to solve the
evolution of non-uniform mesh. The Runge-Kutta method gains much
better accuracy than the modified forward Euler method. However, it
fails for the computations of cuspons.
Using the self-adaptive moving mesh method for the CH equation,
we have obtained interesting numerical computation results
starting with non-exact solutions.
When $\kappa$ is very small, the peakon train is generated from the
non-exact initial condition.

%!
As further topics, it is interesting to construct integrable
discretizations, or, the self-adaptive moving mesh methods for a class of integrable
nonlinear wave equations possessing
soliton solutions with singularities such as peakon,
cuspon or loop solutions. For example,
such equations include
the short pulse equation which
was
derived as a model for the
propagation of ultra-short optical pulses in nonlinear
media \cite{SP},
\begin{equation}
u_{XT}=u+\frac{1}{6}(u^3)_{XX},
\end{equation}
and the Degasperis-Procesi (DP) equation \cite{DP}
\begin{equation}
u_T+3\kappa^3u_X-u_{TXX}+4uu_X=3u_Xu_{XX}+uu_{XXX}.
\end{equation}
It is worth pointing out that the authors have constructed semi-
and full-discretization for the short pulse equation, which is
another example of the self-adaptive moving mesh method \cite{FMO09}, in which we have succeeded in computing the one- and two-loop soliton propagations and interactions.

\section*{Acknowledgments}
The work of B.F. was partially supported by the U.S. Army Research Office
under Contract No. W911NF-05-1-0029. 
The work of Y.O. was partly supported by JSPS Grant-in-Aid
for Scientific Research (B-19340031, S-19104002).
Y.O. and K.M. are grateful for
the hospitality of the Isaac Newton Institute for Mathematical
Sciences (INI) in Cambridge where this article
was completed during the programme Discrete Integrable Systems (DIS).
The authors are grateful to the anonymous referee for valuable comments.


\begin{thebibliography}{Oo}
\def\title#1{{\rm#1}}

\bibitem{CH} R. Camassa, D.D. Holm,
An integrable shallow water equation with peaked solitons, Phys.
Rev. Lett. {\bf 71} (1993) 1661-1664.

\bibitem{FF}  B. Fuchssteiner, A. Fokas, Symplectic structures, their
    B\"{a}cklund transformations and hereditary symmetries, Physica
    D {\bf 4} (1981) 47-66.

\bibitem{CHH}  R. Camassa, D.D. Holm,  J.M. Hyman,
A new integrable shallow water equation, Adv. Appl. Mech. {\bf 31}  (1994) 1-33.

\bibitem{Beals}
R. Beals, D. H. Sattinger and J. Szmigielski,
Multipeakons and the Classical Momemt Problem,
Adv. Math. {\bf 154} (2000) 229-257.

\bibitem{Johnson02} R.S. Johnson,
Camassa-Holm, Korteweg-deVries and related models for waterwaves,
J. Fluid Mech. {\bf 457}  (2002) 63-82.

\bibitem{Constantin08} A. Constantin, D. Lannes, The Hydrodynamical
relevance of the
Camassa-Holm and Degasperis-Procesi
equations, Arch. Rational Mech. Anal. {\bf 192} (2009) 165-186

\bibitem{Johnson03} R.S. Johnson,
The Camassa-Holm equation for water waves moving over a shear flow,
Fluid Dynam. Res. {\bf 33}  (2003) 97-111.

\bibitem{Busuioc} V. Busuioc,
On second grade fluids with vanishing viscosity,
C. R. Acad. Sci. Paris Ser. I {\bf 328}  (1999) 1241-1246.

\bibitem{Dai} H.-H. Dai,
Exact travelling-wave solutions of an integrable equation arising
in hyperelastic rods,
Wave Motion {\bf 28}  (1998) 367-381.

\bibitem{Constantin}
A. Constantin, On the scattering problem for the Camassa-Holm
equation, {Proc. R. Soc. London A} {\bf 457} (2001) 953-970.

\bibitem{Constantin2}
A. Constantin, V. S. Gerdjikov and R. I. Ivanov,
Inverse Scattering Transform for the Camassa-Holm equation,
{Inv. Prob.} {\bf 22} (2006)
2197-2207.

\bibitem{Schiff}  J. Schiff, The Camassa-Holm equation:
A loop group approach, Physica D {\bf 121} (1998) 24-43

\bibitem{Kraenkel2}
M. C. Ferreira, R. A. Kraenkel,  A. I. Zenchuk, Soliton-cuspon
interaction for the Camassa-Holm equation, {J. Phys. A: Math. Gen.}
{\bf 32} (1999) 8665-8670.

\bibitem{Johnson}  R.S. Johnson,
On solutions of the Camassa-Holm equation, Proc. R. Soc. London {\bf
A459} (2003) 1687-1708.

\bibitem{Li}  Y. Li, J.E. Zhang,
The multiple-soliton solution of the Camassa-Holm equation,
Proc. R. Soc. London {\bf A460}
    (2004) 2617-2627.

\bibitem{Parker}  A. Parker, On the Camassa-Holm equation and a
direct method of solution. I. Bilinear form and solitary waves,
Proc. R. Soc. London {\bf A460}
     (2004) 2929-2957.

\bibitem{Parker2}  A. Parker, On the Camassa-Holm equation and a
direct method of solution. II. Soliton solutions, Proc. R. Soc.
London {\bf A461}
    (2005) 3611-3632.

\bibitem{DaiLi} H.-H. Dai, Y. Li,
The interaction of the $\omega$-soliton and the $\omega$-cuspon of
the Camassa-Holm equation, J. Phys. A: Math. Gen. {\bf 38}
    (2005) L685-L694.

\bibitem{Matsuno}
Y. Matsuno, Parametric Representation for the Multisoliton Solution
of the Camassa-Holm Equation, {J. Phys. Soc. Jpn.} {\bf 74} (2005)
1983-1987.

\bibitem{Kalisch}
H. Kalisch, J. Lenells, Numerical study of traveling-wave solutions
for the Camassa-Holm equation, {Chaos, Solitons \& Fractals} {\bf
25}  (2005) 287-298.

\bibitem{Holden1}
H. Holden, X. Raynaud, Convergence of a Finite Difference Scheme for
the Camassa-Holm Equation, {SIAM J. Numer. Anal.}  {\bf 44}
 (2006) 1655-1680.

\bibitem{Coclite1}
G. M. Coclite, K. H. Karlsen, N. H. Risebro, A Convergent Finite
Difference Scheme for the Camassa-Holm Equation with General $H^ 1$
Initial Data, {SIAM J. Numer. Anal.} {\bf 46} (2008) 1554-1579.

\bibitem{Artebrant}
R. Artebrant, H. J. Schroll, Numerical simulation of Camassa-Holm
peakons by adaptive upwinding, {Appl. Numer. Math.} {\bf 56} (2006)
695-711.

\bibitem{Chiwang} Y. Xu, C.-W. Shu,
A local discontinuous Galerkin method for
 the Camassa-Holm equation, {SIAM J. Numer. Anal.}  {\bf 46}  (2008) 1998-2021.

\bibitem{Matsuo} T. Matsuo, H. Yamaguchi,
An energy-conserving Galerkin scheme for a class of
nonlinear dispersive equations, {J. Comput. Phys.}  {\bf 228} (2009)
    4336-4358.

\bibitem{Matsuo2} T. Matsuo,
A Hamiltonian-conserving Galerkin scheme for the Camassa-Holm,
 {J. Comput. Appl. Math.} (2009) In press.

\bibitem{Cohen} D. Cohen, B. Owren,  X. Raynaud,
Multi-symplectic integration of the Camassa-Holm equation, {J. Comp.
Phys.}  {\bf 227}  (2008) 5492-5512.

\bibitem{Camassa1}
R. Camassa, Characteristics and the initial value problem of a
completely integrable shallow water equation, Discrete Cont. Dyn.-B
{\bf 3} (2003) 115-139.

\bibitem{Lee1}
R. Camassa, J. Huang,  L. Lee, On a completely integrable numerical
scheme for a nonlinear shallow-water wave equation, {J. Nonlinear
Math. Phys.} {\bf 12}  (2005) 146-162.

\bibitem{Lee2}
R. Camassa, J. Huang,  L. Lee, Integral and integrable algorithms
for a nonlinear shallow-water wave equation, {J. Comp. Phys.} {\bf
216}  (2006) 547-572.


\bibitem{Lee3}
R. Camassa, L. Lee, Complete integrable particle methods and the
recurrence of initial states for a nonlinear shallow-water wave
    equation,
{J. Comp. Phys.} {\bf 227}  (2008) 7206-7221.


\bibitem{Holden2}
H. Holden, X. Raynaud, A convergent numerical scheme for the
Camassa-Holm equation based on multipeakons, {Discrete Contin. Dyn.
Syst.}  {\bf 14} (2006) 505-523.


\bibitem{Ohta}  Y. Ohta, K. Maruno,  B.-F. Feng,
An integrable semi-discretization of the Camassa-Holm equation and
its determinant solution, J. Phys. A {\bf 41}   (2008) 355205.


\bibitem{HH83} A. Harten, J.M. Hyman,
Self-adjusting grid methods for one-dimensional hyperbolic
conservation laws, J. Comput. Phys. {\bf 50} (1983) 235-315.

\bibitem{Miller} K. Miller, R. N. Miller,
Moving finite elements.I,  SIAM J. Numer. Anal.  {\bf 18} (1981)
1019-1032.

\bibitem{Dorfi} E. A. Dorfi, T. J. Kaper,
Simple adaptive grids for 1-d initial value problems,
 J. Comput. Phys. {\bf 69}  (1987)  175-195.

\bibitem{Brackbill} J. U. Brackbill, An adaptive grid with directional control,
J. Comput. Phys. {\bf 108}   (1993) 38-50.

\bibitem{Huang} W. M. Cao, W. Z. Huang,  R. D. Russell, An r-adaptive
finite element method based upon moving mesh PDEs, J. Comput. Phys.
{\bf 149} (1999) 221-244.


\bibitem{Stockie} J. M. Stockie, J. A. Mackenzie,  R. D. Russell,
A moving mesh method for onedimensional hyperbolic conservation
laws, SIAM J. Sci. Comput.  {\bf 22}  (2001)  1791-1813.

\bibitem{TaoTang}  H. Z. Tang, T. Tang, Adaptive mesh methods
for one- and two-dimensional hyperbolic conservation laws, SIAM J.
Numer. Anal. {\bf 41} (2003) 487-515.

%\bibitem{LeVeque}
%R. Fazio and R. J. LeVeque, Moving-Mesh Methods for One-Dimensional
%Hyperbolic Problems Using
%    CLAWPACK, Computers and Mathematics with
%Applications {\bf 45} (2003) 273-.


%%%\bibitem{Constantin}
%%%A. Constantin, On the scattering problem for the Camassa-Holm
%%%equation {Proc. R. Soc. London A}, {\bf 457}, 953(2001).


%%%\bibitem{Coclite2}
%%%G. M. Coclite, K. H. Karlsen and N. H. Risebro, Numerical schemes
%%%for computing discontinuous solutions of the Degasperis-Procesi
%%%equation, IMA J. Numer. Anal., {\bf 28}, 80 (2008)
%http://imajna.oxfordjournals.org/cgi/content/abstract/28/1/80

%\bibitem{Johnson2} R.S. Johnson: J. Fluid Mech. {\bf 455} (2002) 63.

\bibitem{LiOlver}
Y. Li, P. Olver, Convergence of solitary wave solutions in a
perturbed bi-Hamiltonian dynamical system. I. Compactons and
peakons, {Discrete Contin. Dynam. Syst. A}, {\bf 3}
    (1997) 419-432.

\bibitem{PKMatsuno}
A. Parker, Y. Matsuno, The Peakon Limits of Soliton Solutions of
theCamassa-Holm Equation, {J. Phys. Soc. Jpn.} {\bf 75} (2006)
124001.

\bibitem{MatsunoPeakon}
Y. Matsuno, The Peakon Limit of the $N$-Soliton Solution of the
Camassa-Holm Equation, {J. Phys. Soc. Jpn.} {\bf 76} (2007) 034003.


\bibitem{Kamchatnov}
A. M. Kamchatnov, R. A. Kraenkel, and B. A. Umarov,
Asymptotic soliton train solution of Kaup-Boussinesq equations,
{Wave Motion} {\bf 38} (2003) 355-365.

\bibitem{El}
G. A. El, R. H. J. Grimshaw, and N. F. Smyth,
Asymptotic description of solitary wave trains in fully nonlinear
shallow-water theory,
{Physica D} {\bf 237} (2008) 2423-2435.

\bibitem{Karpman}
V. I. Karpman,
An asymptotic solution of the Korteweg-de Vries equation,
{Phys. Lett.} {\bf 25A} (1967) 708-709.

\bibitem{SP}
T. Sch\"afer, C. E.  Wayne, Propagation of ultra-short optical pulses
in cubic nonlinear media,
 Physica D {\bf 196}  (2004) 90-105.

\bibitem{DP}
A. Degasperis, M. Procesi, Asymptotic integrability, in:  A.
Degasperis and G. Gaeta (Eds.), {Symmetry and Perturbation Theory}
, World Scientific, Singapore (1999) 23-37.

\bibitem{FMO09}
B.-F. Feng, K. Maruno, Y. Ohta, Integrable discretizations of
the short pulse equation, J. Phys. A {\bf 43} (2010) 085203.
\end{thebibliography}
\end{document}